\documentclass[preprint,journal]{IEEEtran}
\usepackage{graphicx}
\usepackage{subfigure}
\usepackage{amsmath}
\usepackage{color}
\usepackage{cite}
\usepackage{amsmath,amssymb,amsfonts}
\usepackage{algorithmic}
\usepackage{textcomp}
\usepackage{booktabs}
\usepackage{epstopdf}
\usepackage{array}
\usepackage{pgffor}
\usepackage{threeparttable}
\usepackage{multicol}
\usepackage{mathtools}
\usepackage[marginal]{footmisc}
\usepackage{hyperref}
\usepackage{cleveref}

\newcommand{\fig}[1]{\mbox{Figure~\ref{#1}}}
\newcommand{\eq}[1]{\mbox{Eq.~(\ref{#1})}}

\newcommand{\tab}[1]{\mbox{Table~\ref{#1}}}

\graphicspath{{./figure/}}

\begin{document}

\title{Physics-informed Supervised Residual Learning for Electromagnetic Modeling}

\author{\IEEEauthorblockN{Tao Shan,~\IEEEmembership{Member,~IEEE}, Jinhong Zeng, Xiaoqian Song, Rui Guo,~\IEEEmembership{Member,~IEEE},
		Maokun Li,~\IEEEmembership{Senior Member,~IEEE}, Fan Yang,~\IEEEmembership{Fellow,~IEEE},
		and Shenheng Xu,~\IEEEmembership{Member,~IEEE}}
	\thanks{\color{red}This preprint has been published in IEEE Transactions on Antennas and Propagation on 01 March 2023. Please cite the final published version as [T. Shan et al., "Physics-Informed Supervised Residual Learning for Electromagnetic Modeling," in IEEE Transactions on Antennas and Propagation, vol. 71, no. 4, pp. 3393-3407, April 2023, doi: 10.1109/TAP.2023.3245281]. The link is \url{https://ieeexplore.ieee.org/abstract/document/10057211}.}
	\thanks{\color{red}Digital Object Identifier 10.1109/TAP.2023.3245281}
	\thanks{This work was supported in part by the China Postdoctoral Science Foundation under Grant 2022M711764, in part by the National Natural Science Foundation of China under Grant 61971263, the Institute for Precision Medicine, Tsinghua University, the Biren Technology, and the BGP Inc.}
	\thanks{Tao Shan, Jinhong Zeng, Rui Guo, Maokun Li, Fan Yang and Shenheng Xu are with  the State Key Laboratory on Microwave and Digital Communications, Beijing National Research Center for Information Science and Technology (BNRist), Department of Electronic Engineering, Tsinghua University, Beijing 100084, China. (e-mail:maokunli@tsinghua.edu.cn). }
	\thanks{Xiaoqian Song is with National Institute of Metrology, Beijing, 100013, China.}}
\IEEEoverridecommandlockouts
\IEEEpubid{\makebox[\columnwidth]{0018-926X \copyright 2023 IEEE \hfill} \hspace{\columnsep}\makebox[\columnwidth]{ }}
\maketitle
\IEEEpubidadjcol
\begin{abstract}
	In this study, physics-informed supervised residual learning (PhiSRL) is proposed to enable an effective, robust, and general deep learning framework for 2D electromagnetic (EM) modeling.
	Based on the mathematical connection between the fixed-point iteration method and the residual neural network (ResNet), PhiSRL aims to solve a system of linear matrix equations.
	It applies convolutional neural networks (CNNs) to learn updates of the solution with respect to the residuals.
	Inspired by the stationary and non-stationary iterative scheme of the fixed-point iteration method, stationary and non-stationary iterative physics-informed ResNets (SiPhiResNet and NiPhiResNet) are designed to solve the volume integral equation (VIE) of EM scattering.
	The effectiveness and universality of PhiSRL are validated by solving VIE of lossless and lossy scatterers with the mean squared errors (MSEs) converging to $\sim 10^{-4}$ (SiPhiResNet) and $\sim 10^{-7}$ (NiPhiResNet).
	Numerical results further verify the generalization ability of PhiSRL.
\end{abstract}
\IEEEpeerreviewmaketitle

\section{Introduction}
Computational electromagnetics (CEM) is focused on efficient and accurate numerical algorithms to model electromagnetic phenomena governed by Maxwell's equations\cite{jin2011theory}. 
CEM has wide applications in the scientific research and engineering\cite{knott2004radar,poljak2007advanced, jin2009finite, nikolova1999microwave}.
There exist a variety of numerical algorithms including the finite difference method (FDM)\cite{jin2011theory}, the finite element method (FEM)\cite{jin2015finite} and the method of moments (MoM)\cite{harrington1993field,Chew2007}.
A common approach of these algorithms is to discretize and convert Maxwell's equations into a system of linear matrix equations that can be solved by fixed-point iteration method\cite{axelsson1996iterative}, conjugate gradient method\cite{golub1996matrix}, etc.
The matrix equations usually have a large number of unknowns that is computationally expensive.
\par
The computational complexity can be reduced by fast algorithms, such as conjugate gradient-fast Fourier transform\cite{sarkar1986application}, adaptive integral method\cite{bleszynski1996aim}, multilevel fast multipole algorithm\cite{rokhlin1985rapid}, etc.
On the other hand, acceleration can be performed by dividing the computation into offline and online parts, such as the reduced basis method\cite{dang2017quasi}, the characteristic basis function\cite{prakash2003characteristic} and the model-order reduction method\cite{schilders2008model}.
Machine learning (ML) techniques are also applied to speed up the online computation\cite{chedid1996automatic,wang1997knowledge,massa2005classification}.
\par 
\begin{figure}
	\centering
	\subfigure[]
	{\includegraphics[width=0.25\linewidth]{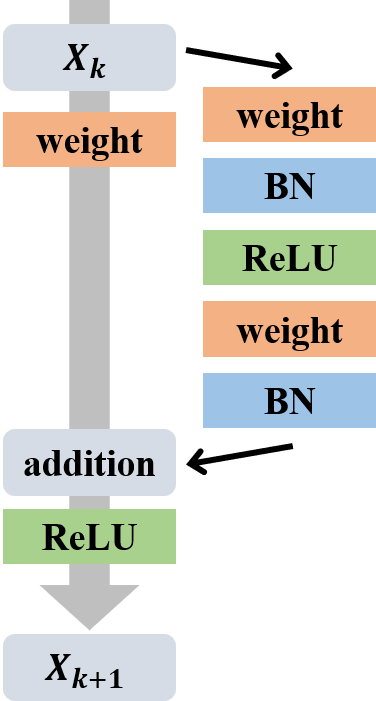}
		\label{resnetblockori}}
	\subfigure[]
	{\includegraphics[width=0.25\linewidth]{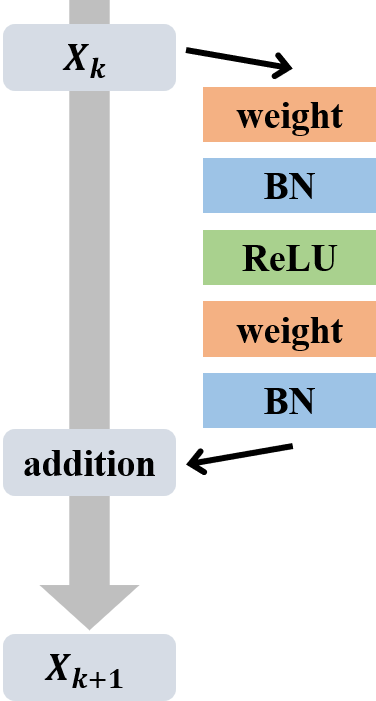}
		\label{resnetblockid}}
	\subfigure[]
	{\includegraphics[width=0.25\linewidth]{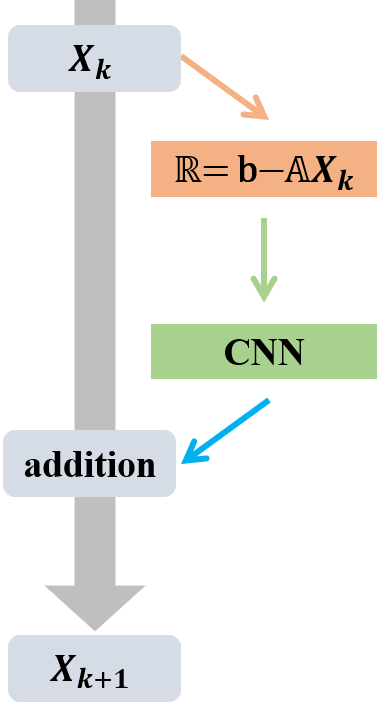}
		\label{residualblock}}
	\caption{Schematics of residual and PhiSRL blocks. (a): the general residual block; (b): the residual block with identity mapping; (c): the proposed PhiSRL block.}
\end{figure}
Recently, deep learning (DL) is developing rapidly\cite{lecun2015deep}. It improves the performance in image\cite{simonyan2014very}, speech\cite{hinton2012deep} and video\cite{ciaparrone2020deep} processing.
It has also been applied to physics and engineering, like intuitive physics\cite{li2016fall, kipf2018neural}, computational fluid dynamics\cite{tompson2017accelerating, xie2018tempogan}, Hamiltonian mechanics\cite{greydanus2019hamiltonian}, simulations of Symplectic integration and Lagrangians\cite{chen2019symplectic, cranmer2020lagrangian}.
In the field of CEM, DL is also studied to improve the computational efficiency by training offline to accelerate online computation\cite{massa2019dnns, yao2018machine, sun2020machine, shan2021application}. 
In EM imaging, various DL-based methods have been proposed to enhance the computational efficiency and accuracy of inversion\cite{salucci2022artificial,shan2022new, wei2018deep,shan2022neural, wei2019dominant,liu2021physical, ma2020learning}.
DL is also integrated to assist the design of microwave devices, such as microwave circuit design\cite{roy2019inverse}, synthesis of array antennas\cite{shan2021phase}, coding programmable metasurfaces\cite{8988246}, metasurface imager\cite{li2019machine}, etc.
Despite the recent progress of applying DL to CEM, it is still challenging to design effective deep neural network (DNN) architectures for EM computation.
Most works are fully data-driven and considering DNN as "black-box" approximators without the inner reasoning  \cite{chen2020review, massa2019dnns}. 
The architectures of DNNs are determined by either hyperparameter optimization\cite{hernandez2016general} or trial and error method, which is practical but unexplainable and costly\cite{hernandez2016general, ruthotto2019deep}.
\par
Research has been reported to interpret DNNs by relating them to the theory of ordinary differential equations (ODEs) or partial differential equations (PDEs)\cite{weinan2017proposal,haber2017stable}.
On one hand, DL can help reduce the curse of dimensionalities in solving PDEs, such as Schr\"odinger equation\cite{hermann2020deep}, Navier-stokes equation\cite{raissi2018hidden}, Poisson's equation\cite{9062556}, Burger's equation\cite{sirignano2018dgm}, etc.
The automatic differentiation of DL can also be used to solve PDEs by embedding them into the loss of DNNs\cite{raissi2019physics, lu2021deepxde}.
On the other hand, it allows to better understand the inner reasoning of DNNs inspired by PDEs and corresponding numerical algorithms\cite{lu2018beyond, long2018pde, ruthotto2019deep}.
The design of DNNs is guided by the numerical methods of PDEs to obtain improved performance and generalization ability, such as multi-grid method\cite{greenfeld2019learning, he2019mgnet}, hierarchical matrices\cite{fan2019multiscale}, Runge-Kutta method\cite{zhu2019convolutional}, etc.
The properties of DNNs can be analyzed by PDE theory\cite{weinan2017proposal,li2017deep} and the PDE interpretation of DNNs further bridges the gap between DL and PDEs\cite{lu2018beyond,ruthotto2019deep}.
\par
In this work, we propose the physics-informed supervised residual learning (PhiSRL) that enables an effective, robust, and general framework for solving EM wave equations.
It is based on the ResNet\cite{he2016deep}.
The mathematical link between the fixed-point iteration method and ResNet is  investigated.
PhiSRL is designed based on this link.
In each iteration of PhiSRL, a parameterized function based on CNN is learned to modify the candidate solution to minimize the calculated residual.
PhiSRL can be considered as a learned iterative solver of matrix equations and can also be expanded to solve other PDEs. 
Inspired by the stationary and non-stationary iterative schemes of the fixed-point iteration method, we design two different neural network architectures of PhiSRL including  stationary and non-stationary iterative physics-informed ResNets (SiPhiResNet and NiPhiResNet).
They are validated by solving EM scattering by 2D dielectric scatterers and achieve good computing precisions.
In case of lossless scatterers, the MSEs of SiPhiResNet and NiPhiResNet converge to $3.152 \times 10^{-4}$ and $4.8925 \times 10^{-7}$. For lossy scatterers, the MSEs converge to $1.2775 \times 10^{-4}$ and $1.567 \times 10^{-7}$.
Compared to MoM, by dividing the entire computation into offline and online parts, PhiSRL demonstrates a significant reduction in online computing time after offline training.
Meanwhile, the high-performance computing platform GPU enables massively parallel computation of PhiSRL, which guarantees its computational efficiency.
Furthermore, numerical results demonstrate that PhiSRL can generalize to solve EM scattering by a wide variety of geometries and frequencies different from the ones in the training data set.
The preliminary results have been reported in \cite{shan2021physics}.
\par 
This paper is organized as follows. Section \uppercase\expandafter{\romannumeral2} investigates the mathematical connections between ResNet and fixed-point iteration method.
Section \uppercase\expandafter{\romannumeral3} introduces the volume integral equation and the corresponding numerical methods including MoM and PhiSRL. 
In Section \uppercase\expandafter{\romannumeral4},  the universality of PhiSRL is validated by applying SiPhiResNet and NiPhiResNet to solve VIEs. Furthermore, the generalization abilities are verified.
Observations and discussions are summarized in Section \uppercase\expandafter{\romannumeral5}.
\begin{figure*}
	\centering
	\includegraphics[width=0.7\linewidth]{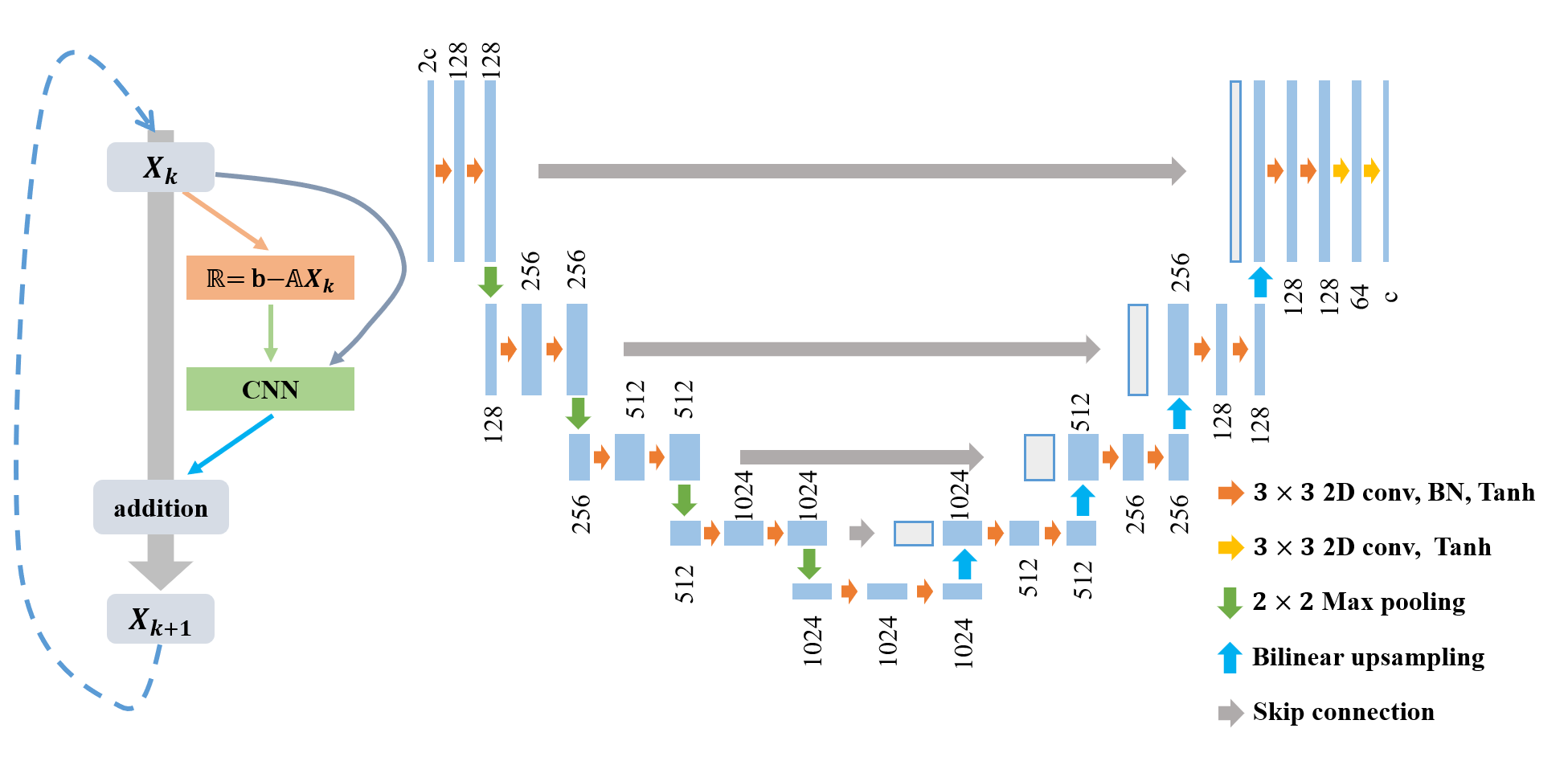}
	\caption{Schematics of stationary iterative physics-informed residual neural network. The U-net is employed recurrently as the CNN for predictions in each iteration.}
	\label{stationaryNN}
\end{figure*}
\begin{figure*}
	\centering
	\includegraphics[width=0.7\linewidth]{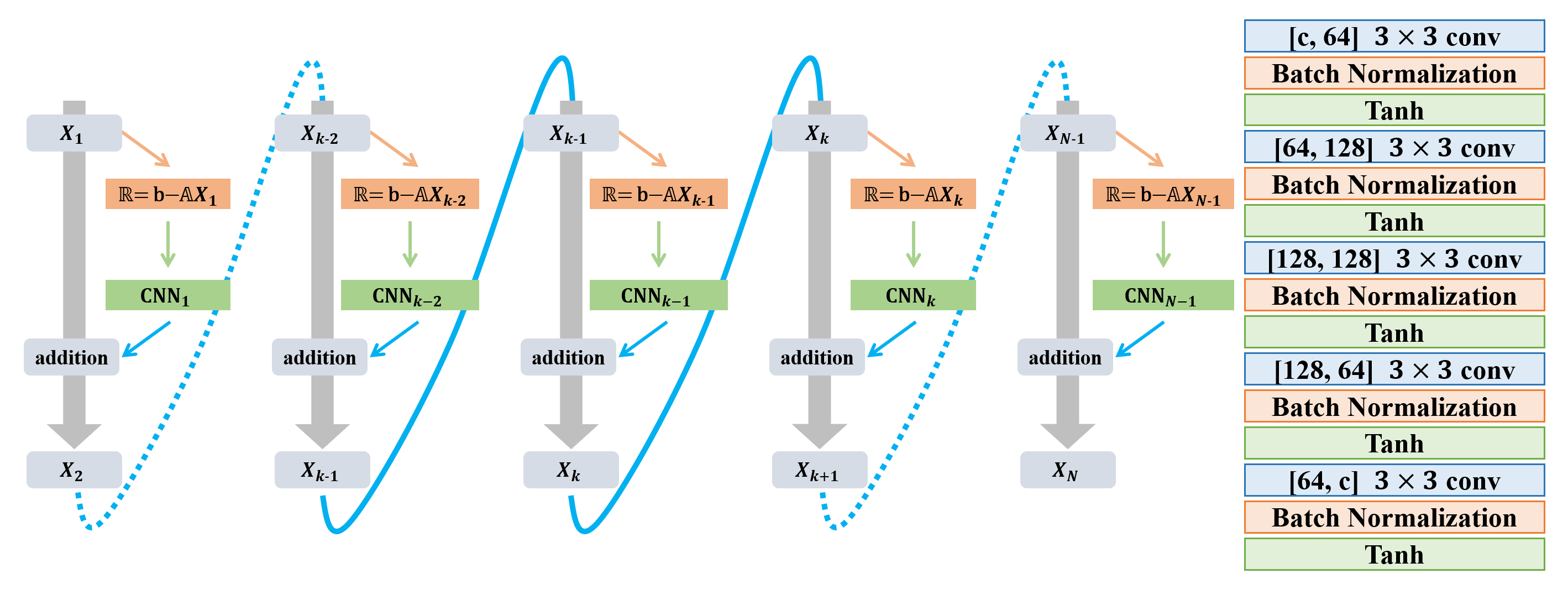}
	\caption{Schematics of non-stationary iterative physics-informed residual neural network. The applied CNN of each iteration is also depicted, which shares the same structure but has independent parameters.}
	\label{nonstationaryNN}
\end{figure*}
\section{Physics-informed Supervised Residual Learning}
The equations in EM modeling can be expressed as:
\begin{equation}
	(\Phi_{\vartheta} u)(x) = f(x), \ x\in D \,,
	\label{eq00}
\end{equation}
where $u: D \to \mathbb{R}$ denotes the field distribution, $x$ is the coordinate defined in $D$, $f: D \to \mathbb{R}$ is a function represents the excitation, $\Phi$ represents integral or differential operator, and $\vartheta: D \to \mathbb{R}$ denotes parameters in the definition of $\Phi$.
Boundary conditions should be included to uniquely determine the solution $u$:
\begin{equation}
	(\Upsilon u) (x) = h(x), \  x \in \partial D \,,
	\label{eq01}
\end{equation}
where $\Upsilon$ denotes an operator on the boundary and $h$ is the value on the boundary.
In EM modeling, \eq{eq00} and \eq{eq01} are converted into a linear system of matrix equations:
\begin{equation}
	\mathbb{A}(\vartheta, \Phi, \Upsilon) \mathrm{x}=\mathrm{b} \,.
	\label{eq1}
\end{equation}
\eq{eq1} can be solved by direct matrix inversion, i.e, LU decomposition or iterative methods such as the conjugate gradient
method \cite{golub1996matrix}. Here we focus on the fixed-point iteration method\cite{axelsson1996iterative}.
\subsection{Fixed-point Iteration Method}
The fixed-point iteration method can be expressed as\cite{golub1996matrix}:
	\begin{equation}
		\mathrm{x}^{a}_{k+1} = \mathrm{x}^{a}_{k} + 
		\mathcal{L} (\mathrm{b} - \mathbb{A}(\vartheta, \Phi, \Upsilon)\mathrm{x}^{a}_{k}, {\mathbb{A}^{a}(\vartheta, \Phi, \Upsilon)}^{-1}) \,,
		\label{eq7}
	\end{equation}
	where $\mathbb{A}^{a}(\vartheta, \Phi, \Upsilon)$ is an approximation of $\mathbb{A}(\vartheta, \Phi, \Upsilon)$ in \eq{eq1}, $\mathrm{x}^{a}_{k}$ denotes the $k$-th candidate solution, and the operator $\mathcal{L}$ denotes the matrix-vector multiplication ${\mathbb{A}^{a}(\vartheta, \Phi, \Upsilon)}^{-1}(\mathrm{b} - \mathbb{A}(\vartheta, \Phi, \Upsilon)\mathrm{x}^{a}_{k})$.
	\eq{eq7} is the stationary iterative scheme of the fixed-pointed iteration method where $\mathbb{A}^{a}(\vartheta, \Phi, \Upsilon)$ is kept the same at each iteration.
	The non-stationary iterative scheme can be written as:
	\begin{equation}
		\mathrm{x}^{a}_{k+1} = \mathrm{x}^{a}_{k} + 
		\mathcal{L} (\mathrm{b} - \mathbb{A}(\vartheta, \Phi, \Upsilon)\mathrm{x}^{a}_{k}, {\mathbb{A}^{a}_k(\vartheta, \Phi, \Upsilon)}^{-1}) \,.
		\label{eq8}
	\end{equation}
	Two widely used iterative methods are Richardson and Jacobi methods. Both of them can be written in the form of \eq{eq7}\cite{axelsson1996iterative}.
\subsection{Physics-informed Supervised Residual Learning}
ResNet is an effective neural network architectures in image processing\cite{he2016deep}.
It is a stacked structure composed of a family of modular blocks named as residual blocks\cite{he2016deep}.
The residual blocks enable ResNet to be easily extended to deep architectures.
ResNet have been analyzed based on the theory of ODEs/PDEs and the corresponding numerical algorithms\cite{chang2018reversible}, such as forward Euler discretization of ODEs\cite{weinan2017proposal, haber2017stable}, transport equation of control problem\cite{li2017deep}, dynamical systems\cite{chang2017multi}.
\par 
The typical structure of a single residual block is depicted in \fig{resnetblockori} that can be expressed as\cite{he2016deep}:
\begin{align}
	\mathrm{y}_{k} &= h(\mathrm{x}_{k})+\mathcal{F}(\mathrm{x}_{k},\mathcal{W}_{k}) \label{eq9} \,, \\
	\mathrm{x}_{k+1} &= \mathcal{N}(\mathrm{y}_{k}) \,, \label{eq10}
\end{align}
where $\mathrm{x}_{k}$ and $\mathrm{x}_{k+1}$ are input and output of the $k$-th residual block; $h$ denotes the linear projection; $\mathcal{N}$ denotes the nonlinear activation; $\mathcal{F}$ denotes the transform function of the residual and $\mathcal{W}_{k}$ is the parameter set of $\mathcal{F}$.
It is pointed out in \cite{he2016identity} that ResNet can achieve better performance and improved generalization ability by taking  $h$ and $\mathcal{N}$ as the identity mappings.
As shown in \fig{resnetblockid}, \eq{eq9} and \eq{eq10} can be combined as:
\begin{equation}
	\mathrm{x}_{k+1} = \mathrm{x}_{k}+\mathcal{F}(\mathrm{x}_k,\mathcal{W}_k) \,.
	\label{eq11}
\end{equation}
\par
It can be observed that the identity residual block \eq{eq11} has a similar update equation as the fixed-point iteration method \cref{eq8,eq9}.
Motivated by this link, PhiSRL is proposed by embedding the fixed-point iteration method into the identity residual block. The stationary scheme can be written as:
\begin{equation}
	\mathrm{x}_{k+1} = \mathrm{x}_{k}+\Psi^{Si}(\mathrm{b} - \mathbb{A}(\vartheta, \Phi, \Upsilon) \mathrm{x}_{k},\Theta) \,,
	\label{eq12}
\end{equation}
and the nonstationary one is:
\begin{equation}
	\mathrm{x}_{k+1} = \mathrm{x}_{k}+\Psi^{Ni}(\mathrm{b} - \mathbb{A}(\vartheta, \Phi, \Upsilon) \mathrm{x}_{k},\Theta_k) \,,
	\label{eq13}
\end{equation}
where $\Psi$ and $\Theta$ denote the CNNs and the corresponding parameters.
\fig{residualblock} illustrates the structure of a PhiSRL block. 
As shown in \fig{residualblock}, the residual $\mathbb{R}_k$ is calculated first regarding the input $\mathrm{x}_{k}$:
\begin{equation}
	\mathbb{R}_k = \mathrm{b} - \mathbb{A}(\vartheta, \Phi, \Upsilon) \mathrm{x}_{k} \,.
	\label{eq14}
\end{equation}
In each iteration of PhiSRL, CNNs are trained to predict the modification of the candidate solution with the calculated residual as input.
The training process enables CNNs in PhiSRL to learn effective updates of the solution.
\par
\subsection{Stationary Iterative Physics-informed ResNets}
SiPhiResNet is designed based on the stationary iterative scheme of the fixed-point iteration method. 
Formulated as \eq{eq12}, the CNNs of SiPhiResNet share the same structure and parameter set, which can be considered as the recurrent neural network (RNN).
The update equation of SiPhiResNet is (as shown in \fig{stationaryNN}):
\begin{equation}
	\begin{split}
		\mathbb{R}_k &= \mathrm{b} - \mathbb{A}(\vartheta, \Phi, \Upsilon) \mathrm{x}_{k} \,, \\
		\Delta_k &= \Psi^{Si}(\mathbb{R}_k \oplus \mathrm{x}_{k},\Theta)\,, \\
		\mathrm{x}_{k+1} &= \mathrm{x}_{k}+ \Delta_k \,,
		\label{eqsi}
	\end{split}
\end{equation}
where $\oplus$ denotes the concatenation operation, $x_k$, $\mathbb{R}_k$ and $\Delta_k$ denote the candidate solution, the corresponding residual and modification.
SiPhiResNet requires the shared CNNs to possess strong learning capacities to handle the different mappings between residuals and modifications in every iteration.

$\Psi^{Si}$ adopts U-Net\cite{ronneberger2015u} as shown in \fig{stationaryNN}.
U-Net is an effective network in image segmentation with a good learning ability\cite{ronneberger2015u}.
U-Net has almost the same computing process with the V-cycle scheme of the multi-grid method\cite{he2019mgnet}.
It is noted that U-Net shares the same parameter set through the iterative process.
With the RNN-like structure, the output $\mathrm{x}_{k}$ is also concatenated with the residual $\mathbb{R}_k$ as input to the next iteration. Here, $\mathrm{x}_{k}$ plays a role of the hidden state that is widely applied in RNNs.
Thus, the channels of input and output can be denoted as $2c$ and $c$. The value of $c$ depends on the linear equation systems, for example, $c$ will be $1$ in the real-valued equations and $2$ in the complex-valued equations.
\subsection{Non-stationary Iterative Physics-informed ResNets}
Inspired by the non-stationary scheme of the fixed-point iteration method, NiPhiResNet is designed by using independent CNNs to update the candidate solution in each iteration. 
The CNNs share the same structure but owns independent and different parameter sets.
It can be considered as the unrolled structure of the SiPhiResNet, as shown in \fig{nonstationaryNN}.
According to \eq{eq13}, the specific iteration update equation of NiPhiResNet can be written as:
\begin{equation}
	\begin{split}
		\mathbb{R}_k &= \mathrm{b} - \mathbb{A}(\vartheta, \Phi, \Upsilon) \mathrm{x}_{k} \,, \\
		\Delta_k &= \Psi^{Ni}(\mathbb{R}_k,\Theta_k) \,, \\
		\mathrm{x}_{k+1} &= \mathrm{x}_{k}+ \Delta_k \,.
		\label{eqni}
	\end{split}
\end{equation}

With $\mathbb{R}_k$ as the input of the $k^{th}$-iteration CNN, the computation can be described as:
\begin{equation}
	\Psi^{Ni}(\cdot,\Theta_k) = (\sigma_H^k \circ \tau_H^k \circ \kappa_H^k \circ \cdots \circ \sigma_{1}^k \circ \tau_1^k \circ \kappa_{1}^k)(\cdot) \,,
\end{equation}
where  $\sigma$, $\tau$, $\kappa$ denote the tanh nonlinear activation, batch normalization and $3 \times 3$ convolution, $H$ is the number of layers of the CNN and $H$ is set as 5 in this work. 
The independent CNN of each iteration tackles the learning task of a single iteration.
Therefore, the CNN does not require a complicated structure.
The specific hyperparameters of the CNN are mostly determined by author’s experience and trials. 
It should be emphasized that the nonlinearity function needs to be capable of providing both positive and negative values.
The output has the same size as the input due to the fully convolutional operations.
\subsection{Stability Analysis of Physics-informed ResNets}
Both SiPhiResNet and NiPhiResNet are built on top of ResNets and they inherit the properties of ResNet. 
The stability analysis of ResNet has been investigated in \cite{he2016identity} and a similar analysis can be applied to SiPhiResNet and NiPhiResNet.
Here, we take NiPhiResNet as an example. 
With the update equation of NiPhiResNet described as \eq{eqni}, we can define the loss function $ \mathcal{E} $ as MSE and $ \mathcal{E} $  can be written as:
\begin{equation}
	\mathcal{E}  = \frac{1}{N}|| \mathrm{x}_{L} - \mathrm{x}^{*}||_{F}^2 \;,
\end{equation}
where  $\mathrm{x}^{*}$ and $\mathrm{x}_{L}$ are ground truth and the prediction at the $L$-th iteration, $||   \cdot  ||_F$ is the Frobenius norm.
Then, based on the chain rule, the back propagation of $ \mathcal{E} $ with respect to the $l$-th input can be written as\cite{he2016identity}:
\begin{equation}
	\begin{aligned}
		\frac{\partial \mathcal{E}}{\partial \mathrm{x}_l} &= \frac{\partial \mathcal{E}}{\partial \mathrm{x}_L}\frac{\partial \mathrm{x}_L}{\mathrm{x}_{L-1}} \cdots \frac{\partial \mathrm{x}_{l+1}}{\mathrm{x}_{l}} \\
		& = \frac{\partial \mathcal{E}}{\partial \mathrm{x}_L} \prod_{i=l}^{L-1} (1 + \frac{\partial \Psi^{Ni}(\mathrm{b} - \mathbb{A}(\vartheta, \Phi, \Upsilon) \mathrm{x}_{i},\Theta_i)}{\partial \mathrm{x}_{i}}) \;,
	\end{aligned}
	\label{chain}
\end{equation}
where $\mathrm{x}_l$ is the input of the $l$-th iteration.
By removing the residual connections, \eq{eqni} can be expressed as:
\begin{equation}
	\mathrm{x}_{k+1} = \Psi^{Ni}(\mathrm{b} - \mathbb{A}(\vartheta, \Phi, \Upsilon) \mathrm{x}_{k},\Theta_k) \,,
\end{equation}
Then the back propagation of $ \mathcal{E} $ can be re-written as:
\begin{equation}
	\frac{\partial \mathcal{E}}{\partial \mathrm{x}_l} = \frac{\partial \mathcal{E}}{\partial \mathrm{x}_L} \prod_{i=l}^{L-1} \frac{\partial \Psi^{Ni}(\mathrm{b} - \mathbb{A}(\vartheta, \Phi, \Upsilon) \mathrm{x}_{i},\Theta_i)}{\partial \mathrm{x}_{i}}
\end{equation}
For gradient vanishing, the following assumption can be made:
\begin{equation}
	\lim_{i \to \infty} \frac{\partial\Psi^{Ni}(\mathrm{b} - \mathbb{A}(\vartheta, \Phi, \Upsilon) \mathrm{x}_{i},\Theta_i))}{\mathrm{x}_{i}} = 0
	\label{assume}
\end{equation}
Adding the residual connections convert \eq{assume} into:
\begin{equation}
	\lim_{i \to \infty} 1+ \frac{\partial\Psi^{Ni}(\mathrm{b} - \mathbb{A}(\vartheta, \Phi, \Upsilon) \mathrm{x}_{i},\Theta_i))}{\mathrm{x}_{i}} = 1
\end{equation}
Then, the limit $\lim_{L \to \infty} \prod_{i=l}^{L-1} (1 + \frac{\partial \Psi^{Ni}(\mathrm{b} - \mathbb{A}(\vartheta, \Phi, \Upsilon) \mathrm{x}_{i},\Theta_i)}{\partial \mathrm{x}_{i}})$ converges. Since $\frac{\partial \mathcal{E}}{\partial \mathrm{x}_L}$ is finite, we can conclude that $\lim_{L \to \infty}\frac{\partial \mathcal{E}}{\partial \mathrm{x}_l}$ converges to a non-zero value according to \eq{chain}. This demonstrates that the residual connections can alleviate the problem of gradient vanishing and improve the training stability\cite{he2016identity}.
The same conclusion can be derived for SiPhiResNet as well.
\section{Volume Integral Equation}
\begin{figure}
	\centering
	\includegraphics[width=0.5\linewidth]{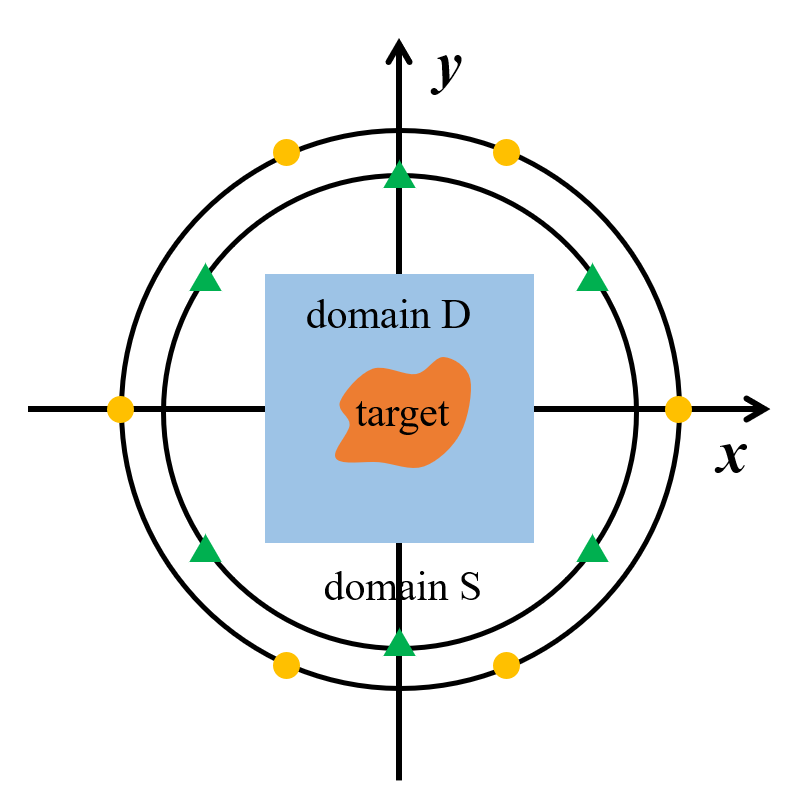}
	\caption{The model setup for volume integral equation. The target is in domain D. Green triangles and yellow squares denote the receivers and transmitters in domain S.}
	\label{viemodel}
\end{figure}
\begin{figure}
	\centering
	\includegraphics[width=0.95\linewidth]{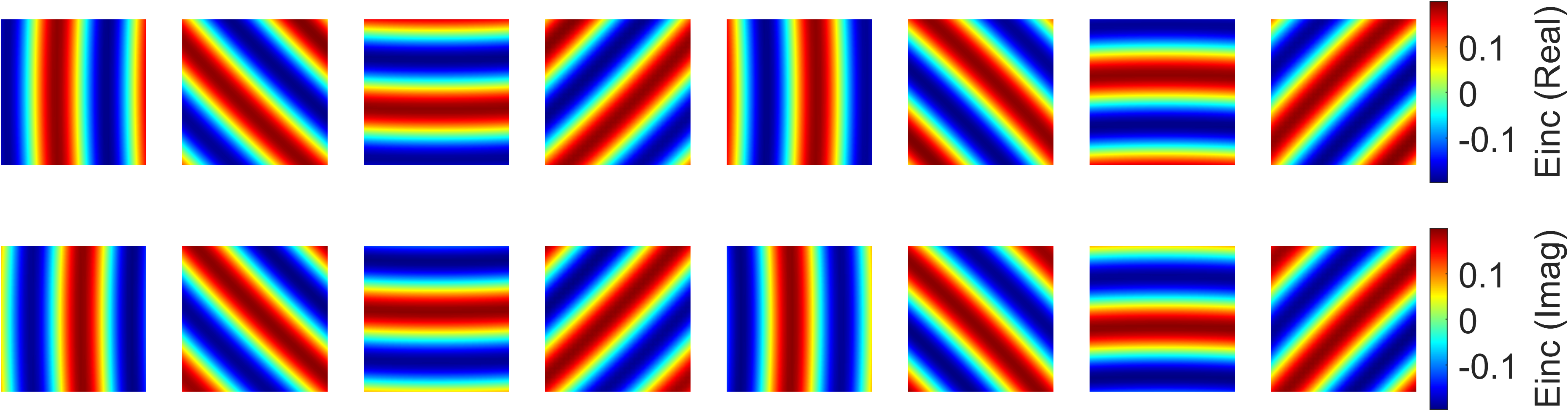}
	\caption{Incident fields in the case of lossless scatterers: the first row is the real part and the second is the imaginary part, the incident angles from left to right are  $[0^{\circ}, 45^{\circ}, 90^{\circ}, 135^{\circ}, 180^{\circ}, 225^{\circ}, 270^{\circ}, 315^{\circ}]$.}
	\label{nonlossypinc}
\end{figure}
\begin{figure}
	\centering
	\includegraphics[width=0.9\linewidth]{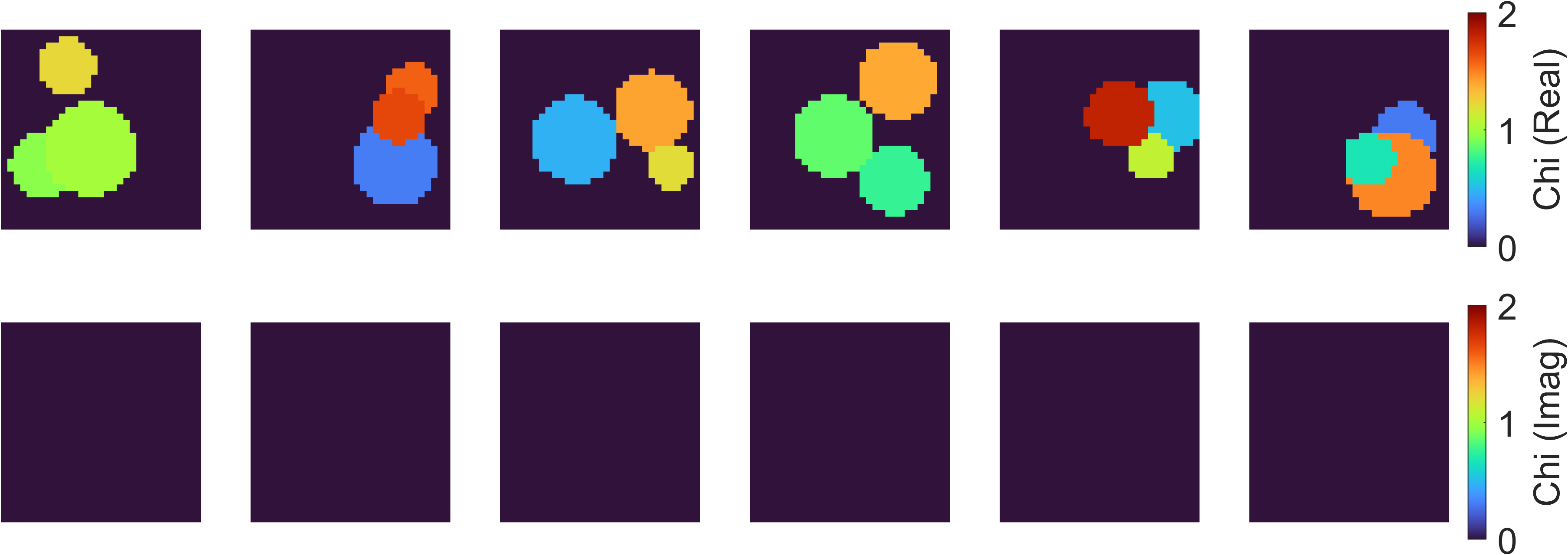}
	\caption{Contrast examples in the case of lossless scatterers: the first row is the real part and the second is the imaginary part.}
	\label{nonlossychi}
\end{figure}
The effectiveness and universality of PhiSRL are verified by solving the VIE\cite{jin2011theory, chen2018computational}.
VIE computes the scattered field of the dielectric scatterers in free space $D$, as shown in \fig{viemodel}. The electrical properties of dielectric scatterers are assumed to only vary along the lateral axes.
The total electric field $E^{tot}(\mathbf{r})$ is induced when the dielectric scatterer is illuminated by the incident field $E^{inc}(\mathbf{r})$.
Taking transverse magnetic (TM) mode into account, the relationship between $E^{tot}(\mathbf{r})$ and $E^{inc}(\mathbf{r})$ satisfies:
\begin{equation}
	\begin{aligned}
		E^{tot}(\mathbf{r})=&E^{inc}(\mathbf{r})\\
		&+k_b^2\int_{D} \mathbf{G}_D \left(\mathbf{r}, \mathbf{r}^{\prime}\right) \chi\left(\mathbf{r}^{\prime}\right) E^{tot}\left(\mathbf{r}^{\prime}\right) d\mathbf{r}^{\prime}, \, \mathbf{r} \in D 
		\label{eq29} 
	\end{aligned}
\end{equation}
where $k_b^2 = \omega \mu_0 \epsilon_0$ is the wavenumber, $\mathbf{G}_D$ is the Green's function defined in $D$\cite{jin2011theory},
and $\chi\left(\mathbf{r}\right)$ is the contrast:
\begin{equation}
	\chi\left(\mathbf{r}\right) = \frac{\varepsilon(\mathbf{r}) - \varepsilon_0}{\varepsilon_0} = \varepsilon_r(\mathbf{r}) -1 - j\frac{\sigma(\mathbf{r})}{\omega \varepsilon_0} \,.
	\label{eq31}
\end{equation}
where $\varepsilon_0$, $\varepsilon_r(\mathbf{r})$, $\sigma(\mathbf{r})$, $\omega$, $\mathbf{r}$ denote vacuum permittivity, relative permittivity, conductivity, angular frequency and the position vector in free space $D$ respectively.
The scattered field can be calculated by:
\begin{equation}
	E^{sca}(\mathbf{r}^{\prime \prime})=\int_{D} \mathbf{G}_S \left(\mathbf{r}^{\prime \prime}, \mathbf{r}^{\prime}\right) \chi\left(\mathbf{r}^{\prime}\right) E^{tot}\left(\mathbf{r}^{\prime}\right) d\mathbf{r}^{\prime}, \, \mathbf{r^{\prime \prime}} \in S 
	\label{eq33}
\end{equation}
where $ \mathbf{r}^{\prime \prime} $ is the position vector of the receiver and $\mathbf{G}_S$ is the Green's function defined in the observation domain $S$\cite{jin2011theory}. 
\subsection{Method of Moments}
To determine the scattered field $E^{sca}$ with respect to the incident field $E^{inc}$, \eq{eq29} needs to be solved by MoM for the total field $E^{tot}$ in $D$.
By discretizing domain $D$ into $M \times N$ subdomains, \eq{eq29} can be written as\cite{jin2011theory, chen2018computational}:
\begin{equation}
	\begin{aligned}
		&E^{tot}(\mathbf{r}_{p})=E^{inc}(\mathbf{r}_{p}) \\
		&+\frac{k_b^2}{4j}  \sum_{q=1}^{M \times N} \chi\left(\mathbf{r}_{q}\right) E^{tot}\left(\mathbf{r}_{q}\right) \int_{D_q} H_0^{(2)}(k_b|\mathbf{r}_{p} - \mathbf{r}_{q}^{\prime}|) d\mathbf{r}_{q}^{\prime} \,,
		\label{eq34}
	\end{aligned}
\end{equation}
where $p$ and $q$ are the indices of subdomains in $D$, $H_0^{(2)}$ is the Hankel function of the second kind.
Then \eq{eq34} can be converted as a linear equation system:
\begin{equation}
	(\mathbf{I} - \mathbb{G}_D \cdot \boldsymbol{\chi}) \cdot \mathbf{E}^{tot}  = \mathbf{E}^{inc} \,,
	\label{eq36}
\end{equation}
where $\mathbf{I}$ is an identity matrix and $\mathbb{G}_D$ is a matrix of size $MN \times MN$.
\subsection{Physics-informed Supervised Residual Learning}
SiPhiResNet is first applied to solve \eq{eq36}. 
The $k$-th update equation of SiPhiResNet refers to \eq{eqsi}. Then, NiPhiResNet is applied to solve \eq{eq36} and its $k$-th update equation follows \eq{eqni}.
It is noted that the calculation of the residual is based on \eq{eq36}:
\begin{equation}
	\mathbb{R}_k = \mathbf{E}^{inc}-	(\mathbf{I} - \mathbb{G}_D \cdot \boldsymbol{\chi}) \cdot \mathbf{E}^{tot}_k \,. 
	\label{eq37}
\end{equation}
In this paper, the $\mathbf{E}^{inc}$ is taken as the initial guess of PhiSRL and it can be regarded as the input of PhiSRL. The $\mathbb{G}_D$ and $\boldsymbol{\chi}$ are needed for the evaluation of $\mathbb{R}_k$ in \eq{eq37}. They can be viewed as the components of PhiSRL. 
The objective function of them is the MSE that can be written as:
\begin{equation}
	\mathcal{E}  = \frac{1}{N}|| \mathbf{E}_{tot}^{\prime} - \mathbf{E}_{tot}^{*}||_{F}^2\,,
	\label{obj}
\end{equation}
where $ \mathbf{E}_{tot}^{\prime}$ is the total field computed by PhiSRL, $ \mathbf{E}_{tot}^{*}$ is the total field by MoM, $||   \cdot  ||_F$ is the Frobenius norm and $N$ is the number of elements in the matrix of $ \mathbf{E}_{tot}$.
\begin{figure}
	\centering
	\includegraphics[width=0.75\linewidth]{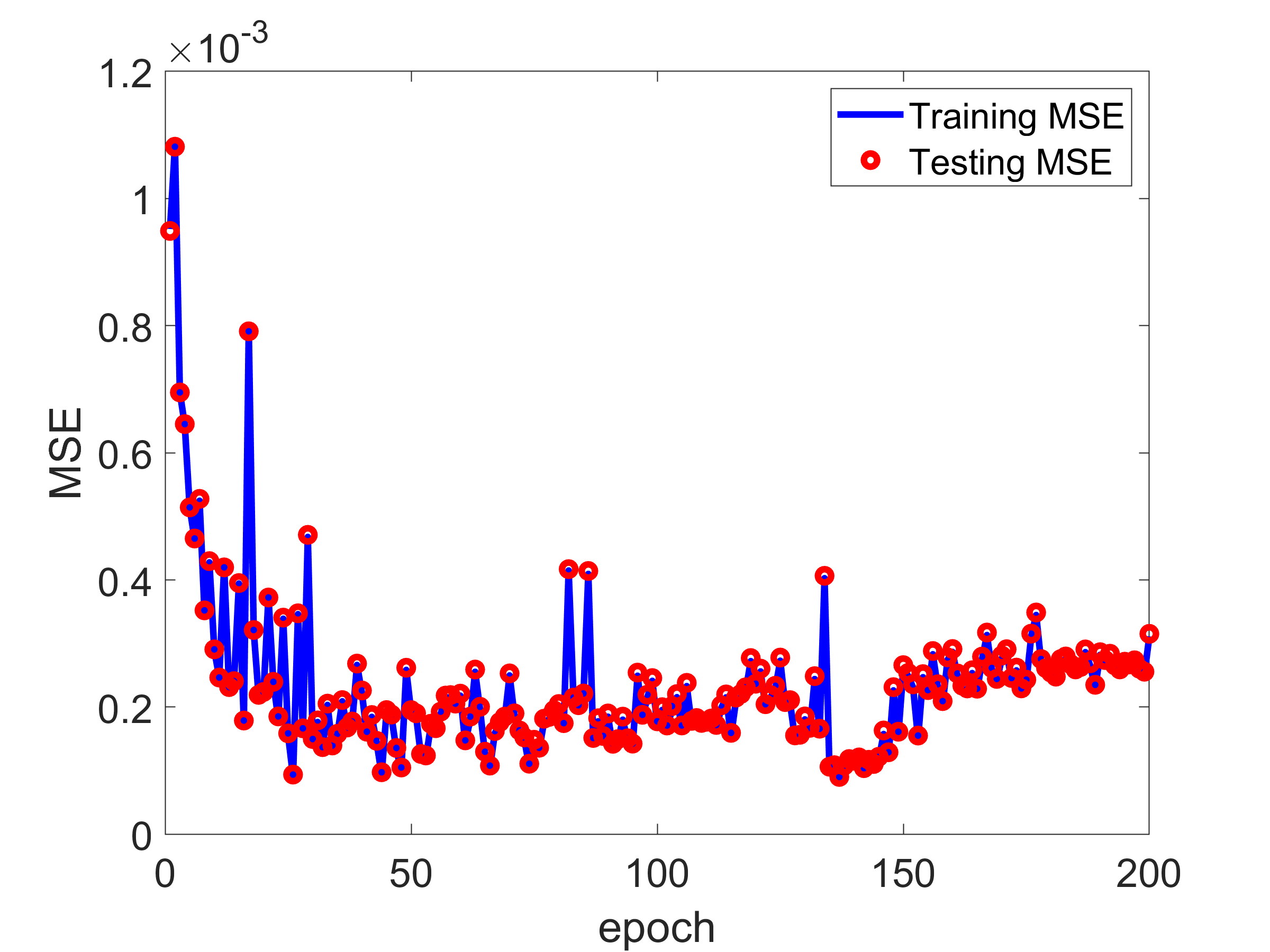}
	\caption{MSE convergence curve of SiPhiResNet for solving VIEs of lossless scatterers. }
	\label{sipisrlnolossy-loss}
\end{figure}
\begin{figure}
	\centering
	\subfigure[ ]{\includegraphics[width=0.99\linewidth]{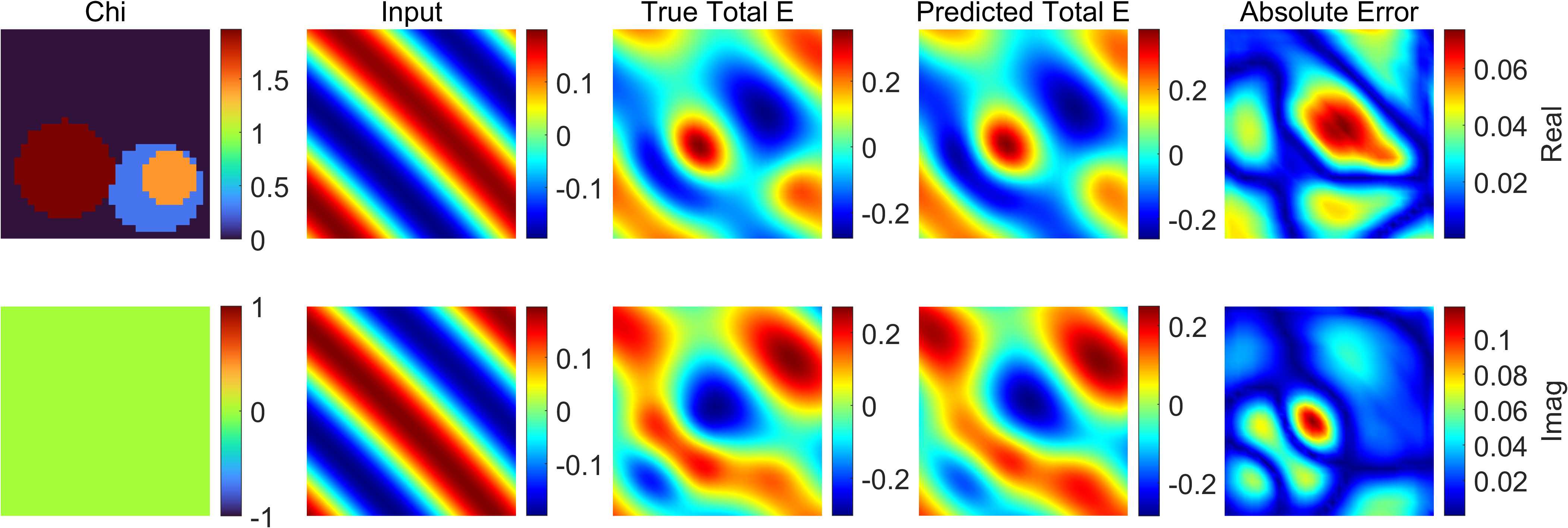}}
	\subfigure[ ]{\includegraphics[width=0.99\linewidth]{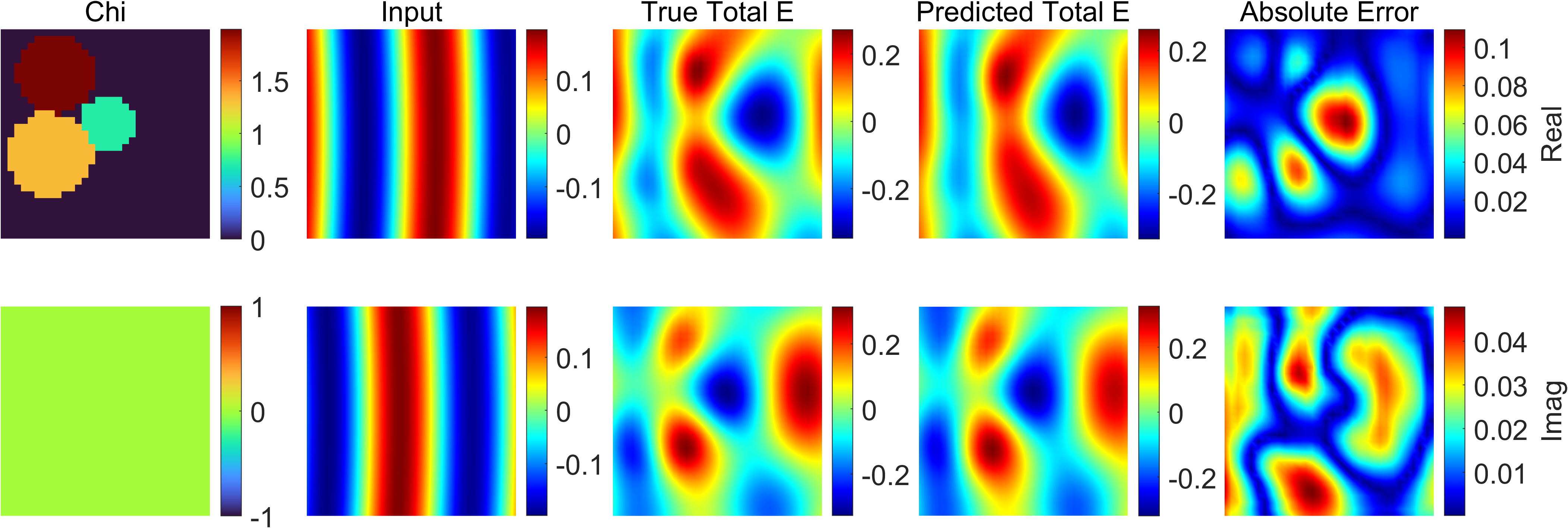}}
	\caption{Lossless scatterer cases: comparisons of total field computed by SiPhiResNet and MoM. (a) and (b) are total fields of VIEs of overlapped and separate cylinders. From left to right: contrast $\chi$, input of SiPhiResNet (initial guess, $E^{inc}$), total field computed by MoM $E^{tot}_{MoM}$, total field computed by SiPhiResNet $E^{tot}_{Si}$, and their absolute error distribution. The first row is the real part and the second is the imaginary part.}
	\label{sipisrlnolossy}
\end{figure}
\begin{figure}
	\centering
	\subfigure[]
	{\includegraphics[width=0.99\linewidth]{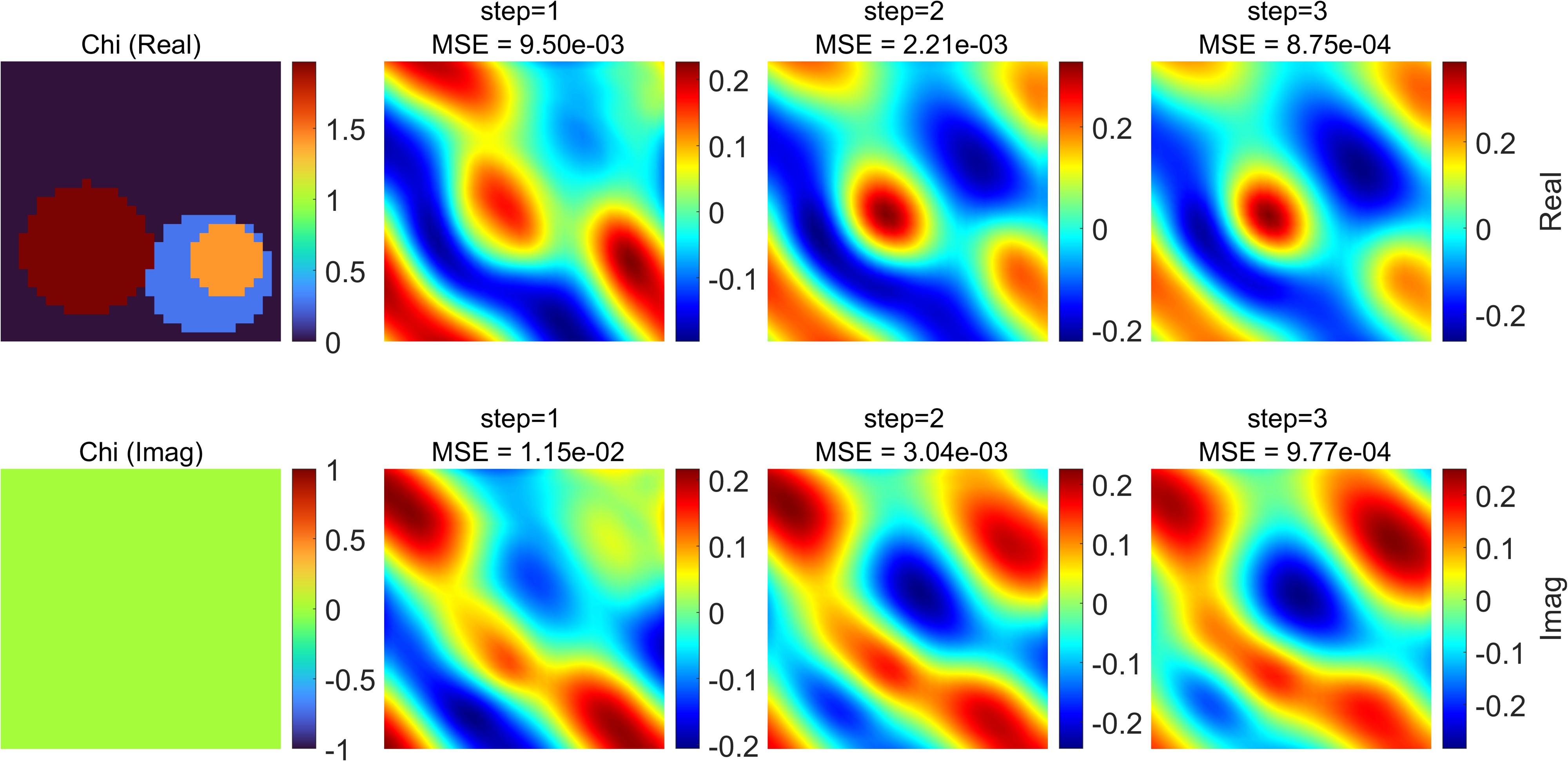}}
	\subfigure[]
	{\includegraphics[width=0.99\linewidth]{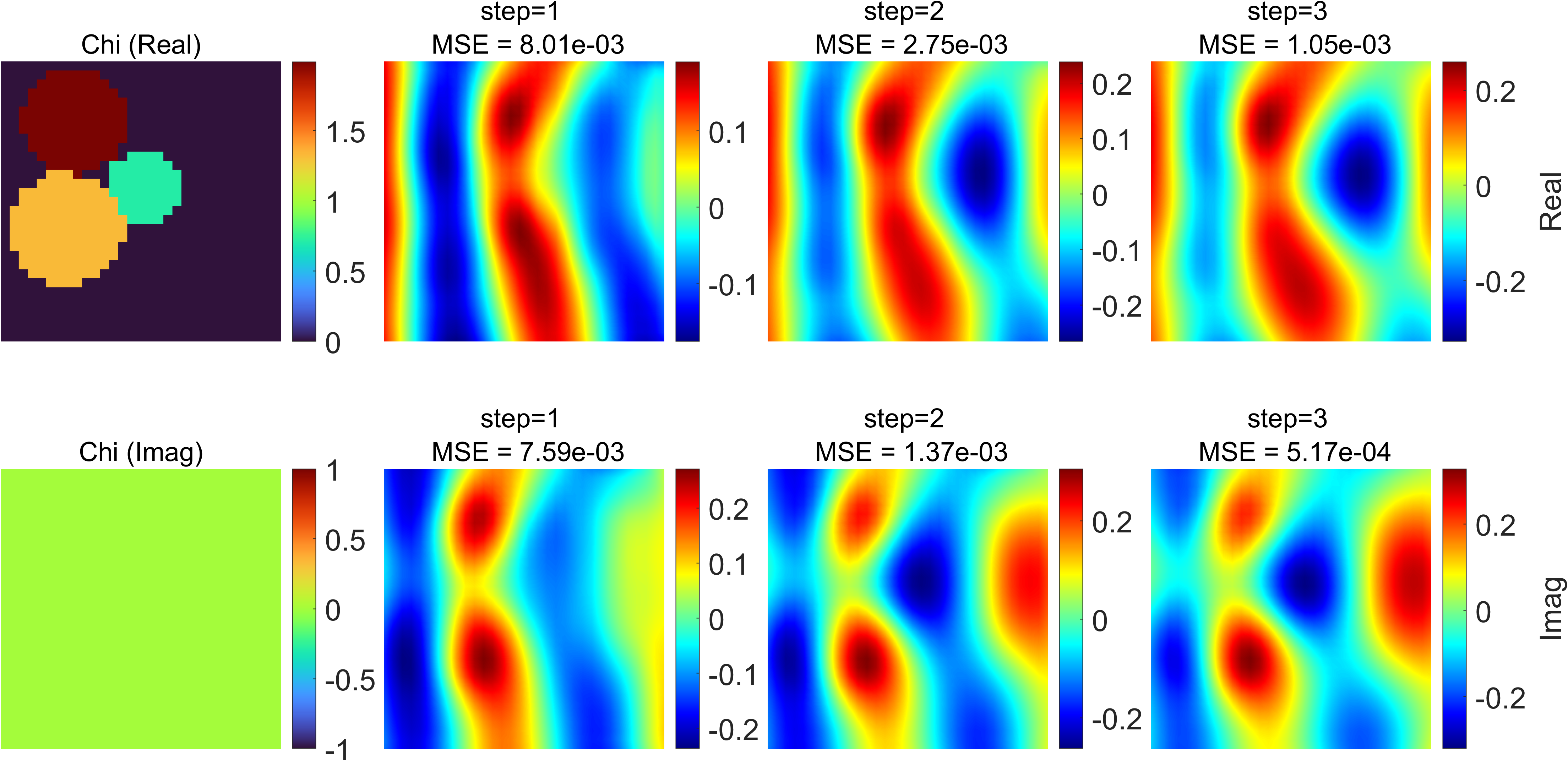}}	
	\caption{Lossless scatterer cases:  updated total fields in each iteration computed by SiPhiResNet. (a) and (b) are updated total fields corresponding to \fig{sipisrlnolossy}. From left to right: contrast $\chi$, total fields computed by SiPhiResNet $E^{tot}_{Si}$ in the first, second and third iteration. The first row is the real part and the second is the imaginary part.}
	\label{sipisrlnolossyall}
\end{figure}
\section{Numerical Results and Analysis}
In this section, we verify PhiSRL by applying both SiPhiResNet and NiPhiResNet to solve 2D VIE. 
The model setup is illustrated in \fig{viemodel} and the size of domain $D$ is $0.15m \times 0.15m$. $D$ is discretized with $32\times 32$ uniform grids. One transmitter is placed outside $D$ and its distance from the center of $D$ is 1.67$m$. The frequency of incident field is 3GHz and different incident angles are taken into account.
It is noted that the coefficient matrix and right-hand side term of \eq{eq36} vary with  $\boldsymbol{\chi}$ and the direction of incident wave $ \mathbf{E}^{inc}$.
\subsection{Lossless scatterers}\label{lossless}
The angle of incident wave is randomly selected from $0^{\circ}$, $45^{\circ}$, $90^{\circ}$, $135^{\circ}$, $180^{\circ}$, $225^{\circ}$, $270^{\circ}$, $315^{\circ}$, as shown in \fig{nonlossypinc}.
The contrast of lossless scatterers can be derived from \eq{eq31}:
\begin{equation}
	\chi\left(\mathbf{r}\right) = \frac{\varepsilon(\mathbf{r}) - \varepsilon_0}{\varepsilon_0} = \varepsilon_r(\mathbf{r}) -1
	\label{eq47}
\end{equation}
Three cylinders with random positions and radii are located in $D$. 
The real parts of their contrasts vary from 0 to 1, 1 to 2 and 0 to 2 while the imaginary parts are 0, as shown in \fig{nonlossychi}.
\par 
Both SiPhiResNet and NiPhiResNet are implemented in Pytorch, and the computing platform is one Nvidia V100 GPU. The stochastic optimizer of SiPhiResNet and NiPhiResNet is Adam\cite{Kingma2014}. The learning rate is initialized as 0.002 and multiplied by 0.8 every 20 epochs.
MoM is applied to compute the field in $D$. A total of 40000 models are solved and form the data set, of which 32000 are as the training data set and 8000 are as the testing data set.
\begin{figure}
	\centering
	\includegraphics[width=0.95\linewidth]{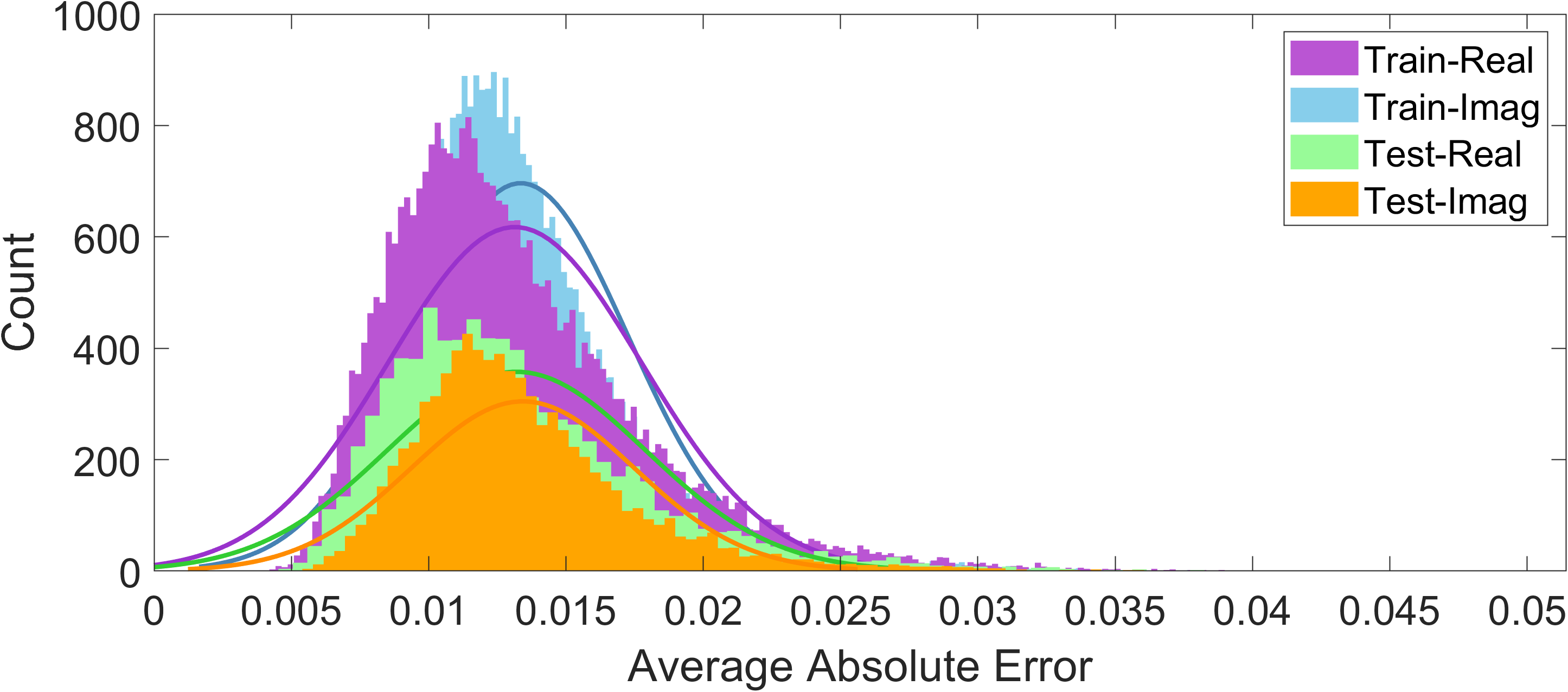}
	\caption{Lossless scatterer cases: histograms of mean absolute error of $E^{tot}_{Si}$. Train-Real and Train-Imag are MAE histograms of real and imaginary parts of $E^{tot}_{Si}$ in the training data set (means are 0.0131 and 0.0134, stds are 0.0046 and 0.0039); Test-Real and Test-Imag are MAE histograms of real and imaginary parts of $E^{tot}_{Si}$ in the testing data set (means are 0.0132 and 0.0134, stds are 0.0047 and 0.0041).}
	\label{sipisrlnolossy-hist}
\end{figure}
\begin{figure}
	\centering
	\includegraphics[width=0.75\linewidth]{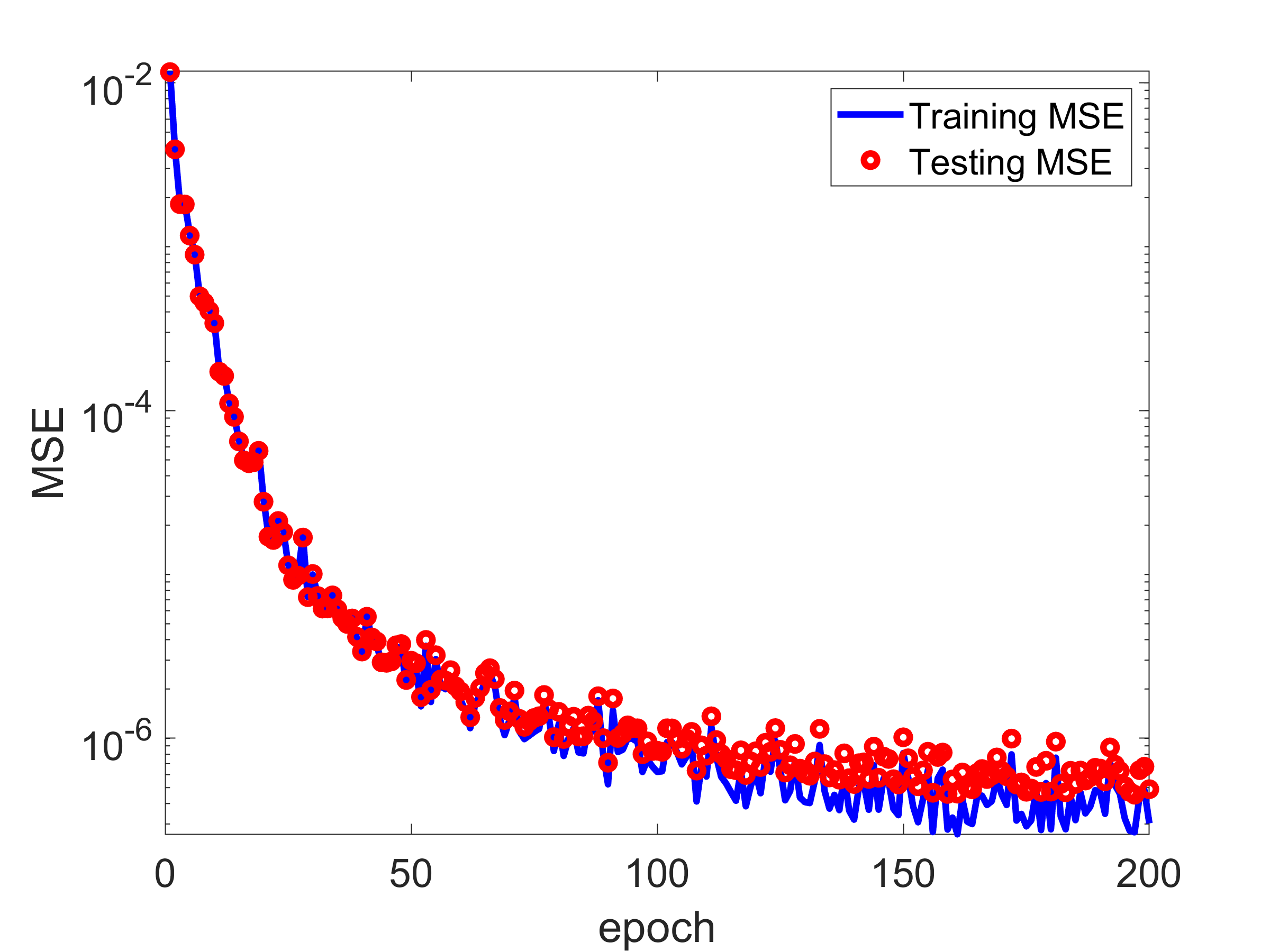}
	\caption{MSE convergence curve of NiPhiResNet for solving VIEs of lossless scatterers. }
	\label{nsipisrlnolossy-loss}
\end{figure}
\begin{figure}
	\centering
	\subfigure[] {\includegraphics[width=0.99\linewidth]{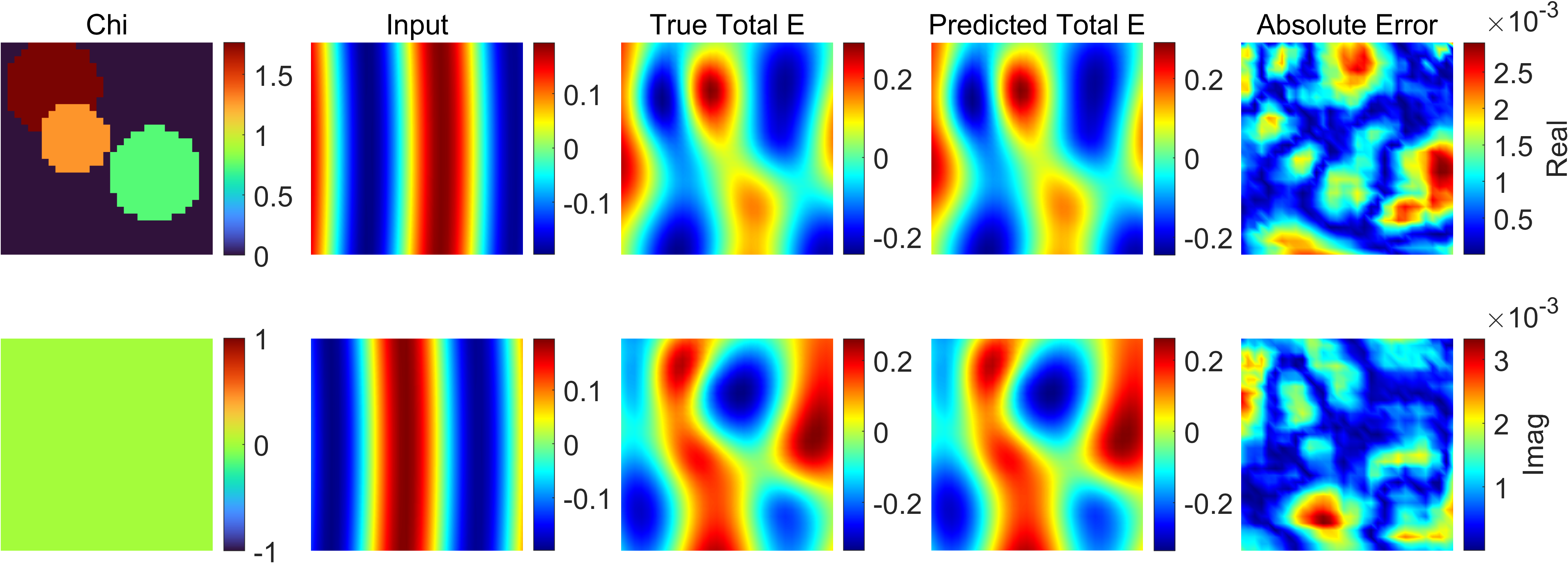}}
	\subfigure[] {\includegraphics[width=0.99\linewidth]{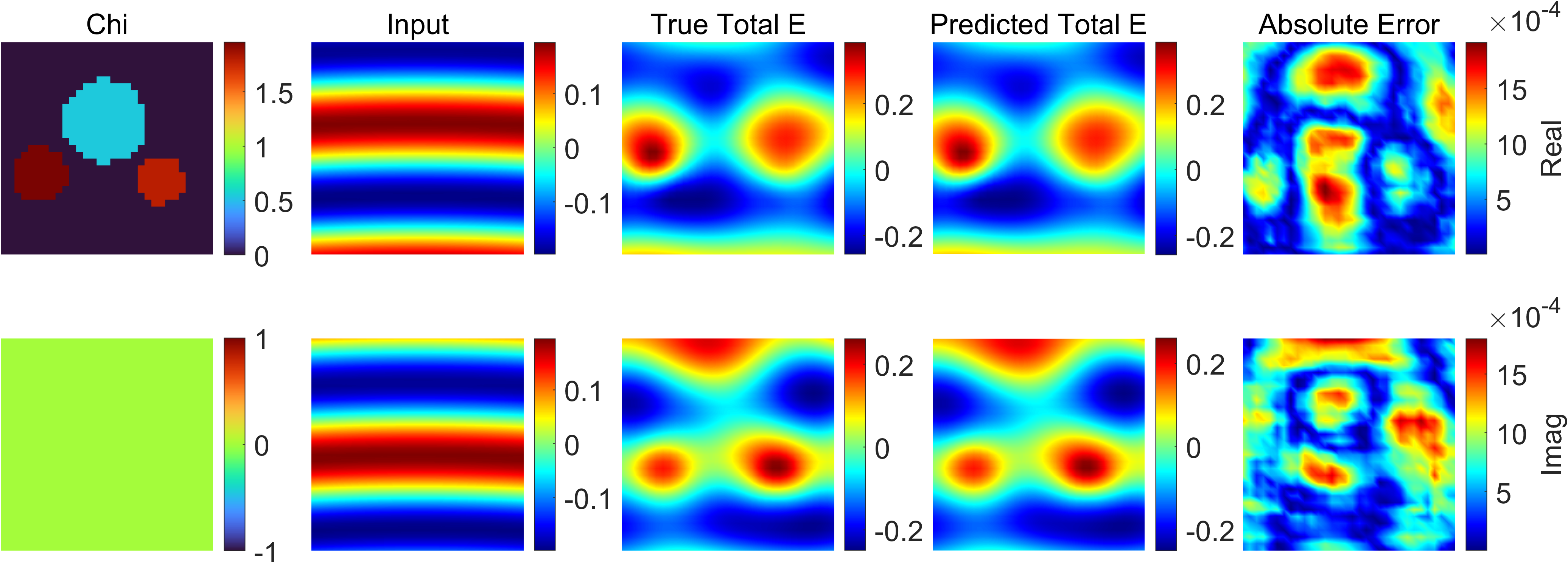}}
	\caption{Lossless scatterer cases: comparisons of total fields computed by NiPhiResNet and MoM. (a) and (b) are total fields of VIEs of overlapped and separate cylinders. From left to right: contrast $\chi$, input of NiPhiResNet (initial guess, $E^{inc}$), total field computed by MoM $E^{tot}_{MoM}$, total field computed by NiPhiResNet $E^{tot}_{Ni}$, and their absolute error distribution. The first row is the real part and the second is the imaginary part.}
	\label{nsipisrlnolossy}
\end{figure}
\begin{figure}
	\centering
	\subfigure[]
	{\includegraphics[width=0.99\linewidth]{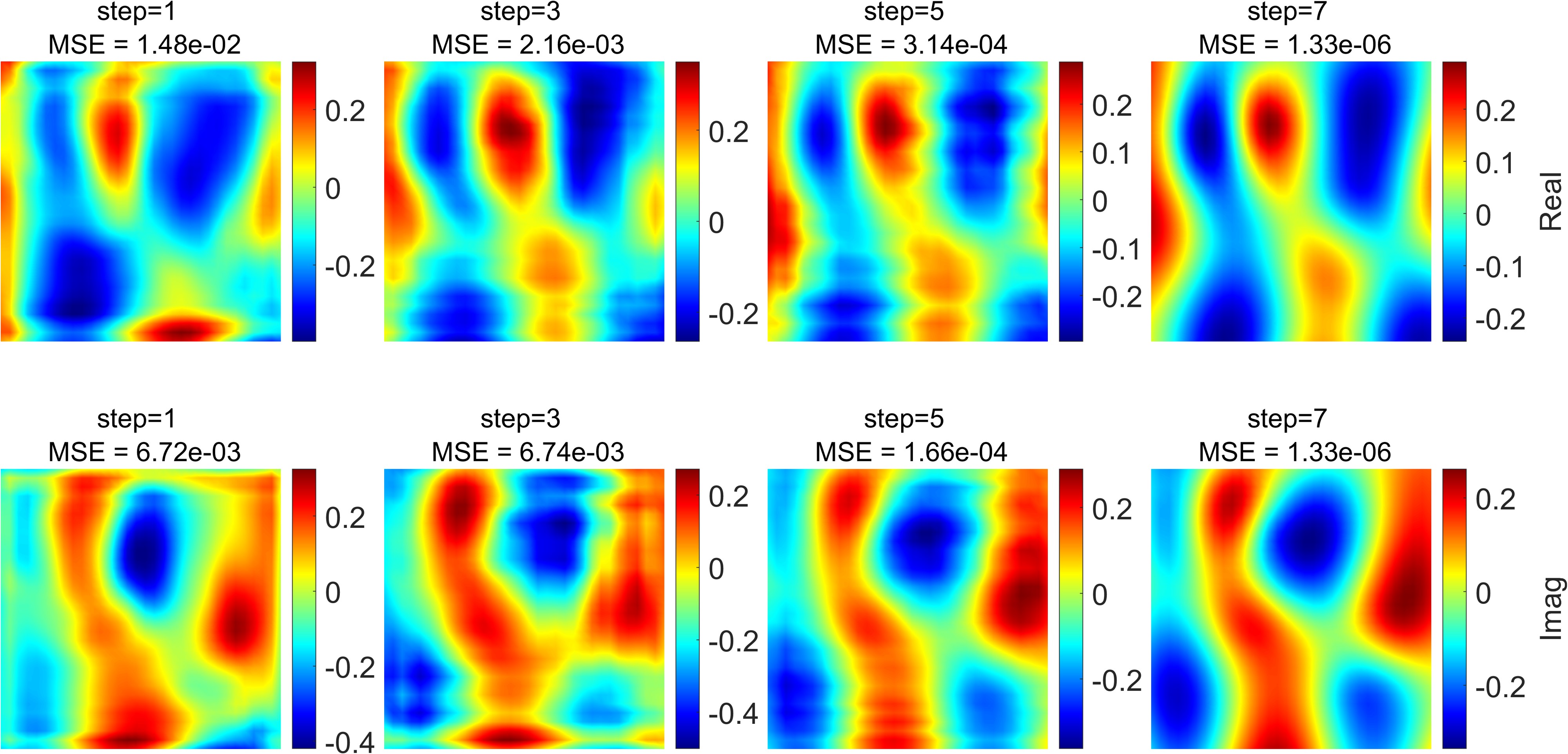}}
	\subfigure[]
	{\includegraphics[width=0.99\linewidth]{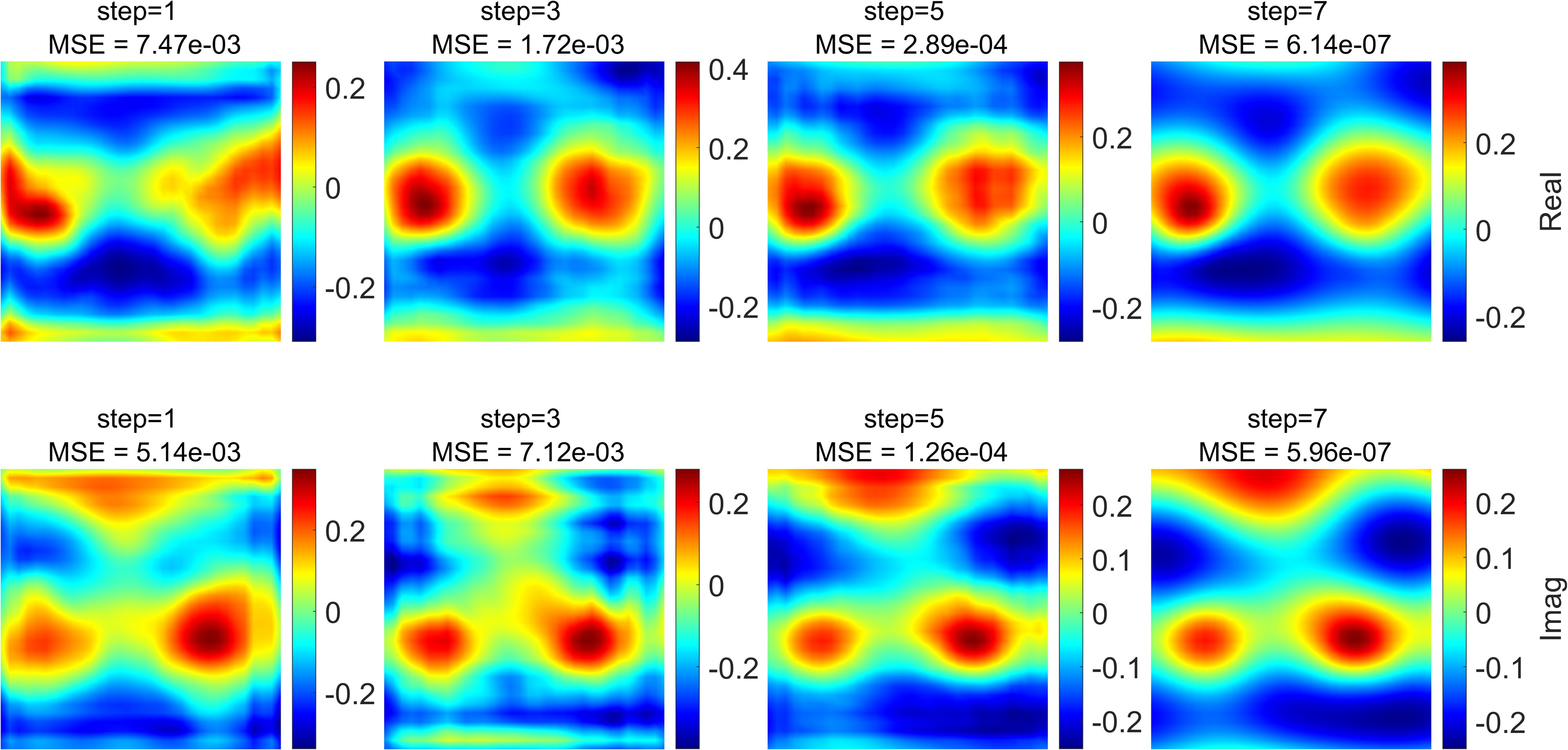}}	
	\caption{Lossless scatterer cases:  updated total fields in each iteration computed by NiPhiResNet. (a) and (b) are update total fields corresponding to \fig{nsipisrlnolossy}. From left to right: total fields computed by NiPhiResNet $E^{tot}_{Ni}$ in the first, third, fifth and seventh iteration. The first row is the real part and the second is the imaginary part.}
	\label{nsipisrlnolossyall}
\end{figure}
\subsubsection{SiPhiResNet}
SiPhiResNet is assumed to have three iterations and the average MSE converges below $3.152 \times 10^{-4}$, as shown in \fig{sipisrlnolossy-loss}. The training and testing MSE agree well with each other, which indicates little overfitting. 
The little overfitting can be reduced by stopping early or adding regularizations.
The U-Net of SiPhiResNet shares the same structure and parameter space throughout the whole iterative process. 
This means that U-Net needs to learn a wide variety of mappings between residuals and modifications in all iterations.
Thus, the optimization of U-Net is not stable enough and fluctuates several times as reflected in \fig{sipisrlnolossy-loss}.
\fig{sipisrlnolossy} demonstrates the detailed comparisons of total field computed by SiPhiResNet ($E^{tot}_{Si}$) and MoM ($E^{tot}_{MoM}$) that are randomly selected from the testing data set.
It can be observed that  $E^{tot}_{Si}$ and $E^{tot}_{MoM}$ are in a good agreement with low level of absolute error.
The updated total field of each iteration is depicted in \fig{sipisrlnolossyall} for a better insight into SiPhiResNet.
The initial guess is the incident field. The difference between  $E^{tot}_{Si}$ and $E^{tot}_{MoM}$ decreases with the increase of iterations. 
The MSE of each iteration is also denoted in \fig{sipisrlnolossyall}.
Furthermore, the histogram of mean absolute error (MAE) between $E^{tot}_{Si}$ and $E^{tot}_{MoM}$ is shown in \fig{sipisrlnolossy-hist}.
The MAE's mean and standard deviation (std) of training and testing data sets agree well with each other that is consistent with the MSE convergence curve in  \fig{sipisrlnolossy-loss}.
\begin{table}
	\renewcommand{\arraystretch}{1.3}
	\caption{Performance of SiPhiResNet and NiPhiResNet on unseen lossless contrast shapes }
	\label{table-nonlossy}
	\centering
	\begin{threeparttable}
		\begin{tabular}{ccc}
			\toprule
			Method & MAE-R\tnote{*} \, (mean/stds)  & MAE-I\tnote{**} \, (mean/stds) \\
			\midrule
			SiPhiResNet & $1.34^{\times10^{-2}}/8.00^{\times 10^{-3}}$ & $1.34^{\times10^{-2}}/5.50^{\times10^{-3}}$\\
			NiPhiResNet & $8.5^{\times 10^{-4}}/1.56^{\times10^{-3}}$ & $8.95^{\times10^{-4}}/1.36^{\times10^{-3}}$ \\
			\bottomrule
		\end{tabular}
		\begin{tablenotes}
			\setlength{\multicolsep}{0cm}
				\item[*] MAE of total field real parts 	
				\item[**] MAE of total field imaginary parts
		\end{tablenotes}
	\end{threeparttable}
\end{table}
\begin{figure}
	\centering
	\includegraphics[width=0.95\linewidth]{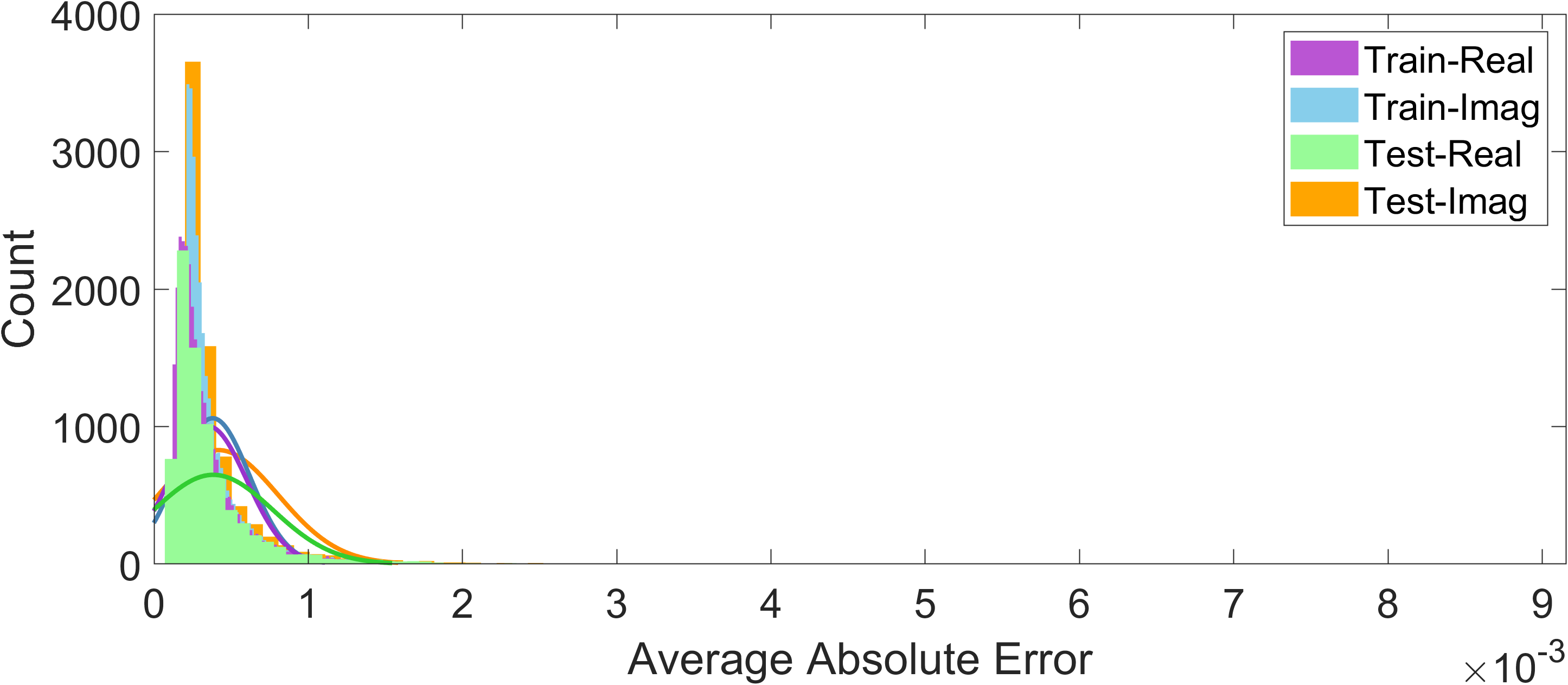}
	\caption{Lossless scatterer cases: histograms of mean absolute error of $E^{tot}_{Ni}$. Train-Real and Train-Imag are MAE histograms of real and imaginary parts of $E^{tot}_{Ni}$ in the training data set (means are $3.49 \times 10^{-4}$ and $3.8 \times 10^{-4}$, stds are $2.53 \times 10^{-4}$ and $2.39\times10^{-4}$); Test-Real and Test-Imag are MAE histograms of real and imaginary parts of $E^{tot}_{Ni}$ in the testing data set (means are $3.84\times10^{-4}$ and $4.16 \times 10^{-4}$, stds are $3.86 \times 10^{-4}$ and $3.89 \times10^{-4}$).}
	\label{nsipisrlnolossy-hist}
\end{figure}
\begin{figure}
	\centering
	\includegraphics[width=0.9\linewidth]{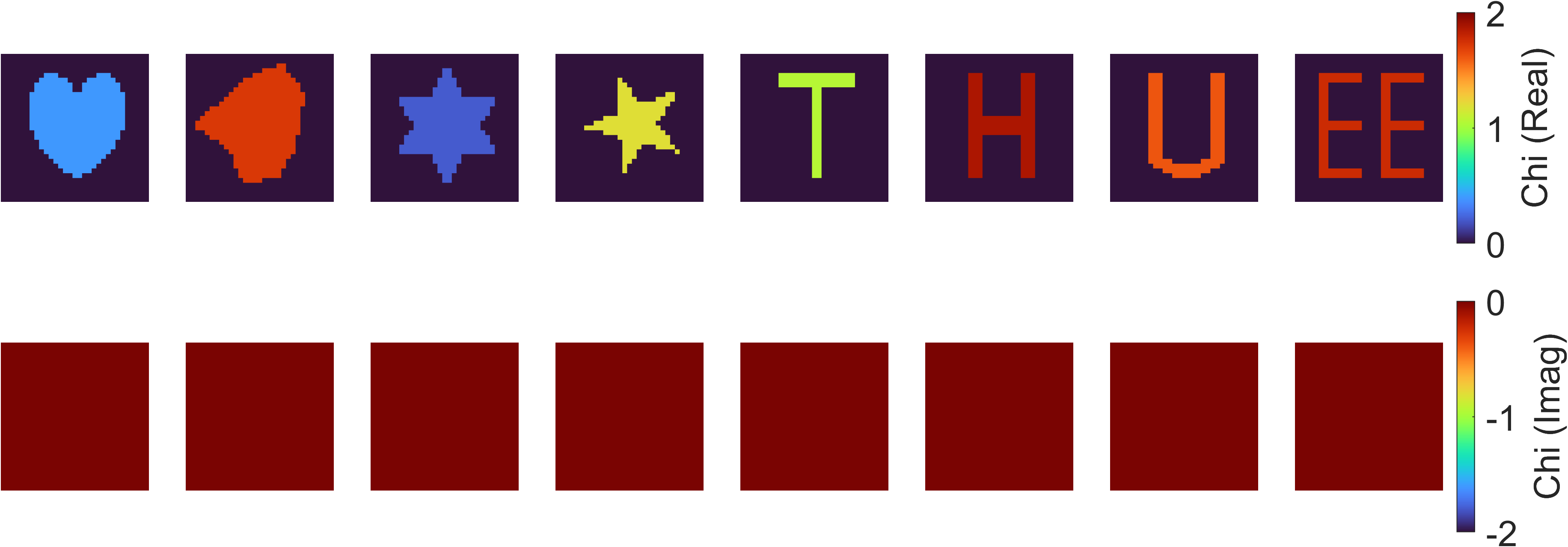}
	\caption{Lossless contrast examples of the generalization validation data set for unseen contrast shapes. The first and second row are real and imaginary parts.}
	\label{gennolossychi}
\end{figure}
\subsubsection{NiPhiResNet}
NiPhiResNet consists of seven independent PhiSRL blocks and the convergence curve of MSE is shown in \fig{nsipisrlnolossy-loss}.
With a good agreement, training and testing MSEs decrease steadily and converge below $4.8925 \times 10^{-7}$.
NiPhiResNet demonstrates better computing precisions than SiPhiResNet.
Although NiPhiResNet has more iterations, the CNN of each iteration owns simpler structure than the U-Net of SiPhiResNet.
It indicates that it is better to use an independent CNN to learn update rules in a single iteration. 
\fig{nsipisrlnolossy} shows two results of total fields solved by NiPhiResNet and they are randomly chosen from the testing data set.
Total fields solved by NiPhiResNet have good precisions with different incident angles and contrast distributions.
\fig{nsipisrlnolossyall} illustrates the updated total field of each iteration.
The total field of the first iteration is a rough approximation of ground truth, and it is refined with the increase of iterations.
\fig{nsipisrlnolossy-hist} plots NiPhiResNet's MAE histogram of training and testing data sets. 
The corresponding means and stds are in a good agreement. Small values of means and stds indicate stable computing precisions of NiPhiResNet.

\subsubsection{Generalization Ability on Contrast Shape}
\begin{figure}
	\centering
	\subfigure[ ]
	{\includegraphics[width=0.99\linewidth]{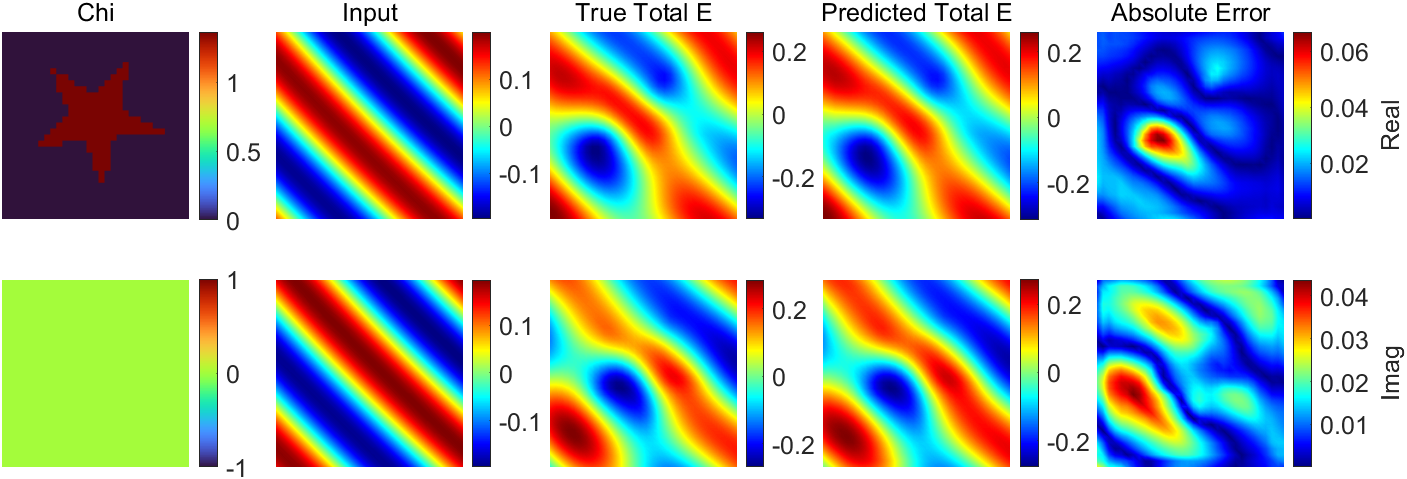}}
	\subfigure[ ]
	{\includegraphics[width=0.99\linewidth]{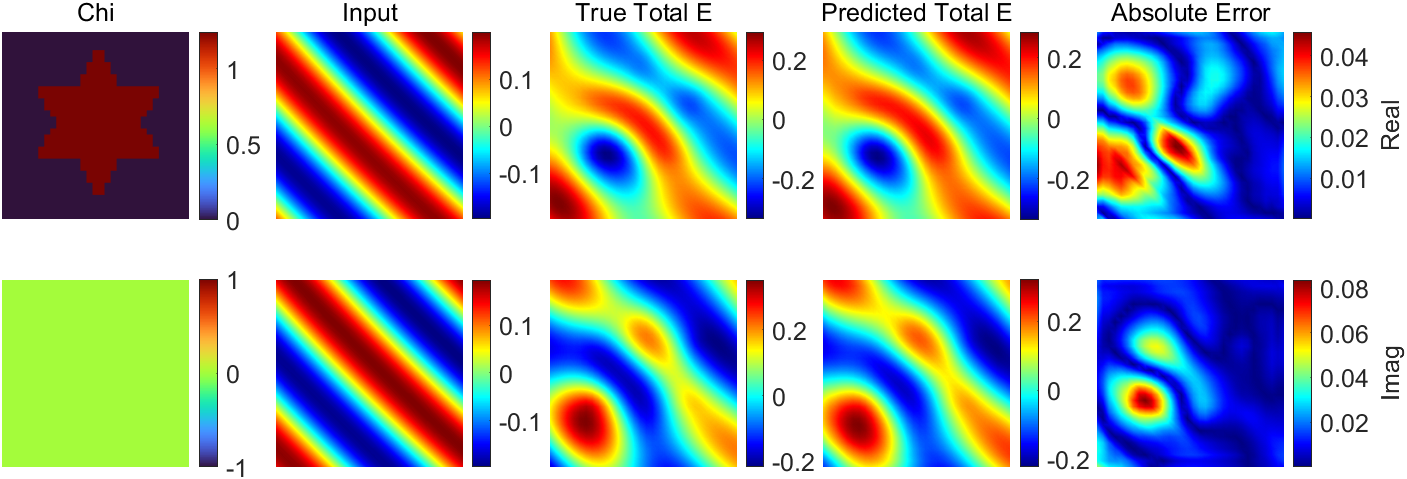}}	
	\centering	
	\subfigure[ ]
	{\includegraphics[width=0.99\linewidth]{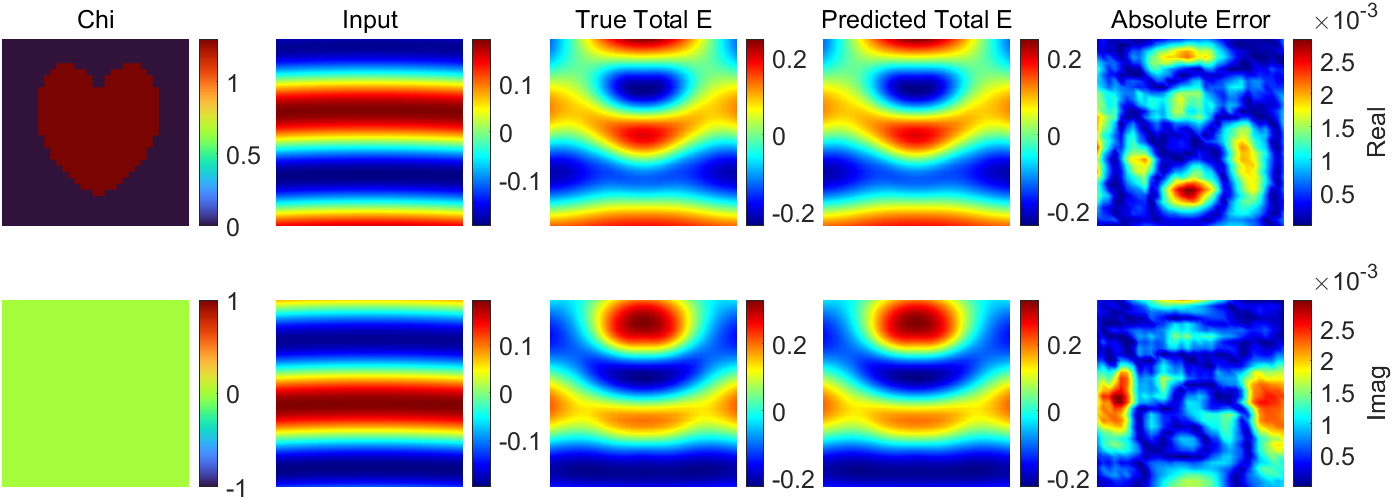}}
	\subfigure[ ]
	{\includegraphics[width=0.99\linewidth]{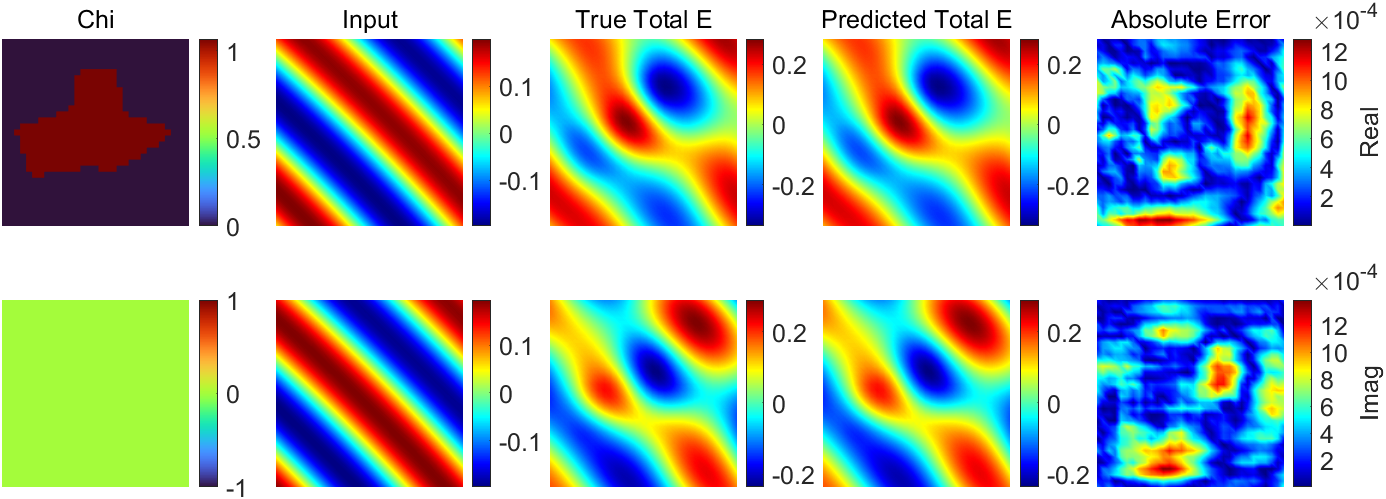}}
	\caption{Lossless scatterer cases: predicted total fields of SiPhiResNet ((a), (b)) and NiPhiResNet ((c), (d)) in the unseen contrast shape generalization validation data set. In each sub-panel, from left to right are contrast, initial guess, true total field,  predicted total field, and their absolute error distribution. The first and second row are real and imaginary parts.}
	\label{nolossy-all-shape}
\end{figure}

The generalization ability of SiPhiResNet and NiPhiResNet on unseen contrast shapes is verified.
Eight types of unseen contrast shapes are considered as shown in \fig{gennolossychi}.
The range of contrasts is $[0,2]$ and 40 samples of each shape are generated for the validation data set.
\tab{table-nonlossy} shows the MAEs of total field solved by SiPhiResNet and NiPhiResNet in the validation data set.
\fig{nolossy-all-shape} plots the samples of total fields solved by SiPhiResNet and NiPhiResNet.
Both of them maintain a small error level on the generalization validation data set.
\subsubsection{Generalization Ability on Incident Frequency}
\begin{figure}
	\centering
	\subfigure[SiPhiResNet MSE]
	{\includegraphics[width=1\linewidth]{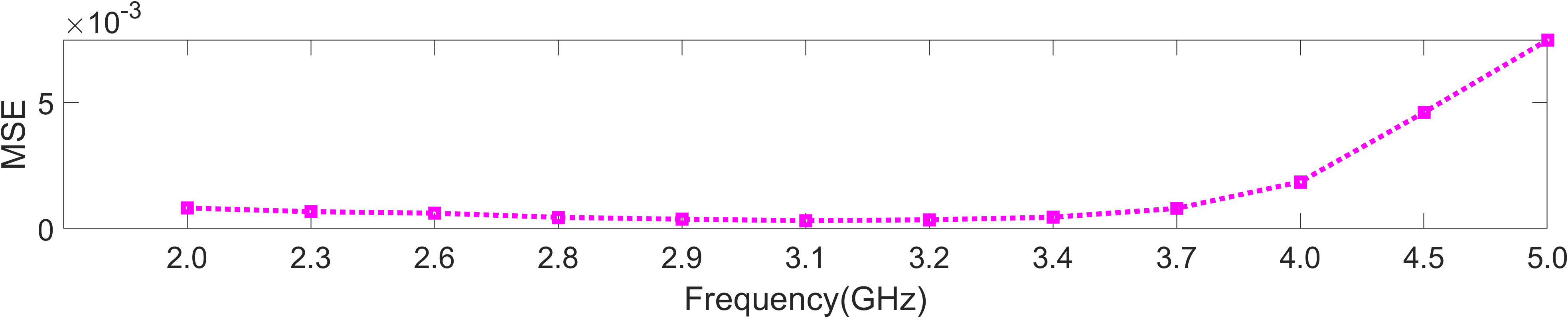}}
	\subfigure[NiPhiResNet MSE]
	{\includegraphics[width=1\linewidth]{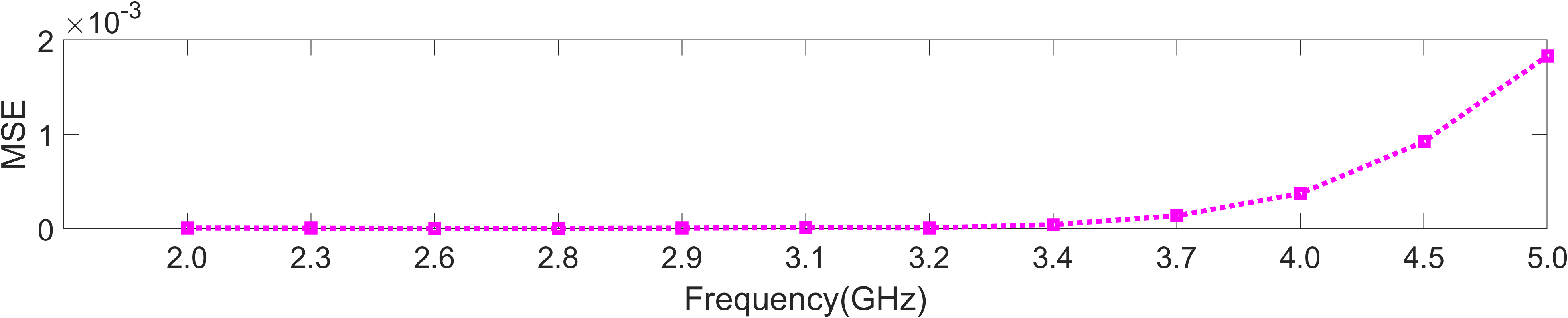}}	
	\caption{MSEs of SiPhiResNet and NiPhiResNet for solving VIEs of lossless scatterers at different frequencies.}
	\label{nolossy_gen_loss}
\end{figure}

The training and testing data set of SiPhiResNet and NiPhiResNet are generated by fixing the incident frequency at 3GHz.
To verify the generalization ability of SiPhiResNet and NiPhiResNet on different incident frequencies, we consider 12 frequencies of incident field: 2GHz, 2.3GHz, 2.6GHz, 2.9GHz, 3.1GHz, 3.2GHz, 3.4GHz, 3.7GHz, 4GHz, 4.5GHz, 5GHz.
Eight types of contrast shapes and eight directions of incident waves are taken into account, as shown in \fig{gennolossychi} and \fig{nonlossypinc}.
We generate 320 samples for each incident frequency.
\fig{nolossy_gen_loss} shows the MSE of SiPhiResNet and NiPhiResNet at different incident frequencies.
Both of them keep good computing precisions across a broad range of incident frequencies, which validates their generalization ability on unseen incident frequencies.
\subsection{Lossy scatterers}\label{lossy}
\begin{figure}
	\centering
	\includegraphics[width=0.9\linewidth]{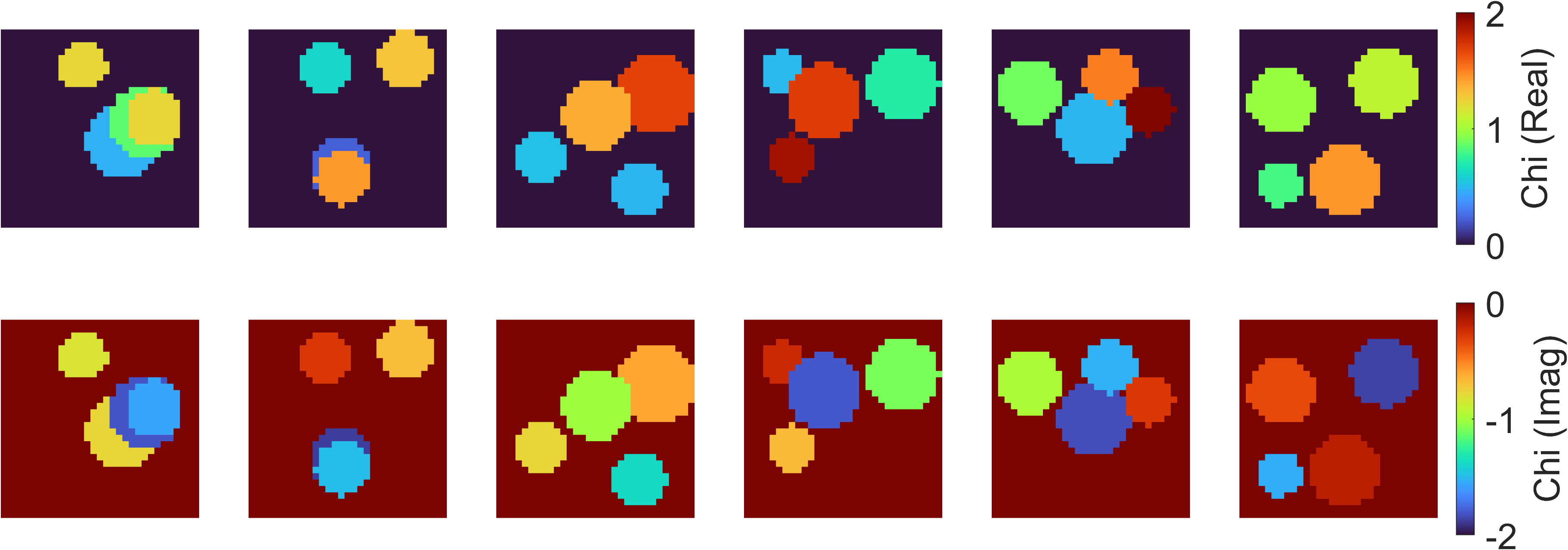}
	\caption{Contrast examples of lossy scatterers: the first row is the real part and the second is the imaginary part.}
	\label{lossychi}
\end{figure}

\begin{figure}
	\centering
	\includegraphics[width=0.8\linewidth]{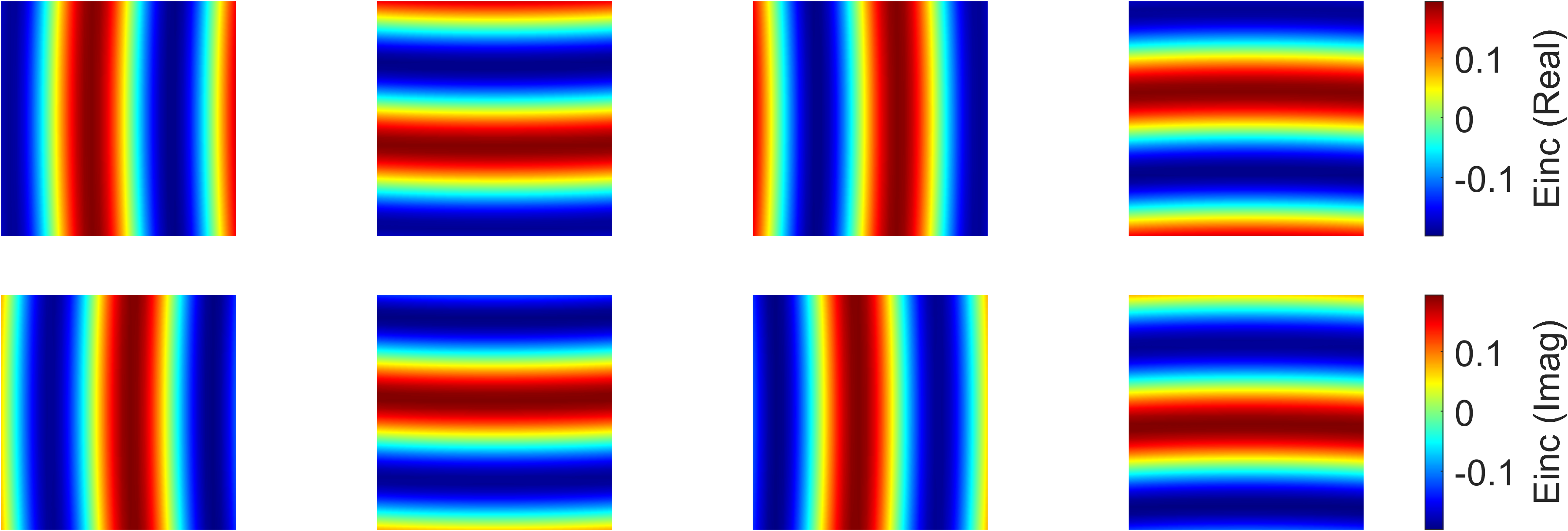}
	\caption{Incident fields in the case of lossy scatterers: the first row is the real part and the second is the imaginary part, the incident angles from left to right are  $[0^{\circ}, 90^{\circ}, 180^{\circ}, 270^{\circ}]$.}
	\label{lossypinc}
\end{figure}

In this section, we apply PhiSRL to solve VIEs with lossy scatterers to verify its universality because lossy scatterers have different interactions with incident fields compared to lossless ones.
Four cylinders with random positions and radii are located in $D$.
The formulation of contrast is \eq{eq31} and the ranges of contrasts can refer to \tab{table01}.
\fig{lossychi} shows six examples.
The incident frequency is fixed at 3GHz and the incident angle is randomly selected in $[0^{\circ}, 90^{\circ}, 180^{\circ}, 270^{\circ}]$, as shown in \fig{lossypinc}.
40000 data samples are generated by MoM of which 80\% and 20\% are for training and testing respectively.
The computing platform and training setup of SiPhiResNet and NiPhiResNet are the same as the ones in the cases of lossless scatterers.
\begin{table}
	\renewcommand{\arraystretch}{1.3}
	\caption{Contrast Values of Lossy Scatterers }
	\label{table01}
	\centering
	\begin{threeparttable}
		\begin{tabular}{ccc}
			\toprule
			Cylinder & Real Part & Imaginary Part \\
			\midrule
			Cylinder 1  & $[0,1]$  & $[-1,0]$  \\
			Cylinder 2  & $[0,1]$  & $[-2,-1]$ \\
			Cylinder 3  & $[1,2]$  & $[-1,0]$  \\
			Cylinder 4  & $[1,2]$  & $[-2,-1]$  \\
			\bottomrule
		\end{tabular}
	\end{threeparttable}
\end{table}
\subsubsection{SiPhiResNet}
\begin{figure}
	\centering
	\includegraphics[width=0.75\linewidth]{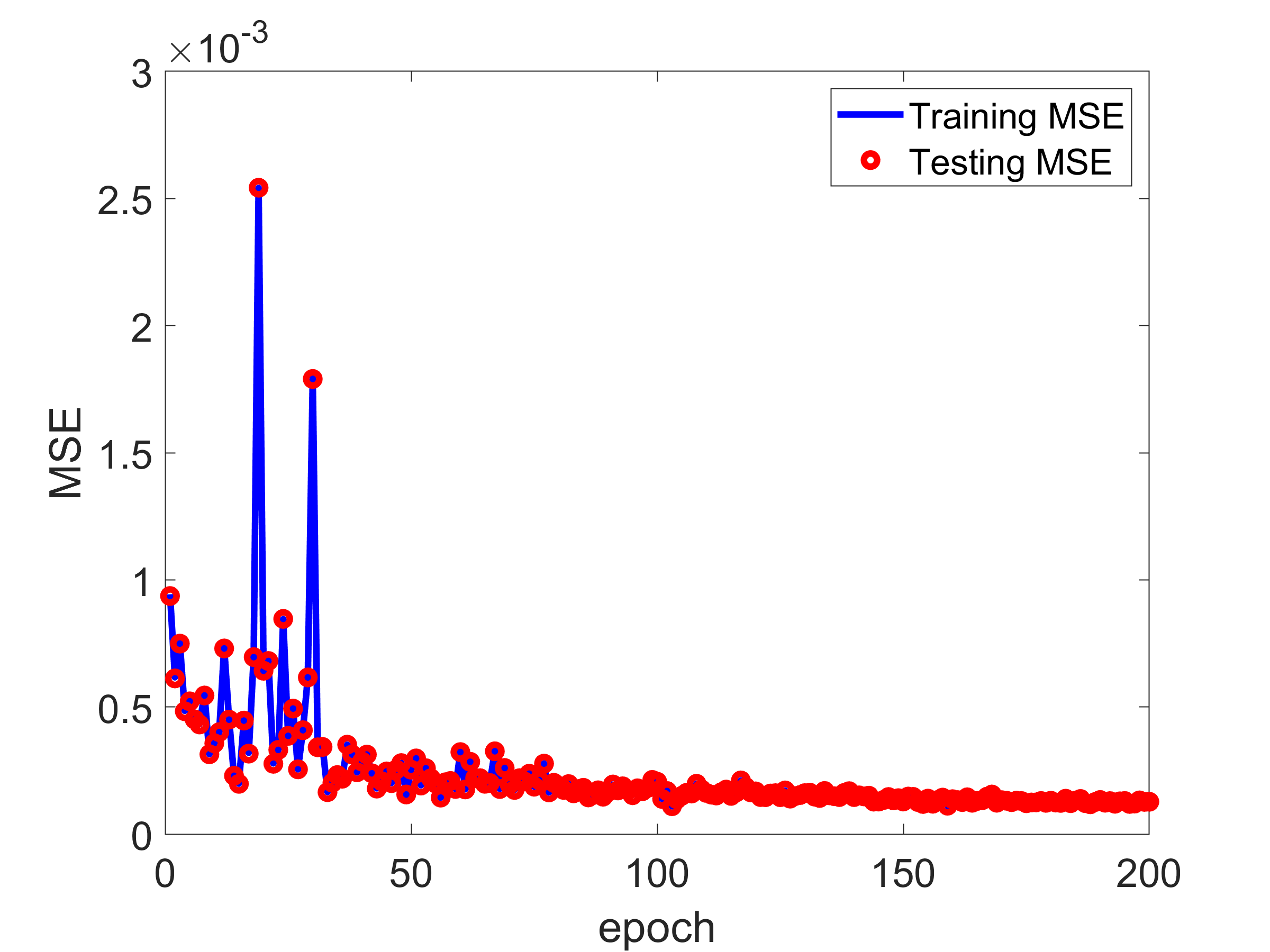}
	\caption{MSE convergence curve of SiPhiResNet for solving VIEs of lossy scatterers. }
	\label{sipisrllossy-loss}
\end{figure}

\begin{figure}
	\centering
	\subfigure[]	
	{\includegraphics[width=0.99\linewidth]{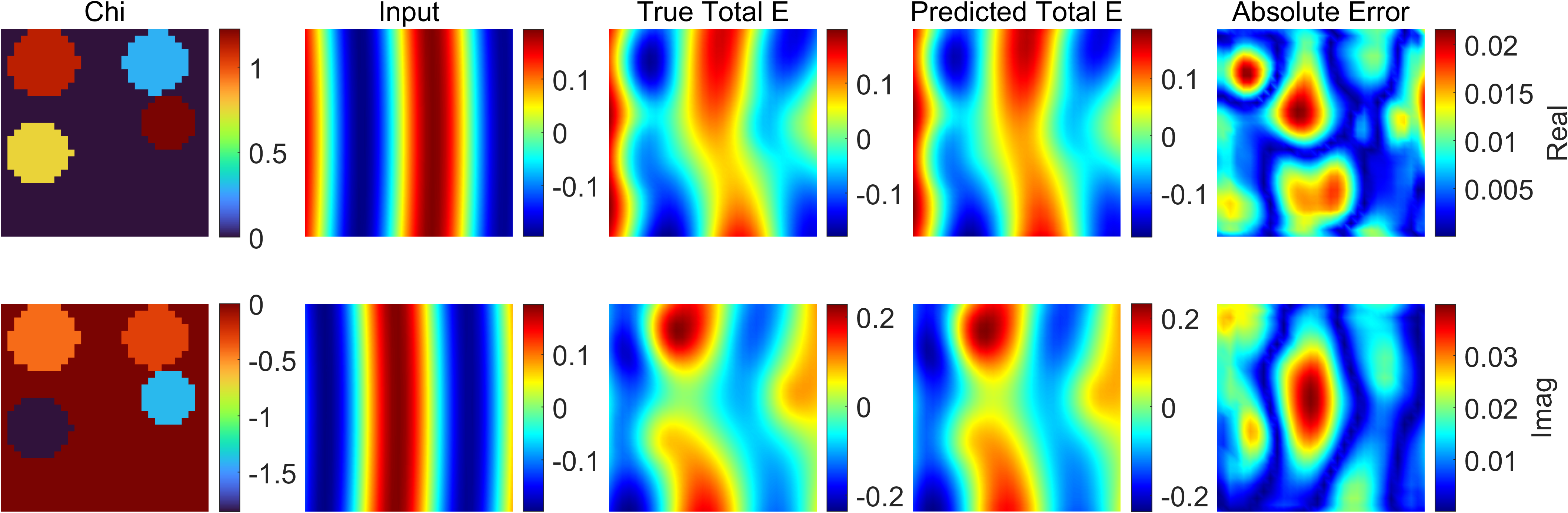}}
	\subfigure[]
	{\includegraphics[width=0.99\linewidth]{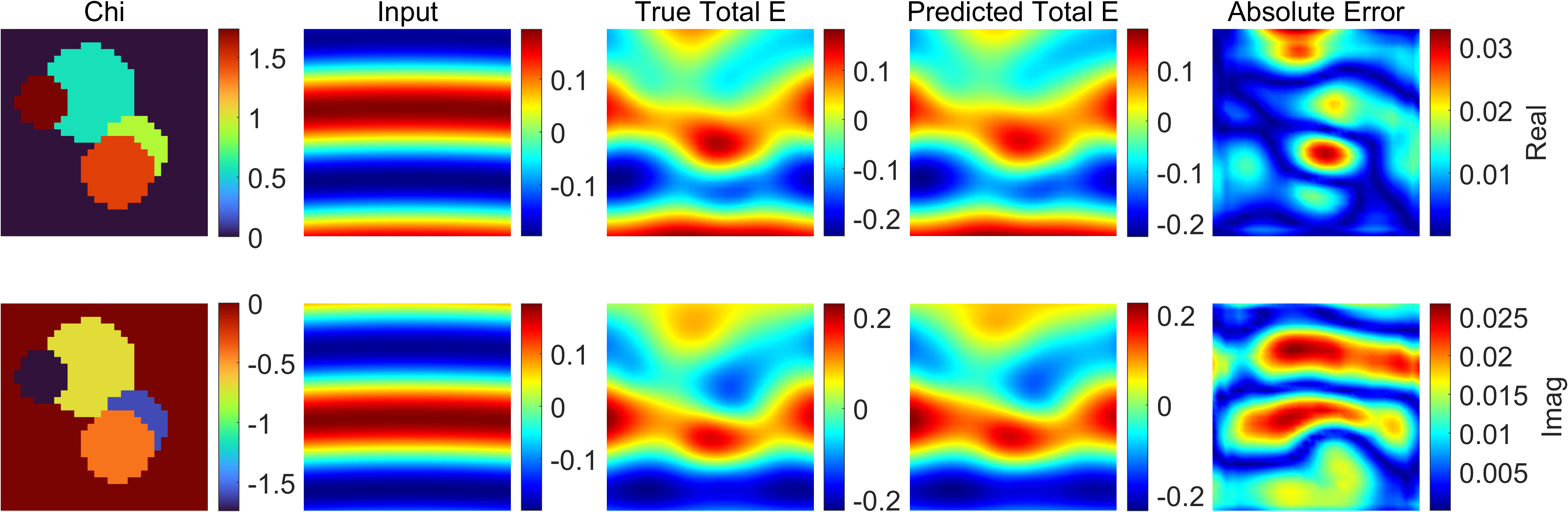}}
	\caption{Lossy scatterer cases: comparisons of total fields computed by SiPhiResNet and MoM. (a) and (b) are total fields of VIEs of separate and overlapped cylinders. From left to right: contrast $\chi$, input of SiPhiResNet (initial guess, $E^{inc}$), total field computed by MoM $E^{tot}_{MoM}$, total field computed by SiPhiResNet $E^{tot}_{Si}$, and their absolute error distribution. The first and second row are the real and imaginary part.}
	\label{sipisrllossy}
\end{figure}

\begin{figure}
	\centering
	\subfigure[]
	{\includegraphics[width=0.99\linewidth]{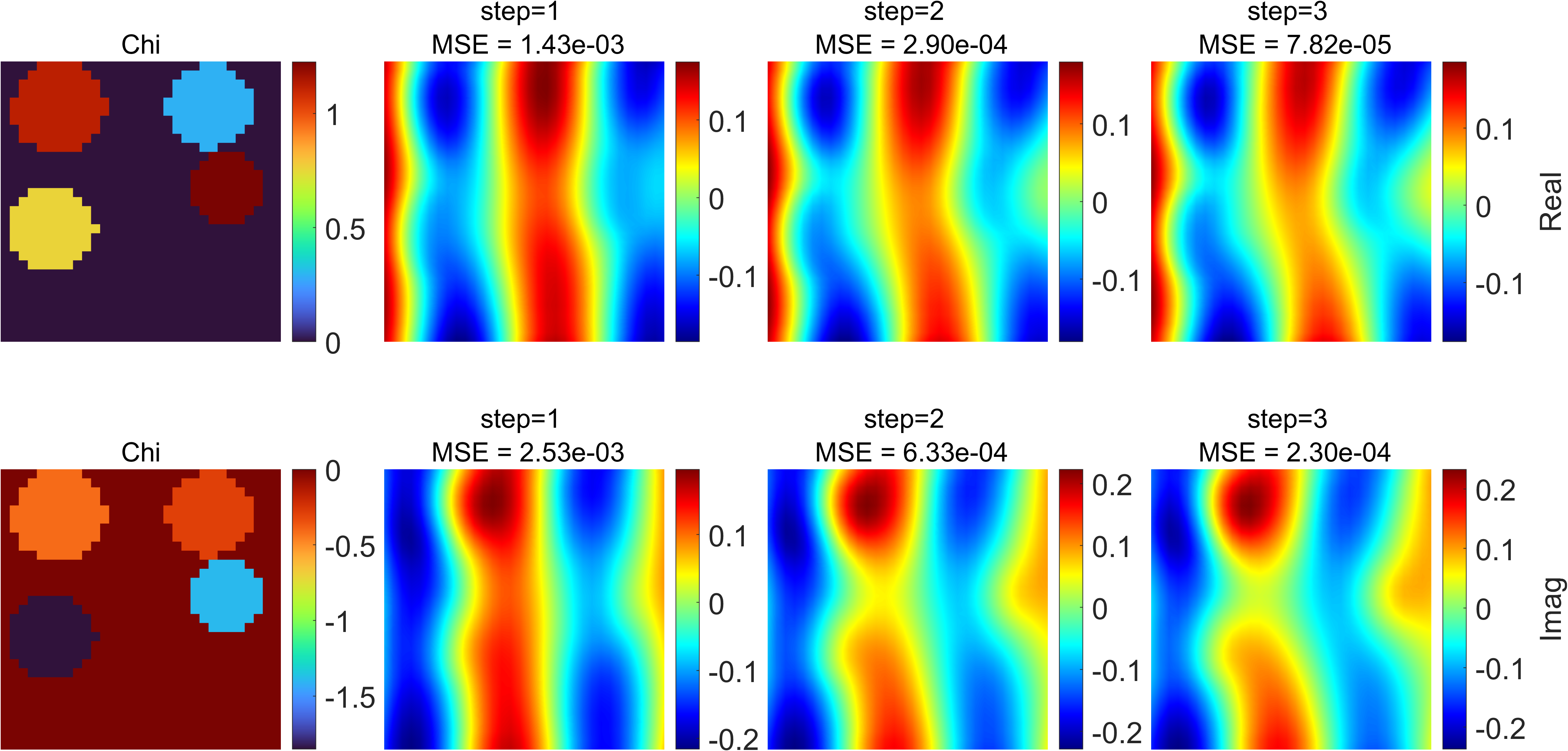}}
	\subfigure[]
	{\includegraphics[width=0.99\linewidth]{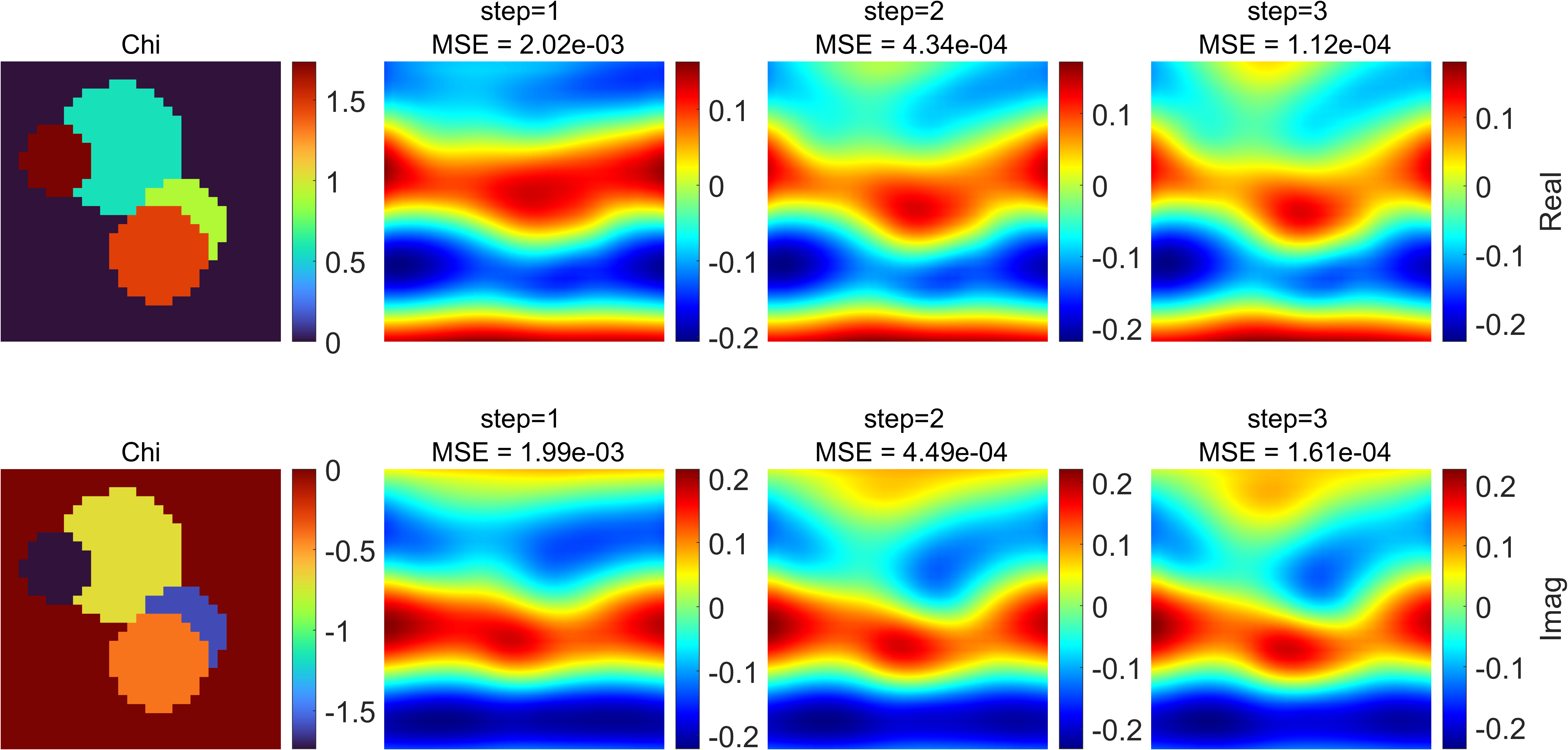}}	
	\caption{Lossy scatterer cases:  updated total fields in each iteration computed by SiPhiResNet. (a) and (b) are updated total fields corresponding to \fig{sipisrllossy}. From left to right: contrast $\chi$, total fields computed by SiPhiResNet $E^{tot}_{Si}$ in the first, second and third iteration. The first  and second row are the real and imaginary part.}
	\label{sipisrllossyall}
\end{figure}

\begin{figure}
	\centering
	\includegraphics[width=0.95\linewidth]{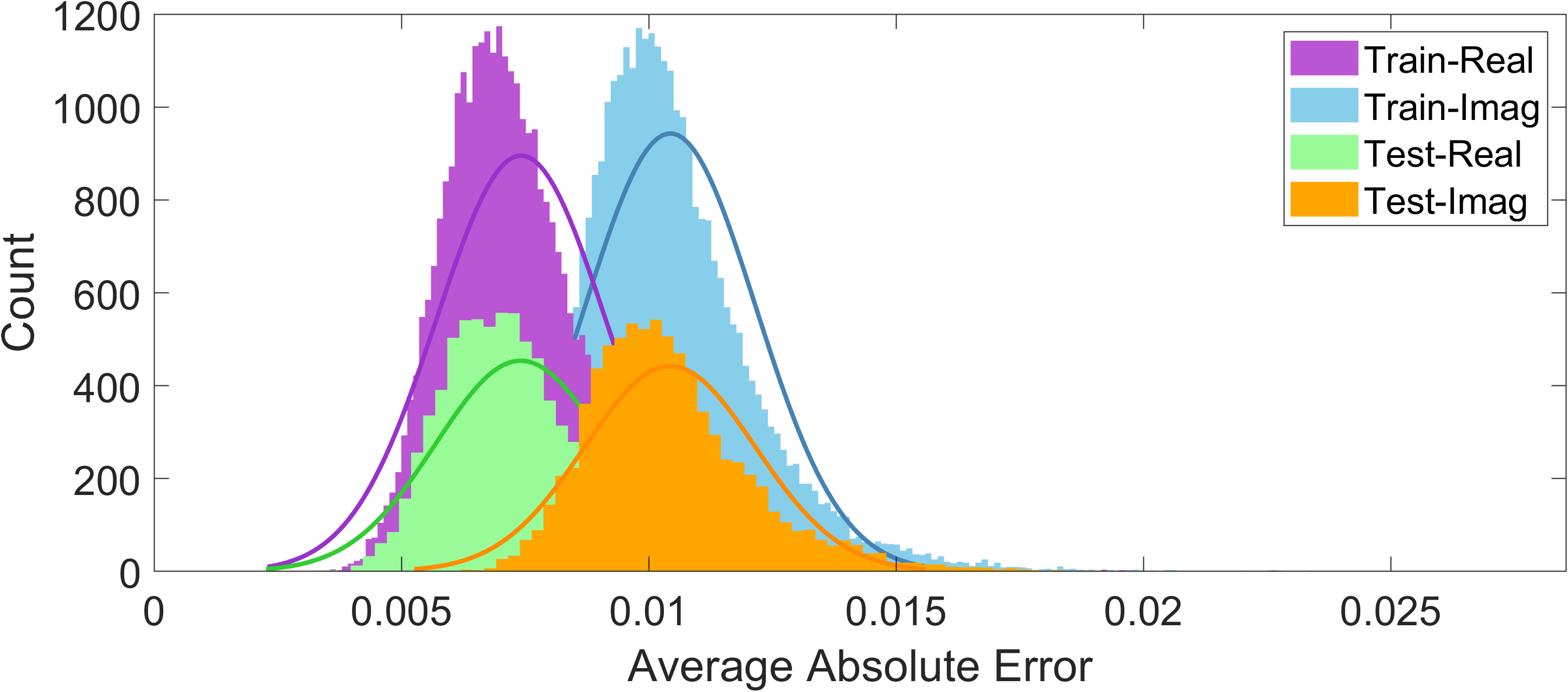}
	\caption{Lossy scatterer cases: histograms of mean absolute error of $E^{tot}_{Si}$. Train-Real and Train-Imag are MAE histograms of real and imaginary parts of $E^{tot}_{Si}$ in the training data set (means are 0.0074 and 0.0104, stds are 0.0017 and 0.0017); Test-Train and Test-Imag are MAE histograms of real and imaginary parts of $E^{tot}_{Si}$ in the testing data set (means are 0.0074 and 0.0104, stds are 0.0017 and 0.0017).}
	\label{sipisrllossy-hist}
\end{figure}

SiPhiResNet has three iterations and \fig{sipisrllossy-loss} shows the convergence curve of MSE with the final average MSE below $1.2775 \times 10^{-4}$.
The training and testing MSEs are in a good agreement.
The training process has a few fluctuations at the beginning and then becomes stabilized gradually.
\fig{sipisrllossy} shows two comparisons between MoM and SiPhiResNet computed total fields ($E^{tot}_{MoM}$ and $E^{tot}_{Si}$).
\fig{sipisrllossyall} shows the updated total field $E^{tot}_{Si}$ of three iterations.
The total field of $1^{st}$ iteration is close to ground truth and it improves gradually with increase of iterations. 
The histogram of MAE between $E^{tot}_{MoM}$ and $E^{tot}_{Si}$ is depicted in \fig{sipisrllossy-hist}. 
The means and stds of both real and imaginary parts of fields are consistent that verifies the MSE convergence curve in \fig{sipisrllossy-loss}.
\subsubsection{NiPhiResNet}
\begin{figure}
	\centering
	\includegraphics[width=0.75\linewidth]{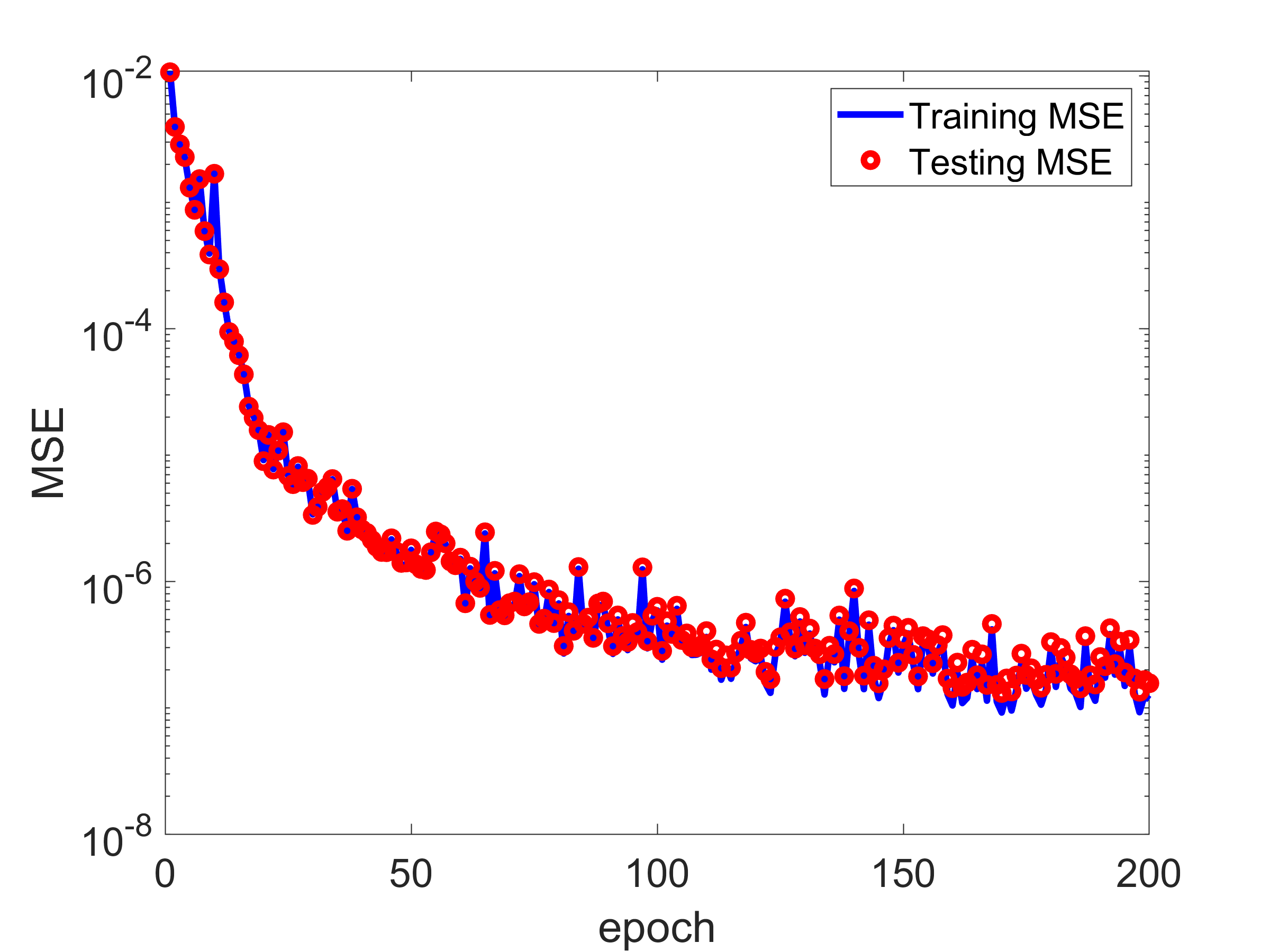}
	\caption{MSE convergence curve of NiPhiResNet for solving VIEs of lossy scatterers. }
	\label{nsipisrllossy-loss}
\end{figure}

\begin{figure}
	\centering
	\subfigure[]	
	{\includegraphics[width=0.99\linewidth]{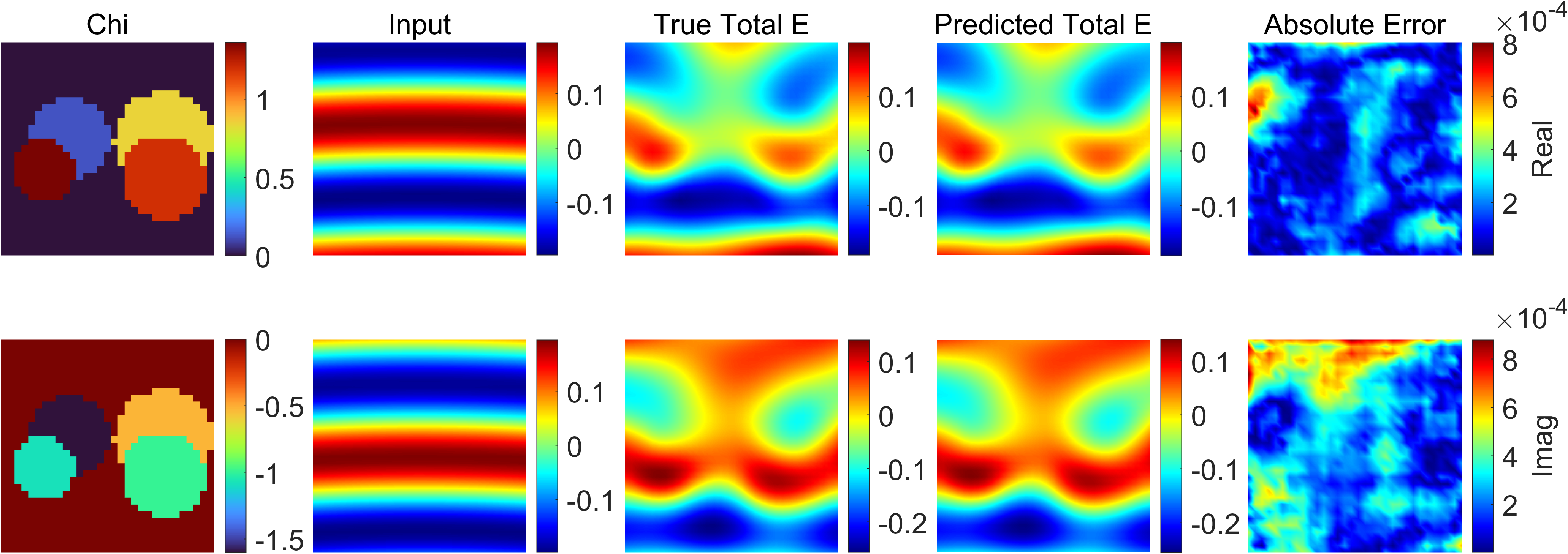}}
	\subfigure[]
	{\includegraphics[width=0.99\linewidth]{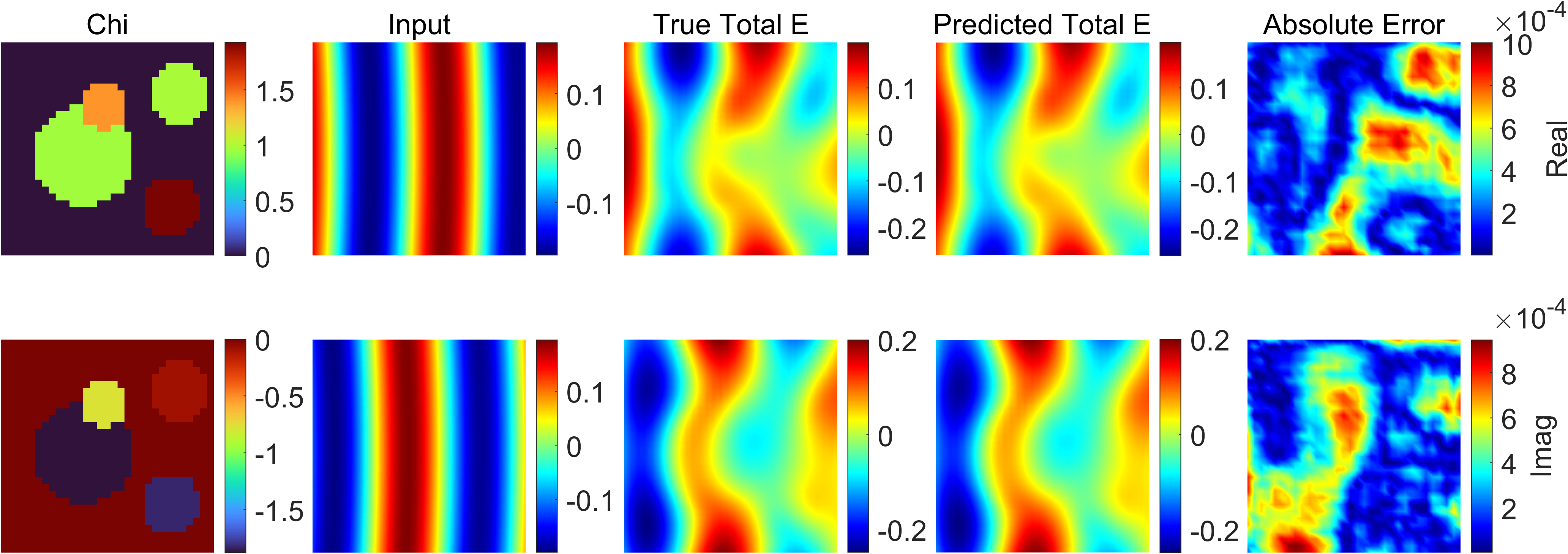}}
	\caption{Lossy scatterer cases: comparisons of total fields computed by NiPhiResNet and MoM. (a) and (b) are total fields of VIEs of overlapped cylinders. From left to right: contrast $\chi$, input of NiPhiResNet (initial guess, $E^{inc}$), total field computed by MoM $E^{tot}_{MoM}$, total field computed by NiPhiResNet $E^{tot}_{Ni}$, and their absolute error distribution. The first and second row are the real and imaginary part.}
	\label{nsipisrllossy}
\end{figure}

\begin{figure}
	\centering
	\subfigure[]
	{\includegraphics[width=0.99\linewidth]{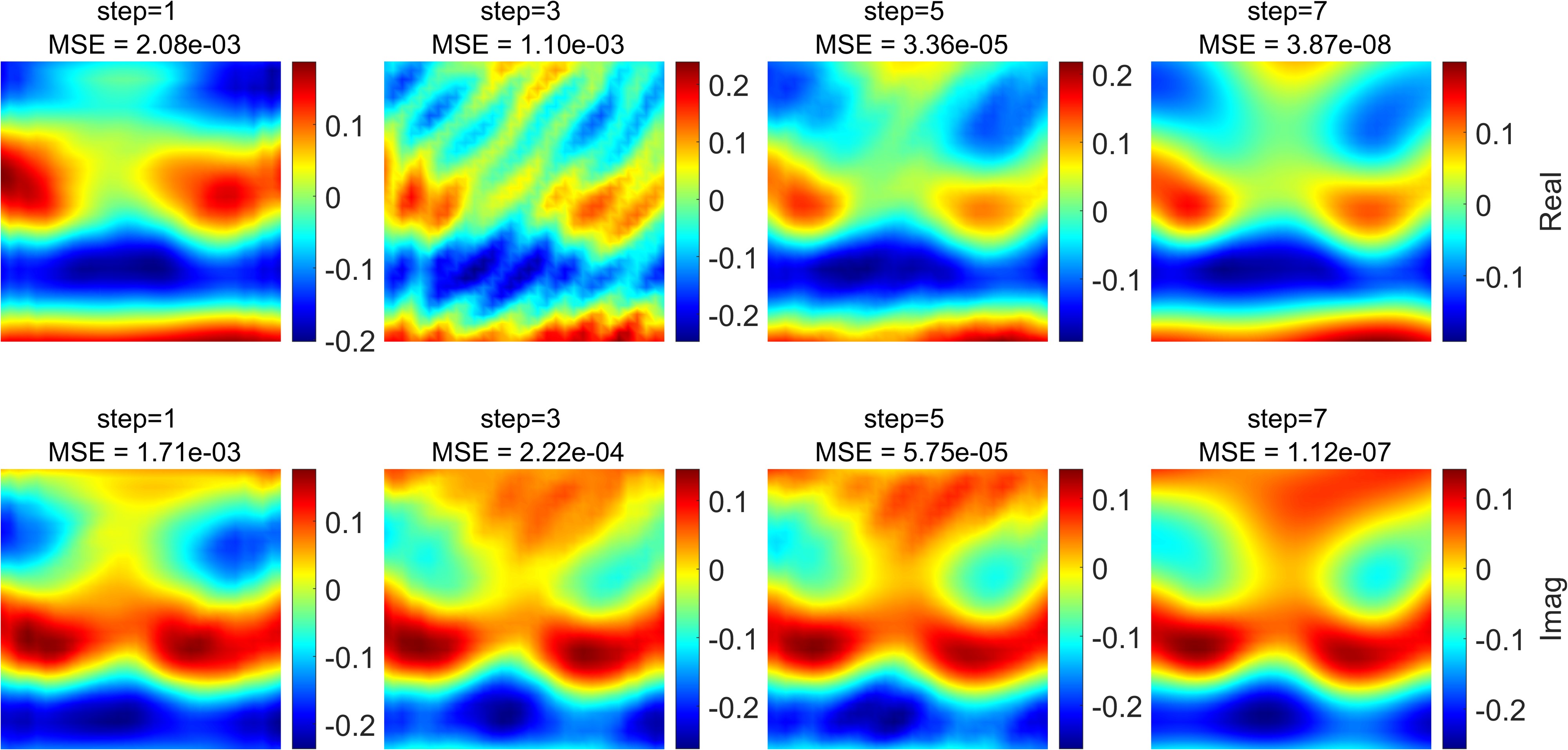}}
	\subfigure[]
	{\includegraphics[width=0.99\linewidth]{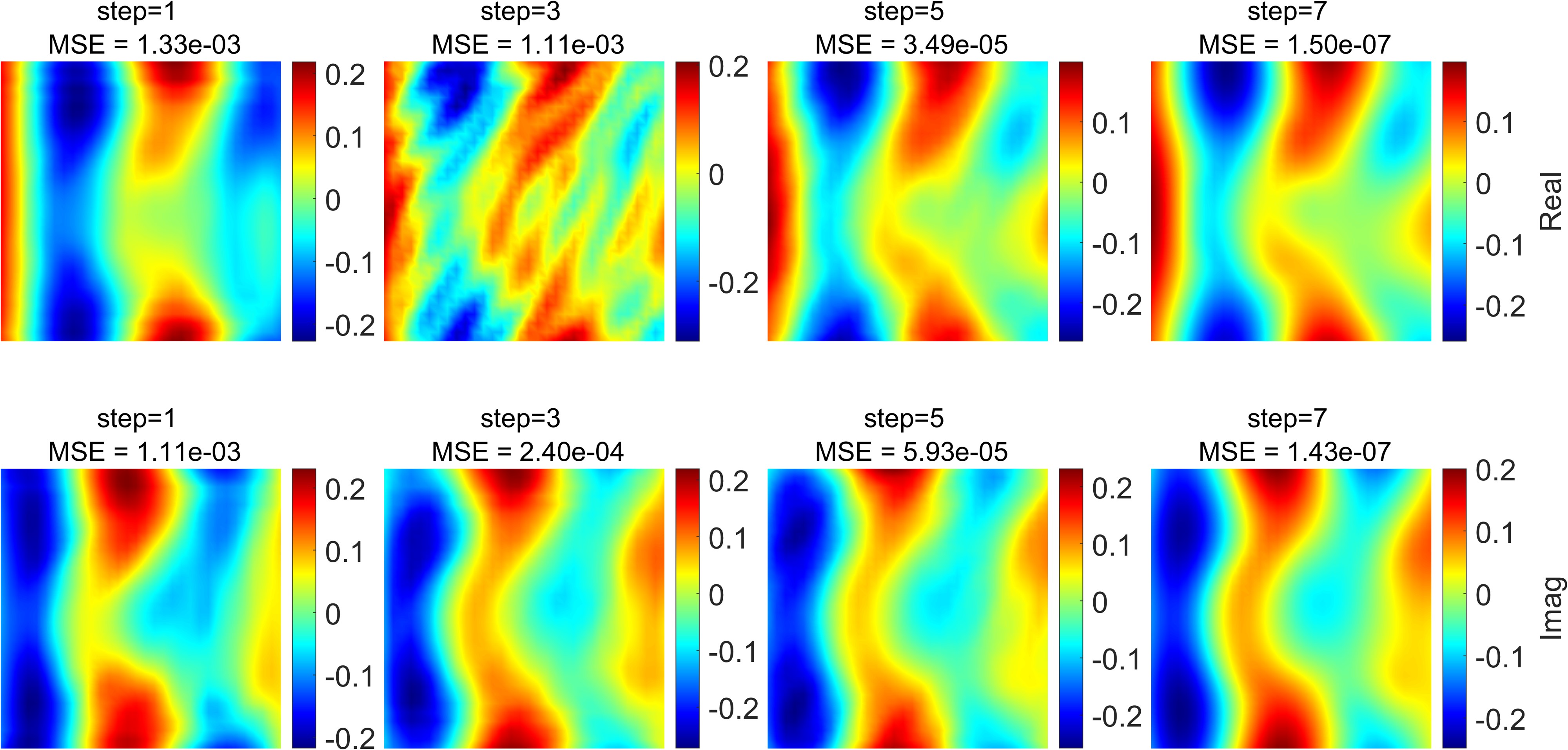}}	
	\caption{Lossy scatterer cases:  updated total fields in each iteration computed by NiPhiResNet. (a) and (b) are updated fields corresponding to \fig{nsipisrllossy}. From left to right: total fields computed by NiPhiResNet $E^{tot}_{Ni}$ in the first, third, fifth and seventh iteration. The first and second row are the real and imaginary part.}
	\label{nsipisrllossyall}
\end{figure}

\begin{figure}
	\centering
	\includegraphics[width=0.95\linewidth]{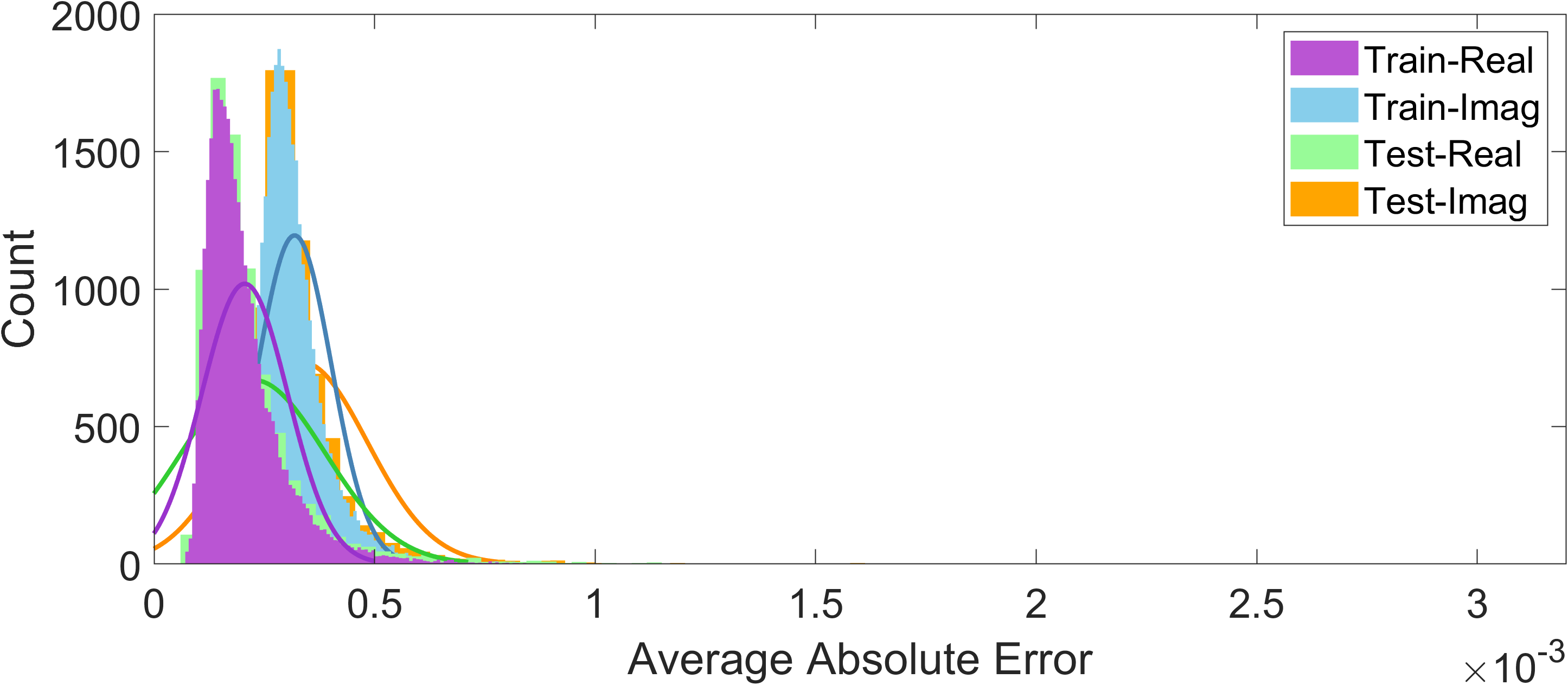}
	\caption{Lossy scatterer cases: histograms of mean absolute error of $E^{tot}_{Ni}$. Train-Real and Train-Imag are MAE histograms of real and imaginary parts of $E^{tot}_{Ni}$ in the training data set (means are $2.06 \times 10^{-4}$ and $3.19 \times 10^{-4}$, stds are $9.8 \times 10^{-5}$ and $8.3 \times 10^{-5}$); Test-Real and Test-Imag are MAE histograms of real and imaginary parts of $E^{tot}_{Ni}$ in the testing data set (means are $2.25 \times 10^{-4}$ and $3.35 \times 10^{-4}$, stds are $1.63 \times 10^{-4}$ and $1.48 \times 10^{-4}$).}
	\label{nsipisrllossy-hist}
\end{figure}

NiPhiResNet has seven iterations and the MSE convergence curve is shown in \fig{nsipisrllossy-loss}.
With the converged average MSE below $1.567 \times 10^{-7}$, NiPhiResNet shows better computing precisions than SiPhiResNet.
\fig{nsipisrllossy} illustrates two results randomly chosen from the testing data set.
The absolute error distributions show small discrepancies between $E^{tot}_{MoM}$ and $E^{tot}_{Si}$.
The field of each iteration is depicted in \fig{nsipisrllossyall}.
The updated total field is refined gradually with  iterations.
\fig{nsipisrllossy-hist} charts the MAE histogram of total fields solved by NiPhiResNet.
The MAE's mean and std of testing data set are a little higher than the ones of training data set while the discrepancy is very small and acceptable.
Small stds of MAE also validates the stable computing precisions of NiPhiResNet.
\subsubsection{Impact of Data Set Size on Performance}
\begin{figure}
	\centering
	\includegraphics[width=0.9\linewidth]{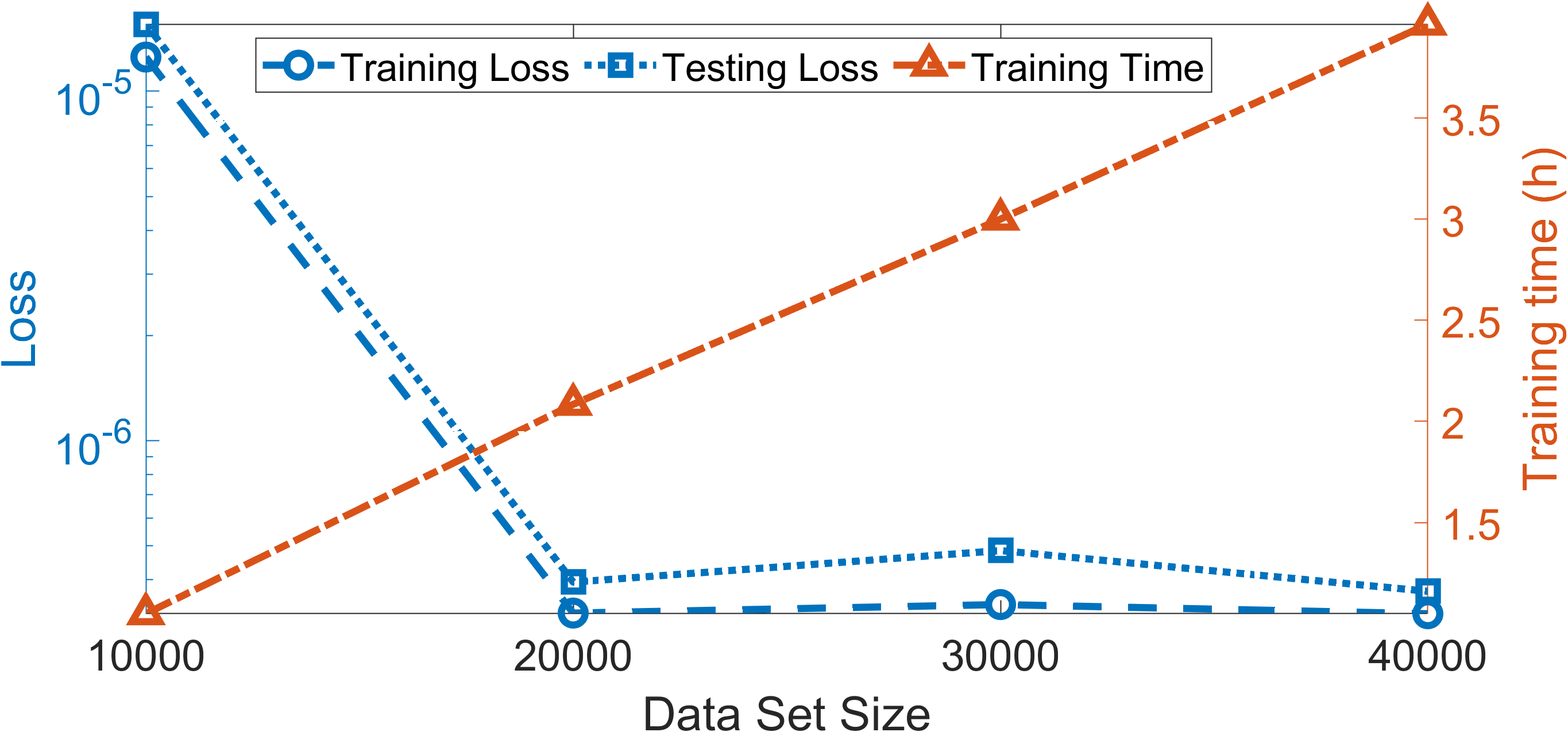}
	\caption{The final losses and training time of NiPhiResNet for solving VIEs of lossy scatterers when trained with different data set sizes.}
	\label{datasetloss}
\end{figure}

NiPhiResNet is taken as an example to investigate the impact of data set size on the performance of NiPhiResNet.
A total of 40000 data samples are generated and divided into 4 data sets of which have 10000, 20000, 30000 and 40000 data samples respectively.
Each data set are separated for training and testing according to 80\%-20\% ratio.
With different data set sizes, NiPhiResNet is trained under the same training configuration.
\fig{datasetloss} shows the corresponding final losses of NiPhiResNet.
Both training and testing losses decrease with the increase of the data set size.
Trained with 40000 data samples, NiPhiResNet demonstrates the best performance, and the difference between training and testing losses is also smallest, which indicates NiPhiResNet is trained efficiently.
The training time of each data set size is also plotted in \fig{datasetloss}.
It can be observed that the training time is generally proportional to the data set size. 
\subsubsection{Generalization Ability on Contrast Shape}
\begin{figure}
	\centering
	\includegraphics[width=0.9\linewidth]{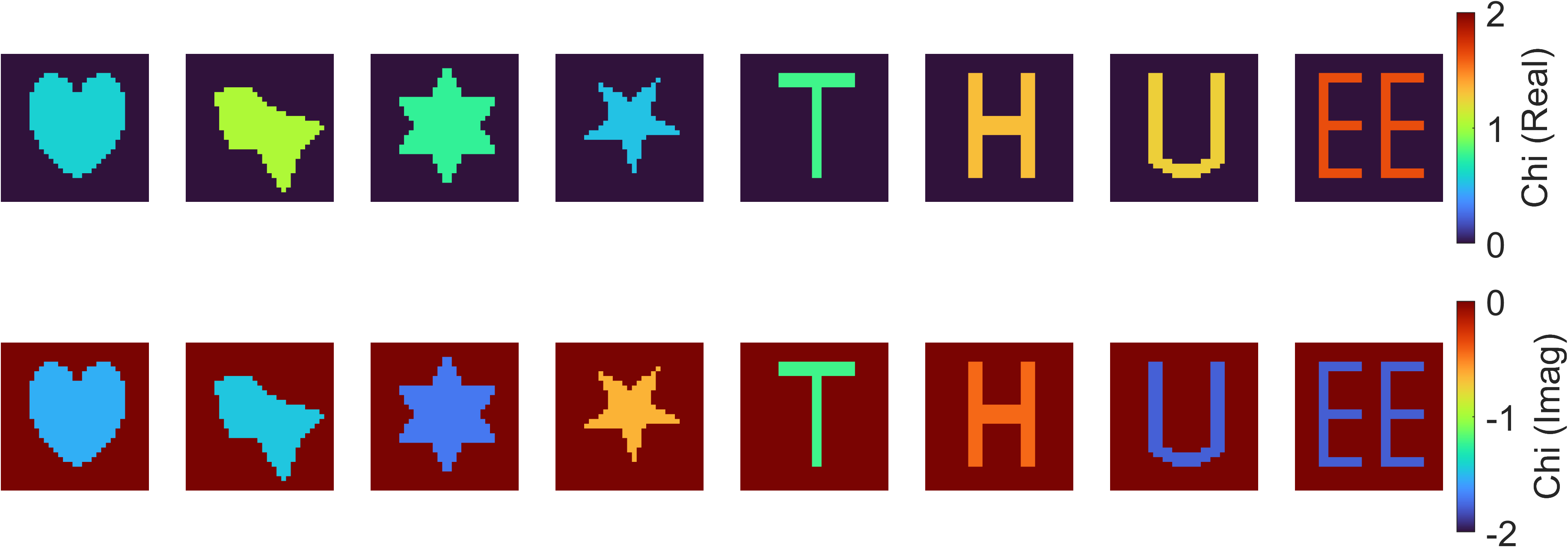}
	\caption{Lossy contrast examples of generalization validation data set for unseen contrast shapes. The first and second row are real and imaginary parts.}
	\label{genlossychi}
\end{figure}

\begin{figure}
	\centering
	\subfigure[ ]
	{\includegraphics[width=0.99\linewidth]{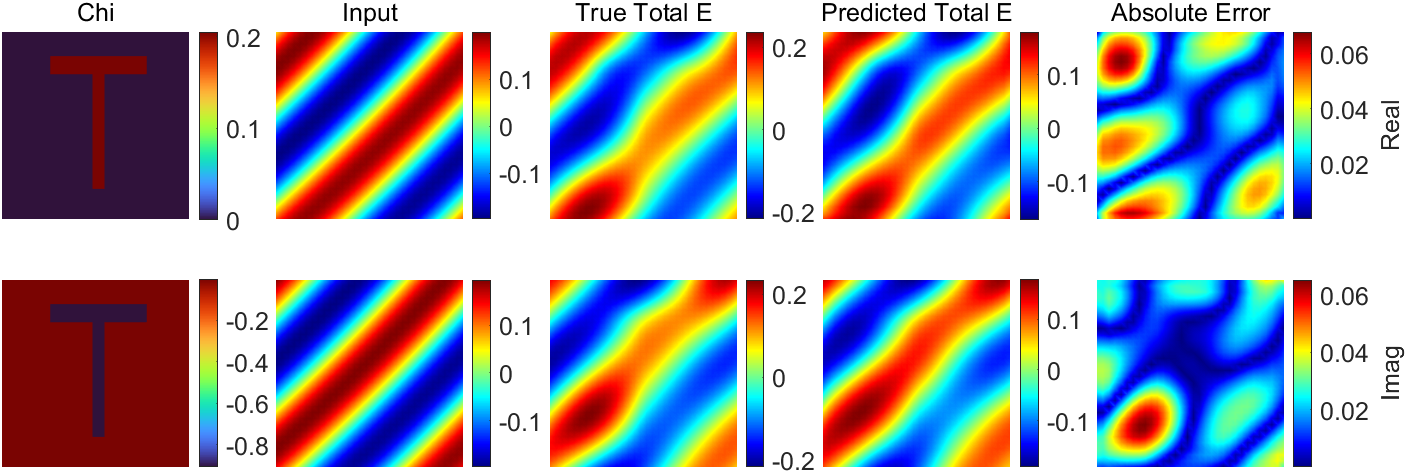}}
	\subfigure[ ]
	{\includegraphics[width=0.99\linewidth]{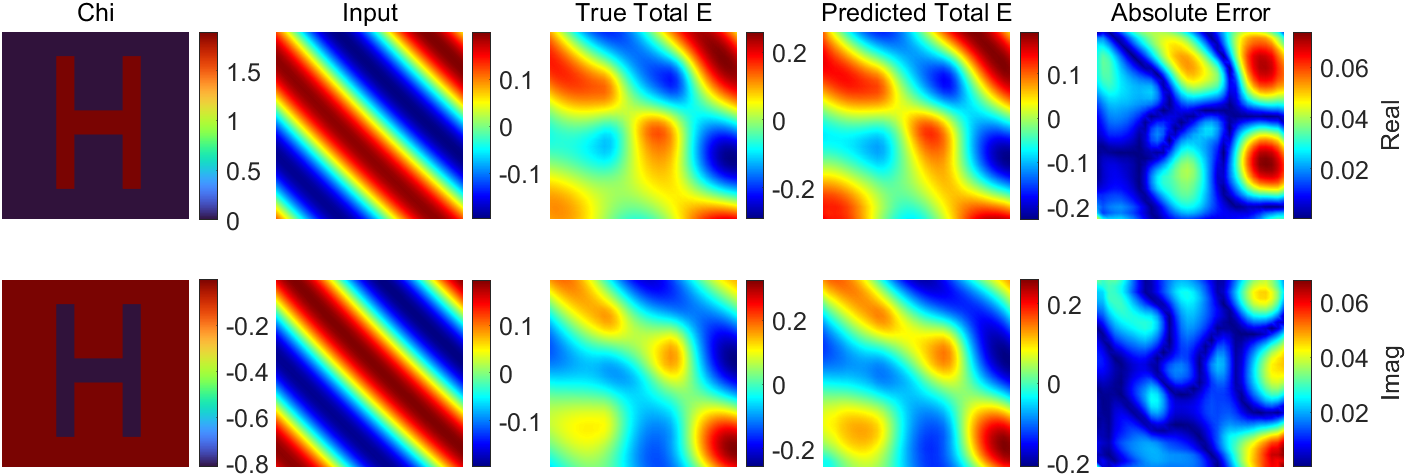}}
	\centering
	\subfigure[ ]
	{\includegraphics[width=0.99\linewidth]{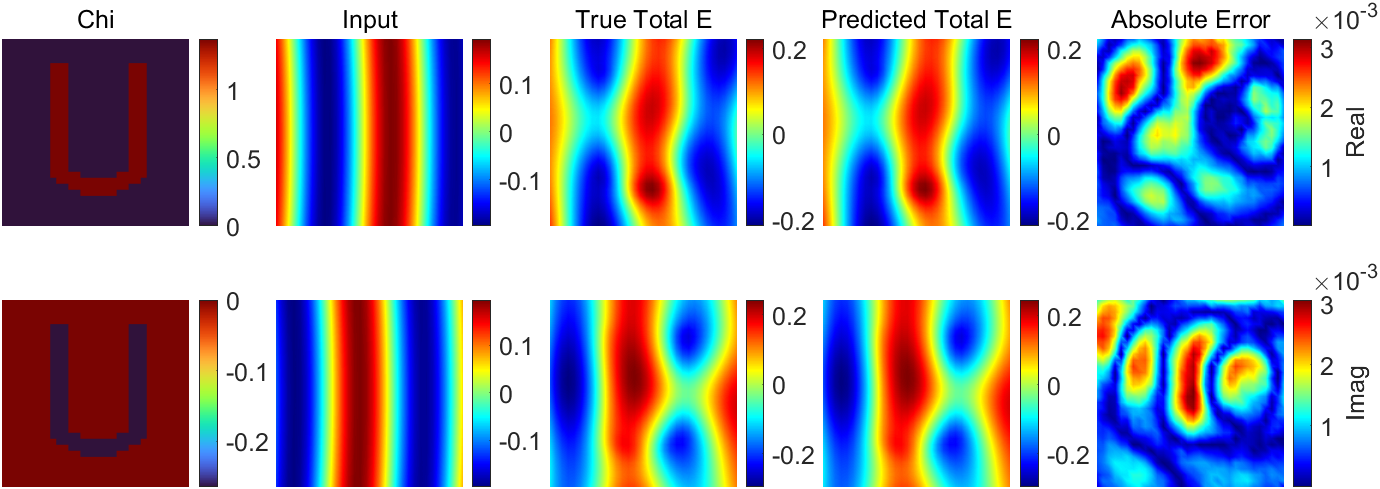}}
	\subfigure[ ]
	{\includegraphics[width=0.99\linewidth]{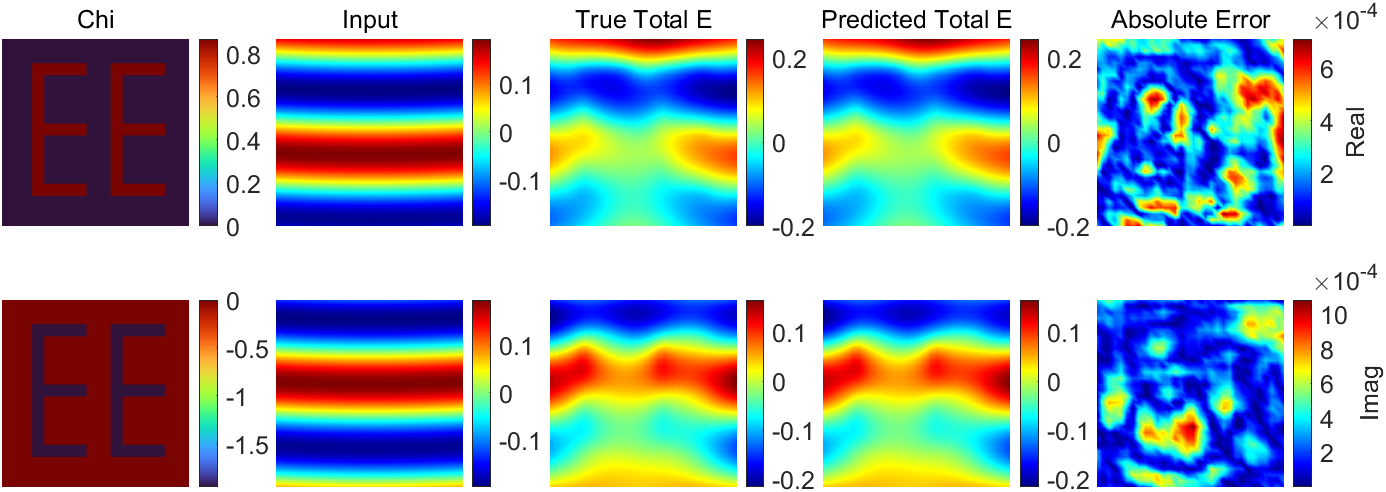}}
	\caption{Lossy scatterer cases: predicted total fields of SiPhiResNet ((a), (b)) and NiPhiResNet ((c), (d)) in the unseen contrast shape generalization validation data set. In each sub-panel, from left to right are contrast, initial guess, true total field,  predicted total field, and their absolute error distribution. The first and second row are real and imaginary part.}
	\label{lossy-all-shape}
\end{figure}

The generalization abilities of SiPhiResNet and NiPhiResNet on the unseen contrast shapes are verified.
Same as the cases of lossless scatterers, 8 types of unseen contrast shapes are taken into account, as shown in \fig{genlossychi}.
The real and imaginary parts of contrasts are randomly selected in $[0,2]$ and $[-2, 0]$ respectively.
The validation data set has a total of 320 data samples and 40 data samples are generated for each shape type.
The MAEs of total fields solved by SiPhiResNet and NiPhiResNet are compared in \tab{table-lossy}.
The MAE's stds of SiPhiResNet are higher than the ones of NiPhiResNet that indicates NiPhiResNet has more stable computing precisions on unseen contrast shapes.  
\fig{lossy-all-shape} illustrates the randomly chosen total fields solved by SiPhiResNet and NiPhiResNet.
\begin{table}
	\renewcommand{\arraystretch}{1.3}
	\caption{Performance of SiPhiResNet and NiPhiResNet on unseen lossy contrast shapes}
	\label{table-lossy}
	\centering
	\begin{threeparttable}
		\begin{tabular}{ccc}
			\toprule
			Method & MAE-R\tnote{*} \, (mean/stds)  & MAE-I\tnote{**} \, (mean/stds) \\
			\midrule
			SiPhiResNet & $1.38^{\times10^{-2}}/7.83^{\times10^{-3}}$ & $ 1.25^{\times10^{-2}}/4.15^{\times10^{-3}}$\\
			NiPhiResNet & $5.06^{\times 10^{-4}}/1.11^{\times10^{-3}}$ & $6.53^{\times10^{-4}}/1.33^{\times10^{-3}}$ \\
			\bottomrule
		\end{tabular}
		\begin{tablenotes}
			\setlength{\multicolsep}{0cm}
				\item[*] MAE of total field real parts 	
				\item[**] MAE of total field imaginary parts
		\end{tablenotes}
	\end{threeparttable}
\end{table}
\subsubsection{Generalization Ability on Incident Frequency}
\begin{figure}
	\centering
	\subfigure[SiPhiResNet MSE]
	{\includegraphics[width=1\linewidth]{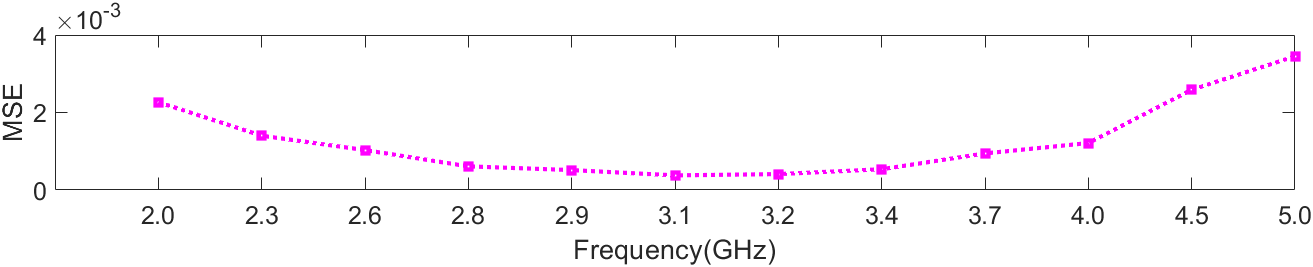}}
	\subfigure[NiPhiResNet MSE]
	{\includegraphics[width=1\linewidth]{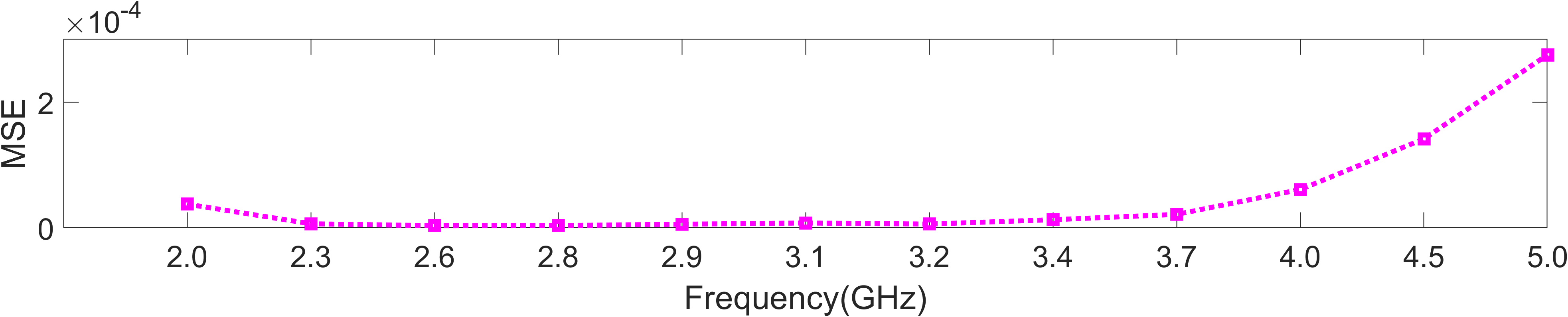}}	
	\caption{MSEs of SiPhiResNet and NiPhiResNet for solving VIEs of lossy scatterers at different frequencies.}
	\label{lossy_gen_loss}
\end{figure}

In this section, the generalization abilities of SiPhiResNet and NiPhiResNet are verified on incident frequencies unseen at training time.
The incident frequency is fixed at 3GHz when generating training and testing data sets.
Here, we consider 12 incident frequencies from 2GHz to 5GHz with a step of 0.3GHz.
The incident angles vary in $[0^{\circ}, 90^{\circ}, 180^{\circ}, 270^{\circ}]$, as shown in \fig{lossypinc}.
For each incident frequency, we generate 320 data samples, and the contrast shapes keep consistent with the contrast generalization validation data set, as shown in \fig{genlossychi}.
The MSEs of SiPhiResNet and NiPhiResNet at different incident frequencies are plotted in \fig{lossy_gen_loss}.
Both of them maintain a good level of computing precision across a wide band of incident frequencies. 
\par
\begin{figure}
	\centering
	\subfigure[ ]
	{\includegraphics[width=0.99\linewidth]{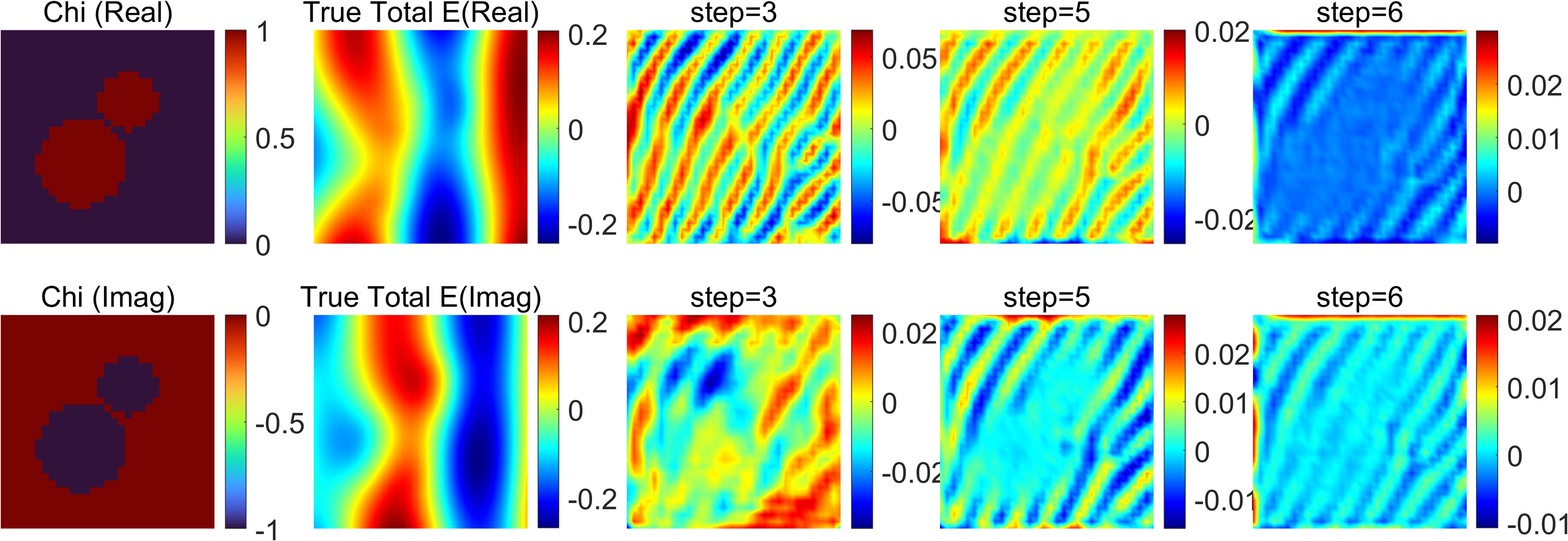}}
	\subfigure[ ]
	{\includegraphics[width=0.99\linewidth]{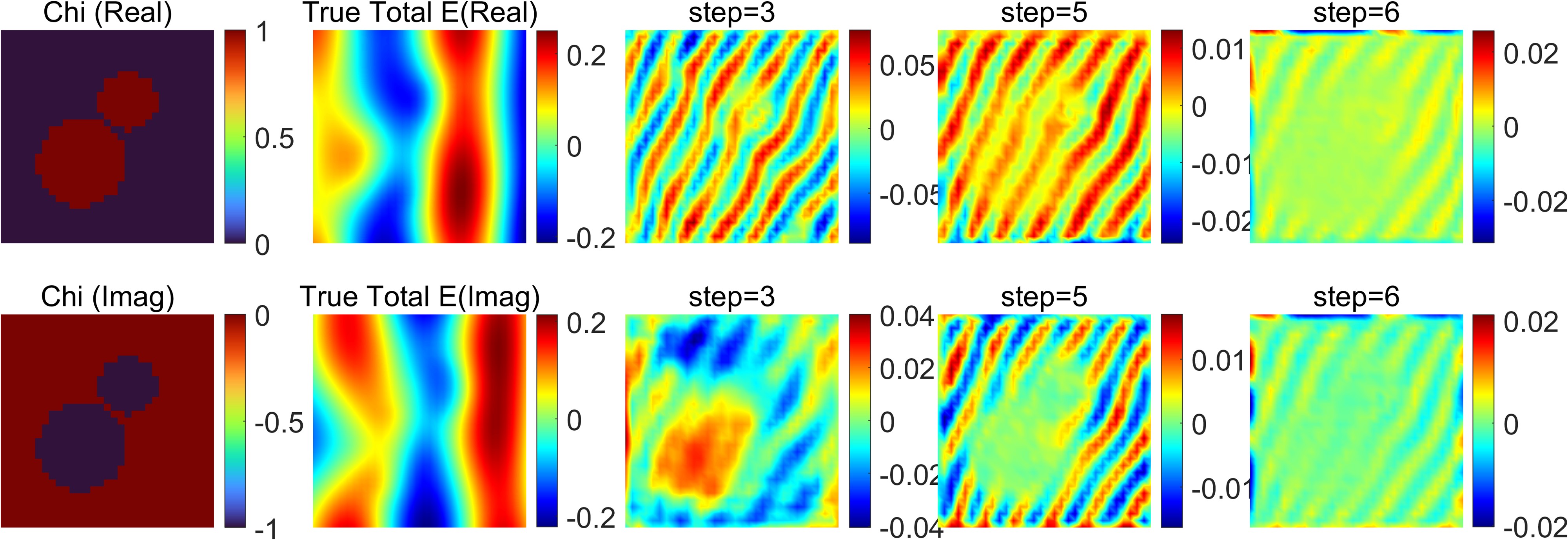}}
	\caption{Contrasts, true total fields and updated residuals at incident frequency of $2.6$GHz (a) and $2.9$GHz (b). In each sub-panel, the first and second row are real and imaginary parts; from left to right, each column is contrast, true total field, residual at the second, fifth and sixth iteration.}
	\label{freqres}
\end{figure}

 
Furthermore, we study the reasons behind the good generalization ability of PhiSRL on different incident frequencies. 
Two cylinders are considered with their contrasts fixed as $1-j1$. The incident angle is set as $0^{\circ}$ and two different frequencies are employed including $2.6$GHz and $2.9$GHz. 
NiPhiResNet is applied to perform predictions and \fig{freqres} illustrates the corresponding contrasts, true total fields, and updated residuals at different iterations.
In PhiSRL, the residuals are input of CNNs and can be calculated via $\mathbb{R}=\mathbf{E}^{inc}-	(\mathbf{I}-\mathbb{G}_D\cdot\boldsymbol{\chi})\cdot\mathbf{E}^{tot}$.
The calculated residuals can effectively alleviate the effects caused by the varying incident frequencies, which is further verified in \fig{freqres}.
It can be observed in \fig{freqres} that the updated residuals share similar distributions at the same iteration despite the different incident frequencies.
This simplifies the learning burden of CNNs and enables a better generalization ability on a wide range of incident frequencies.
\subsection{Generalization on Contrast Value}
\begin{figure}
	\centering
	\subfigure[real parts of contrast, true and predicted total fields ]
	{\includegraphics[width=0.9\linewidth]{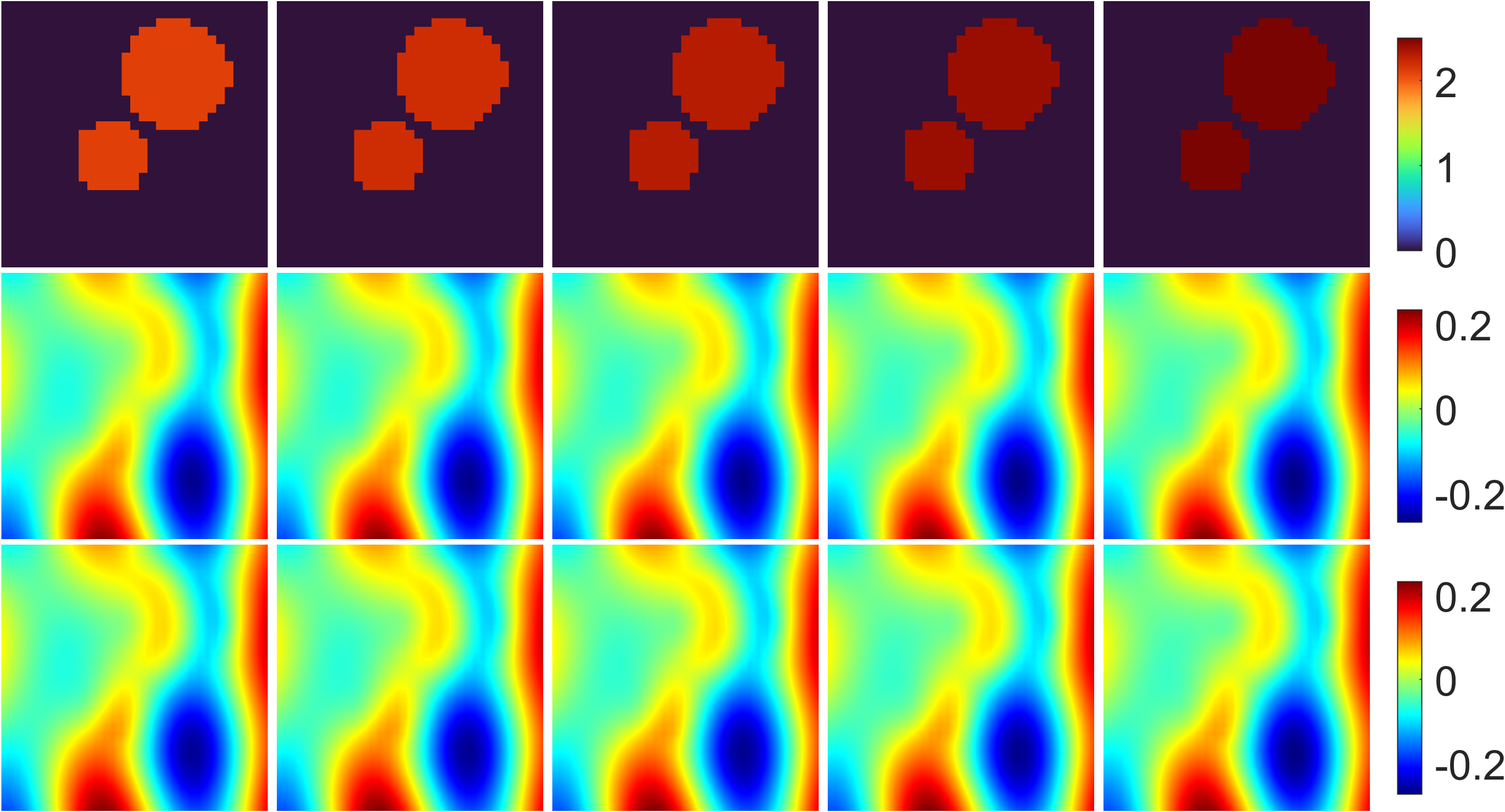}}
	\subfigure[imaginary parts of contrast, true and predicted total fields ]
	{\includegraphics[width=0.9\linewidth]{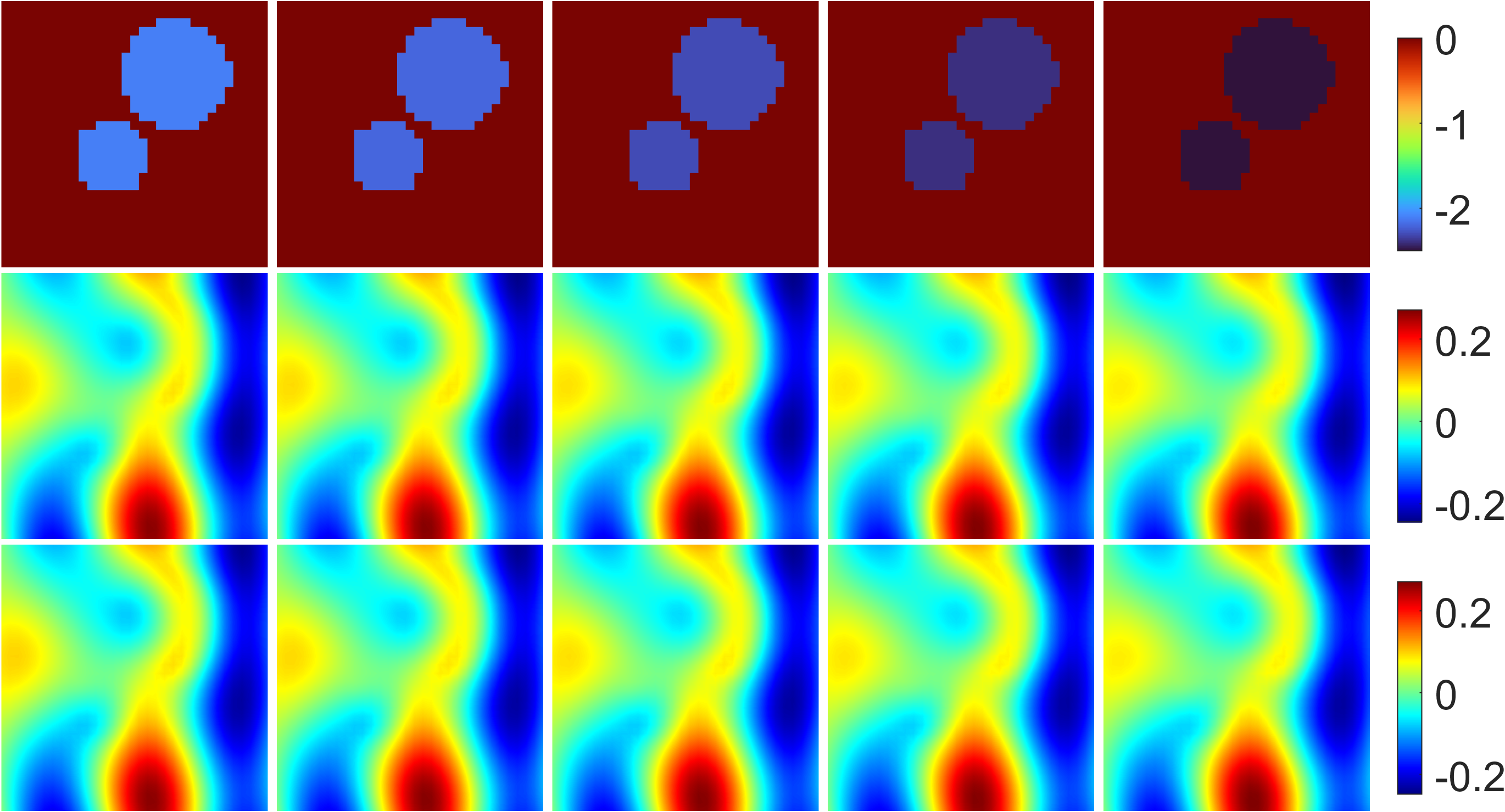}}
	\centering
	\subfigure[MSE of different contrast values ]
	{\includegraphics[width=1\linewidth]{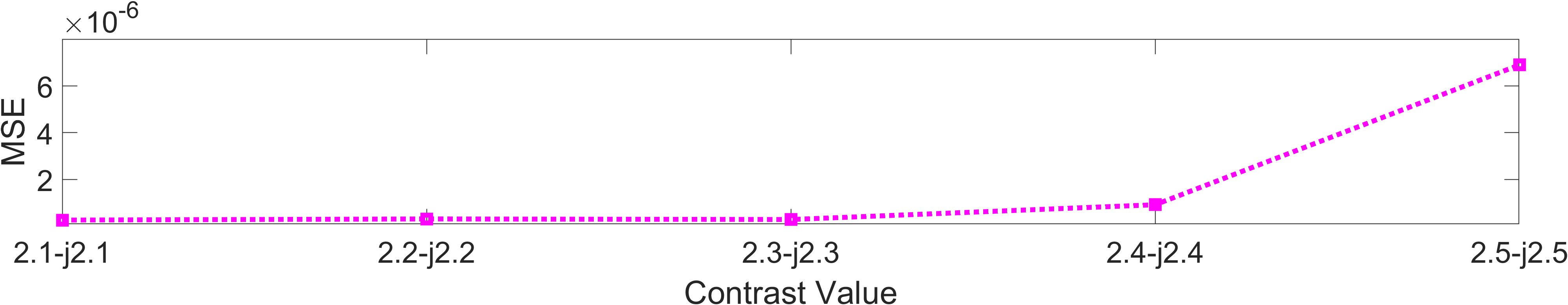}}
	\caption{Generalization of NiPhiResNet on different contrast values. In (a) and (b), the first, second and third row are contrast, true and predicted total fields, the contrast value of each column is: (from left to right) $2.1-j2.1$, $2.2-j2.2$, $2.3-j2.3$, $2.4-j2.4$, $2.5-j2.5$, and the MSEs between true and predicted total fields are plotted in (c).}
	\label{chitest}
\end{figure}

The generalization ability of PhiSRL is verified on the different contrast values that are unseen at training time.
Here, we take NiPhiResNet as an example. 
When generating training data, the range of real and imaginary parts of contrast are $[0, 2]$ and $[-2, 0]$ respectively.
Two cylinders are taken into account, of which the contrast values are out of range, including $2.1-j2.1$, $2.2-j2.2$, $2.3-j2.3$, $2.4-j2.4$ and $2.5-j2.5$. The incident angle is fixed as $0^{\circ}$. \fig{chitest} compares ground truth and NiPhiResNet predictions. NiPhiResNet demonstrates a stable performance on the out-of-range contrast values, which further validates its generalization ability.
\subsection{Comparison with the vanilla ResNet}
\begin{figure}
	\centering
	\includegraphics[width=0.9\linewidth]{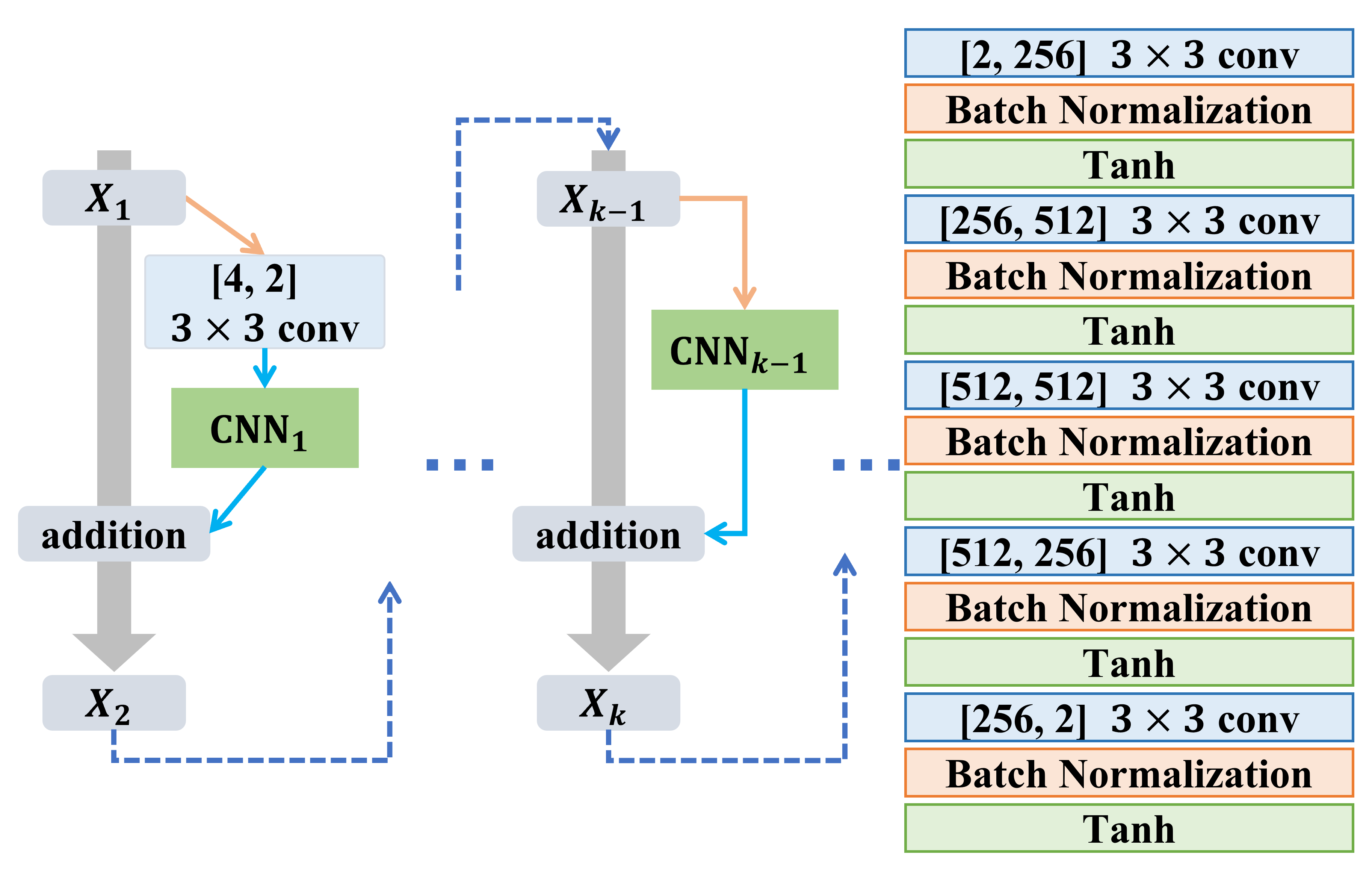}
	\caption{Schematics of the vanilla ResNet with identity mappings. The applied CNN of each iteration is also depicted, which has the same structure but different parameters.}
	\label{resnetonly}
\end{figure}

The performance of PhiSRL is also compared to the ResNet with identity mappings which is called as the vanilla ResNet for brevity. Due to that the vanilla ResNet applies an independent CNN in each block, NiPhiResNet is taken as an example here. The vanilla ResNet has 7 iterations and the employed CNN shares the same structure as the one in PhiSRL. It takes as input the concatenation of the contrast and incident field arrays.
The input channel is 4 but the channel of CNN prediction needs to be 2. Thus, a $3\times3$ convolution is added in the first block to change the channel from 4 to 2, as shown in \fig{resnetonly}. It should be emphasized that the main difference between the vanilla ResNet and NiPhiResNet is whether to incorporate the numerical calculation of $\mathbb R$.
The training configuration and data set of the vanilla ResNet are also kept the same as the ones of NiPhiResNet. After 200 training epochs, the final training and testing MSEs of the vanilla ResNet are $2.74\times10^{-5}$ and $3.67\times10^{-5}$ respectively, while those for NiPhiResNet are $1.16\times10^{-7}$ and $1.57\times10^{-7}$.
Such comparison not only validates the efficacy of PhiSRL, but also demonstrates the benefits of incorporating the numerical computation of $\mathbb R$.
\begin{table}
	\renewcommand{\arraystretch}{1.3}
	\caption{Comparisons of Computing Performance }
	\label{table02}
	\centering
	\begin{threeparttable}
		\begin{tabular}{cccc}
			\toprule
			Scatterers & Method & MAE (real/imag)  & Time(s)/Reduction \\
			\midrule
			lossless  & BiCG2 & $1.85^{\times10^{-2}}/1.85^{\times10^{-2}}$ & $4.40^{\times 10^{-3}}$\\
			lossless  & BiCG3 & $8.06^{\times10^{-3}}/8.03^{\times10^{-3}}$ & $5.18^{\times 10^{-3}}$\\
			lossless  & BiCG3\tnote{GPU} & $8.06^{\times10^{-3}}/8.03^{\times10^{-3}}$ & $1.42{\times 10^{-2}}$\\
			lossless  & SiPhiResNet  & $1.31^{\times 10^{-2}}/1.34^{\times10^{-2}}$ & $2.30^{\times10^{-3}}/55.6\%$  \\
			lossless  & BiCG6 &  $3.84^{\times10^{-4}}/3.85^{\times10^{-4}}$ & $7.52^{\times10^{-3}}$  \\
			lossless  & BiCG7&  $1.43^{\times10^{-4}}/1.44^{\times10^{-4}}$ & $8.65^{\times10^{-3}}$  \\
			lossless  & BiCG7\tnote{GPU} &  $1.43^{\times10^{-4}}/1.44^{\times10^{-4}}$ & $2.21^{\times10^{-2}}$  \\
			lossless  & NiPhiResNet  & $3.56^{\times10^{-4}}/3.87^{\times10^{-4}}$  & $5.08^{\times10^{-4}}/94.1\%$\\
			lossy  & BiCG2 & $2.00^{\times10^{-2}}/1.97^{\times10^{-2}}$  & $4.54^{\times10^{-3}}$\\
			lossy  & BiCG3 & $8.83^{\times10^{-3}}/8.76^{\times10^{-3}}$  & $6.36^{\times10^{-3}}$\\
			lossy  & BiCG3\tnote{GPU} & $8.83^{\times10^{-3}}/8.76^{\times10^{-3}}$  & $1.37^{\times10^{-2}}$\\
			lossy  & SiPhiResNet  & $7.42^{\times10^{-3}}/1.04^{\times10^{-2}}$ & $2.30^{\times10^{-3}}$/63.8\% \\
			lossy  & BiCG6 & $2.44^{\times10^{-4}}/2.44^{\times10^{-4}}$ & $7.92^{\times10^{-3}}$ \\
			lossy  & BiCG7 & $6.95^{\times10^{-5}}/6.99^{\times10^{-5}}$ & $8.38^{\times10^{-3}}$ \\
			lossy  & BiCG7\tnote{GPU} & $6.95^{\times10^{-5}}/6.99^{\times10^{-5}}$ & $2.11^{\times10^{-2}}$ \\
			lossy  & NiPhiResNet  & $2.09^{\times10^{-4}}/3.22^{\times10^{-4}}$  & $5.08^{\times10^{-4}}/93.9\%$ \\
			\bottomrule
		\end{tabular}
		\begin{tablenotes}
			\item BiCG2,3,6,7 denote BiCGSTAB with 2, 3, 6, 7 iterations. BiCG3\tnote{GPU} and BiCG7\tnote{GPU} denote BiCGSTAB  computed on GPU. The time reductions of SiPhiResNet and NiPhiResNet are compared to BiCG3 and BiCG7 respectively.
		\end{tablenotes}
	\end{threeparttable}
\end{table}
\subsection{Comparisons of Computing Performance}
In this section, we compare the computing performance of SiPhiResNet, NiPhiResNet and biconjugate gradient stabilized method (BiCGSTAB).
BiCGSTAB is a widely used Krylov subspace method for solving linear matrix equations\cite{saad2003iterative}. 
It is applied to evaluate the computing performance of SiPhiResNet and NiPhiResNet.
Here, we adopt the data sets employed in Section \ref{lossless} and Section \ref{lossy} and each of them consists of 40000 data samples.
In BiCGSTAB, the relative residual is an important stop criterion of solving matrix equations $Ax=b$ which can be written as:
\begin{equation}
	\zeta = \frac{||Ax-b||}{||b||}\,.
\end{equation}
The total fields solved by BiCGSTAB are taken as ground truth with relative residuals $\zeta$ below $1\times 10^{-8}$.
As SiPhiResNet and NiPhiResNet have 3 and 7 iterations respectively, BiCGSTAB is applied with $3$ and $7$ iterations.
Then we evaluate the MAEs between the ground truth and the total fields solved by SiPhiResNet, NiPhiResNet and BiCGSTAB, as shown in \tab{table02}.
Both SiPhiResNet and NiPhiResNet maintain a good level of computing precisions compared to BiCGSTAB.
\tab{table02} also demonstrates the computing time of SiPhiResNet, NiPhiResNet and BiCGSTAB.
The computing platform of BiCGSTAB is Intel(R) Xeon(R) Gold 5118 CPU @ 2.30GHz.
The BiCGSTAB owns better precisions than SiPhiResNet and NiPhiResNet, while both SiPhiResNet and NiPhiResNet demonstrate faster computing speed than BiCGSTAB.
NiPhiResNet needs less computing time than SiPhiResNet due to that NiPhiResNet applies simpler and smaller CNN architectures.

Furthermore, the computing time of PhiSRL is compared to BiCGSTAB when they achieve the same MAE level. The BiCGSTAB of 2 and 6 iterations can achieve the similar MAE level with SiPhiResNet and NiPhiResNet respectively. The corresponding computational time is summarized in \tab{table02}. It can be observed that SiPhiResNet and NiPhiResNet still maintain a faster computing speed.
Besides, we implement BiCGSTAB on one Nvidia V100 GPU that is the same computing platform for PhiSRL.
The problem scale is small in this paper and the communication between CPU and GPU is time-consuming. The BiCGSTAB on GPU does not show the significant acceleration, as shown in \tab{table02}.
\subsection{Discussions}
The word "physics-informed" in PhiSRL is intended to convey that the supervised residual learning is informed of physics by incorporating physical operators. Such incorporation not only guarantees the interpretability of PhiSRL, but also simplifies and specifies the learning tasks of the applied CNNs. 
The universality of PhiSRL is validated by solving VIEs of lossless and lossy scatterers.
Two different architectures of PhiSRL are discussed including SiPhiResNet and NiPhiResNet. They both demonstrate a good computational precision after adequate training.
NiPhiResNet shows better computing precisions than SiPhiResNet.
It can be tentatively concluded that it is recommended to apply independent CNNs to learn the update rules in PhiSRL.
Numerical results also shows that PhiSRL has a good generalization ability on contrast shape, contrast value and incident frequency.
Benefiting from the parallel computing on GPUs, PhiSRL shows a great potential to achieve a significant reduction in computing time. 
However, the generation of data samples and the training of PhiSRL are time-consuming. The numerical calculation of residuals is based on the dense matrix-vector multiplication in this paper and it needs to be further accelerated by fast Fourier transform in the case of large-scale problems.
\section{Conclusion}
In this paper, we propose the physics-informed supervised residual learning to enable an effective, robust and general deep learning framework for electromagnetic modeling. 
Motivated by the mathematical link between ResNet and fixed-point iteration method, 
PhiSRL is designed on top of fixed-point iteration method by applying CNNs to learn update rules at each iteration.
SiPhiResNet and NiPhiResNet are proposed based on the stationary and non-stationary scheme of fixed-point iteration method.
With the aim of solving a linear system of matrix equations, PhiSRL is not limited to a specific EM forward modeling problem and can be easily extended to other scenarios. 
The universality of PhiSRL is validated by solving VIE of lossless and lossy scatterers.
The MSEs of SiPhiResNet and NiPhiResNet achieve below $3.152 \times 10^{-4}$ and $4.8925 \times 10^{-7}$ when solving VIEs of lossless scatterers, and converge below $1.2775 \times 10^{-4}$ and $1.567 \times 10^{-7}$ when solving VIEs of lossy scatterers.
PhiSRL further proves its generalization ability on the contrast shapes, incident frequencies, out-of-range contrast values that are unseen at training time.
Numerical experiments demonstrate that the non-stationary scheme of PhiSRL (NiPhiResNet) can achieve better computing precisions with an independent CNN learning update rule of single iteration. 

\bibliographystyle{IEEEtran}
\bibliography{IEEEabrv,ref}
\end{document}